\documentclass[prb,amsmath,amssymb,floatfix,superscriptaddress,showpacs]{revtex4}
\usepackage[toc,page]{appendix}
\usepackage{graphicx,pdflscape,epsf,epsfig,amsmath}
\usepackage{alltt,dsfont,amsmath,amssymb,bm}
\usepackage{float}

\usepackage{verbatim}

\usepackage{hyperref}
\usepackage{color}
\hypersetup{
    colorlinks,
    citecolor=red,
    filecolor=cyan,
    linkcolor=blue,
    urlcolor=magenta
}

\newcommand{\ag}[1]{\textcolor{black}{#1}}

\newcommand{\rzjm}[1]{\textcolor{black}{#1}}

\newcommand{\omegabar}{\bar{\omega}\,}
\newcommand{\domega}{\delta\omega}
\newcommand{\sigmir}{\sigma_0}
\newcommand{\rhomir}{\rho_0}

\newcommand{\si}{\sigma}
\newcommand{\sib}{\bar{\sigma}}

\newcommand{\tJ}{\ $t$-$J$ \ }

\newcommand{\nn}{\nonumber}

\newcommand{\barray}{\begin{eqnarray}}
\newcommand{\earray}{\end{eqnarray}}
\newcommand{\beq}{\begin{eqnarray}}
\newcommand{\eeq}{\end{eqnarray}}

\newcommand{\disp}[1]{Eq.~(\ref{#1})}
\newcommand{\refdisp}[1]{Ref.~(\onlinecite{#1})}
\newcommand{\figdisp}[1]{Fig.~(\ref{#1})}

\newcommand{\condth}{\kappa_{\mathrm{th}}}

\newcommand{\vq}{\vec{q}}
\newcommand{\vR}{\vec{R}}
\newcommand{\vk}{\vec{k}}

\begin{document}

\title{ Transport and Optical Conductivity in the Hubbard Model:\\ 
A High-Temperature Expansion Perspective.}

\author{Edward Perepelitsky}
\affiliation{Coll\`ege de France, 11 place Marcelin Berthelot, 75005 Paris, France}
\affiliation{Centre de Physique Th\'eorique, \'Ecole Polytechnique, CNRS, Universit\'e Paris-Saclay, 91128 Palaiseau, France}
\author{Andrew Galatas}
\affiliation{Physics Department, University of California,  Santa Cruz, CA 95064 }
\author{Jernej Mravlje}
\affiliation{Jo\v{z}ef Stefan Institute, Jamova~39, Ljubljana, Slovenia}
\author{Rok \v{Z}itko}
\affiliation{Jo\v{z}ef Stefan Institute, Jamova~39, Ljubljana, Slovenia}
\author{Ehsan Khatami}
\affiliation{Department of Physics and Astronomy, San Jos\'e State University, San Jos\'e, CA 95192}
\author{ B Sriram Shastry }
\affiliation{Physics Department, University of California,  Santa Cruz, CA 95064 }
\author{Antoine Georges}
\affiliation{Coll\`ege de France, 11 place Marcelin Berthelot, 75005 Paris, France}
\affiliation{Centre de Physique Th\'eorique, \'Ecole Polytechnique, CNRS, Universit\'e Paris-Saclay, 91128 Palaiseau, France}
\affiliation{Department of Quantum Matter Physics, University of Geneva, 24 Quai Ernest-Ansermet, 1211 Geneva 4, Switzerland}

\begin{abstract}
We derive  analytical expressions for the spectral moments of the dynamical response functions of the Hubbard model using
the high-temperature series expansion. We consider generic dimension $d$ as well as the infinite-$d$ limit, arbitrary electron density $n$,
and both finite and infinite repulsion $U$. 
{We use moment-reconstruction methods to obtain the one-electron spectral function}, the self-energy, and the optical conductivity. 
They are all smooth functions at high-temperature {and, at large-$U$, they are featureless 
with characteristic widths of order the lattice hopping parameter $t$}. 
In the infinite-$d$ limit we compare the series expansion results with accurate numerical renormalization group
and interaction expansion quantum Monte-Carlo results. We find excellent agreement {down to surprisingly low temperatures, 
throughout most of the bad-metal regime which applies for $T\gtrsim (1-n)D$, the Brinkman-Rice scale.}
The resistivity increases linearly in $T$ at high-temperature without saturation. This results from the $1/T$ behaviour of the 
compressibility or kinetic energy, which play the role of the effective carrier number. In contrast, the scattering time (or 
diffusion constant) saturate at high-$T$. 
We find that $\sigma(n,T) \approx (1-n)\sigma(n=0,T)$ to a very good approximation for all $n$, 
with $\sigma(n=0,T)\propto t/T$ at high temperatures. The saturation at small $n$ occurs due
to a compensation between the
density-dependence of the effective number of carriers and that of the
scattering time.  
{The $T$-dependence of the resistivity displays a knee-like feature which signals}    
a cross-over to the intermediate-temperature regime where the diffusion constant (or scattering time) start increasing
with decreasing $T$. At high-temperatures, {the thermopower obeys the Heikes formula}, 
while the Wiedemann-Franz law is violated with the Lorenz number vanishing as $1/T^2$.
\ag{The relevance of our calculations to experiments probing high-temperature transport in materials with strong electronic 
correlations or ultra-cold atomic gases in an optical lattice is briefly discussed.}
\end{abstract}


\maketitle

\section{Introduction}

Electronic transport is one of the most poorly understood properties
of strongly-correlated electron systems.  A universally observed
characteristic is the absence of resistivity
saturation~\cite{gunnarsson_saturation_rmp}.  In contrast to systems
dominated by the electron-phonon
coupling~\cite{gunnarsson_saturation_rmp,CalandraGunnarsson,calandra03epl,berg_badmetal_2016},
most correlated systems are `bad metals' at high
temperature~\cite{ emery_kivelson_prl_1995,hussey_phil_mag_2004}. Their
resistivity exceeds the value which corresponds in the Drude-Boltzmann
picture to a mean-free path $l$ of the order of the lattice spacing
($k_F l \approx 1$). This characteristic Mott-Ioffe-Regel (MIR)
resistivity $\rho_\mathrm{MIR}$ (typically of the order of
100-300~$\mu\Omega$cm in oxides) is smoothly crossed at
high temperature and the resistivity remains metallic-like with
$d\rho/dT >0$. 

Early discussions in the context of underdoped cuprate superconductors
extrapolated the bad-metal behavior at high temperatures to a
low-temperature state that is an unusual metal without
quasiparticles~\cite{emery_kivelson_prl_1995}.  While the situation
remains controversial for
cuprates~\cite{mirzaei_fermiliquid_cuprates_pnas_2013,barisic_hidden_2015},
there is ample evidence that other transition metal-oxides which are
bad metals at high temperature do become good Fermi liquids at low
temperatures, the best characterized case being
Sr$_2$RuO$_4$~\cite{maeno_badmetal_prb_1998,Mackenzie-2003}.  
%

The bad-metal behavior and its implications for optical spectroscopy
and photoemission have been discussed by Deng et al.~\cite{Badmetal}
within the dynamical mean-field theory (DMFT\cite{georges_rmp_1996})
(see also Refs.~\onlinecite{MerinoMcKenzie,xu_hidden_prl_2013}). It
was demonstrated that quasiparticle excitations disappear only at
temperatures well above the low temperature Fermi liquid scale
$T_{\mathrm{FL}}$ below which the resistivity is quadratic in
temperature. For the large-$U$ doped Hubbard model, the scale at which
the MIR value is reached was identified as the Brinkman-Rice scale, 
of order $(1-n) D \gg T_{\mathrm{FL}}$ with $1-n$ the doping level
counted from half-filling and $D$ the (half-)bandwidth. 
The asymptotic high-temperature state was not, however, fully
characterized in that work 
\ag{(for an early study of high-$T$ transport within DMFT, see Refs.~\onlinecite{palsson_thermo_1998_prl,palsson_phd})
.}
Vu\v{c}i\v{c}evi\'{c} et al. \cite{vucicevic_badmetal_prl}, also
working within the DMFT framework, proposed a connection between bad
metallic behaviour and Mott quantum criticality and argued that the
temperature at which the MIR value is reached coincides with the
quantum Widom line associated with the metal-insulator transition
occurring at low temperatures.
Recently, incoherent transport within the bad-metal regime became the subject of renewed attention in the context of `holographic'  
approaches to hydrodynamics and transport in quantum fluids, see e.g. Refs.~\onlinecite{hartnoll_lectures} for reviews 
and Ref.~\onlinecite{Hartnoll, pakhira15} for a recent discussion of the incoherent regime of transport in this framework. 

In this article, we address these issues, and more generally the behaviour of transport and response functions, 
from a high-temperature perspective. We setup a general formalism for the high-temperature expansion 
of single-particle Green's functions and two-particle response functions and apply this formalism to the Hubbard model. 
The presented formalism allows for the analytical evaluation of moments of these correlation functions and 
allows us to make general statements on the behaviour of transport and optical conductivity in the high-temperature 
bad-metal regime, which shed light on transport mechanisms in this regime. High-temperature series for thermodynamic properties of the Hubbard model have been developed and applied by several 
authors \cite{Plischke,KuboTada,KuboTada2,PanWang,Bartkowiak,PLS,Scarolaetal,hiTDMFT}, but remarkably little previous work has been devoted to high-temperature 
series for dynamical response functions and transport\cite{Pairault}. In \refdisp{Ehsan}, the high-temperature series was applied to high orders to the single-particle spectral function of the two-dimensional infinite-$U$ Hubbard model. In \refdisp{hiTECFL}, the results were compared favorably to the ``Extremely Correlated Fermi Liquid" (ECFL) theory \cite{ECFL,ECFL2ndorder,ECFLcutoff} for the $\tJ$ model. {This method may also have applications to understanding the conductivity in models that display Many-body Localization \cite{MBLT1,MBLT2}}.

In the limit of infinite dimensions, we are able to obtain quantitative results, using moment-reconstruction methods, for the resistivity, thermal 
transport coefficients and frequency-dependence of the optical conductivity of the $U=\infty$ Hubbard model.  
These results are successfully compared to solutions of the DMFT equations using the numerical renormalization-group method. 

This article presents the formalism and its applications in details. In order to facilitate its reading, we 
provide in Sec.~\ref{sec:outline} an overview of its organization and of the main results.

\section{Overview of main results and outline}
\label{sec:outline}

\subsection{General formalism for high-temperature expansion of dynamic correlations.}


This article is based on a general formalism for expanding correlation functions as a series in inverse powers of temperature. 
The general formula for the spectral density $\chi_{O.O}^{''}(\vk,\omega)$ associated with the correlation function of a two-particle 
operator $\hat{O}_{\vk}$ reads, when $\hat{O}_{\vk}(\tau)$ is dimensionless: 
\beq 
\frac{1}{\omega}\chi_{O.O}^{''}(\vk,\omega)\,=\,\frac{1}{t^2}\sum_{i=1}^\infty \left(\frac{t}{T}\right)^i f^{(i)}(\vk,\frac{\omega}{t}).
\label{eq:general_beta_series_2pcf}
\eeq
In this expression, $t$ is an energy scale which can be conveniently chosen to be the hopping amplitude from the non-interacting 
part of the Hamiltonian. Unless specified otherwise we set $k_B=\hbar=1$ .  The $f^{(i)}$'s are dimensionless functions of momentum and normalized frequency. 
For the $U=\infty$ Hubbard model, they depend solely on the density $n$, while for the finite-$U$ model they also
depend on the dimensionless coupling $U/t$. 
The full frequency dependence of the functions $f^{(i)}$ cannot be derived in general from a high-temperature series approach,  
reflecting the fact that the low-frequency long-time regime $\omega < T$ is not directly accessible in this framework. Instead, we derive in Sec.~\ref{sec:2pcf} a 
general formula
for the {\it moments} of these functions, namely 
$\int d\omegabar\, \omegabar^{2n} f^{(i)}(\vk,\omegabar)$. 
Note that being a bosonic correlator, $\chi_{O.O}^{''}(\vk,\omega)/\omega$ is an even function 
of frequency, and therefore odd moments are zero.
For cases where enough moments can be calculated, we will attempt to
approximately reconstruct the frequency-dependence of $f^{(i)}(\vk,\frac{\omega}{t})$.

The expression of the moments is obtained at a fixed value of the electron density, which is defined per site as 
$n\equiv \langle \sum_\sigma c^\dagger_{i\sigma}c_{i\sigma}\rangle$
($n=1$ at half-filling). The chemical potential 
has a high-temperature expansion: 
\begin{equation}
\mu\,=\,T\,\sum_{i=0}^{\infty} \left(\frac{t}{T}\right)^i \bar{\mu}^{(i)}
\end{equation}
Since $\mu\propto T$ at high temperatures, 
\begin{equation}
\bar{\mu}\equiv \mu/T
\end{equation}
has a finite high-$T$ limit. The dominant term $\bar{\mu}^{(0)}$ is 
given by the atomic limit. For $U=\infty$ it reads 
\begin{equation}
\bar{\mu}^{(0)}=\lim_{T\rightarrow\infty}\mu/T = \ln \frac{n}{2(1-n)},
\end{equation}
while for finite $U$ it is 
\begin{equation}
\bar{\mu}^{(0)}= \ln \frac{n}{2-n}.
\end{equation}

A similar high-$T$ expansion can be performed for the single-particle Green's function and self-energy. The expansion applies for 
$\omega$ in the vicinity of $-\mu=-T\bar{\mu}$, and has the general form:
\beq
\rho_G(\vk,-\mu + \domega) = \frac{1}{t} \sum_{i=0}^\infty  \left(\frac{t}{T}\right)^i  g^{(i)}(\vk,\frac{\domega}{t})\,\,\,,\,\,\,
\rho_\Sigma(\vk,-\mu + \domega) = t \sum_{i=0}^\infty  \left(\frac{t}{T}\right)^i   h^{(i)}(\vk,\frac{\domega}{t}).
\label{eq:general_beta_series_1pcf}
\eeq
In this expression $\rho_G(\vk,\omega)\equiv -\mathrm{Im}G(\vk,\omega+i0^+)/\pi$ and 
$\rho_\Sigma(\vk,\omega)\equiv -\mathrm{Im}\Sigma(\vk,\omega+i0^+)/\pi$ are the spectral functions 
associated with the Green's function and self-energy, respectively. In
Sec.~\ref{sec:2pcf} we derive a general expression
for the moments $\int dx\, x^m g^{(i)}(\vk,x)$, from which the moments
of $h^{(i)}$ can also be obtained. 
The spectral functions $\rho_G$ and $\rho_\Sigma$ have  a non-trivial frequency-dependence at $T=\infty$, 
given by $g^{(0)}$ and $h^{(0)}$. Hence, the lower and upper Hubbard bands in the single-particle spectrum have a non-trivial 
shape and a finite width in this limit and do not simply reduce to the atomic limit, 
as previously pointed out by Palsson and Kotliar~\cite{palsson_phd,palsson_thermo_1998_prl}.  

Using these general formulas, we have derived explicit analytical expressions for several moments of the current-current correlation 
function (conductivity) $\sigma(\omega)$, $\rho_G$ and $\rho_\Sigma$. In the limit of infinite dimensions we managed to derive
a larger number of moments, allowing for approximate reconstruction of some of the dynamical correlation functions 
and comparison to numerical solutions of the DMFT equations (Sec.~\ref{comparetoDMFT}). A summary of the moments calculated is given in Table~\ref{table_summary}.
\begin{table}[b] 
    \begin{tabular} {| p{2.0cm}  | p{2.0cm}  | p{2.0cm} |p{1.7 cm} | p{2.4cm} |}
    \hline
    Qunatity & Lattice & dimension& U&highest moment  \\ \hline
    $\rho_G(\vk,\omega)$& hypercubic &$d$&infinite&4  \\ \hline
    $\rho_G(\vk,\omega)$ &hypercubic&infinite &infinite&9\\ \hline
    $\rho_G(\vk,\omega)$&Bethe&infinite &infinite&9\\ \hline
    $\rho_G(\vk,\omega)$&hypercubic&$d$ &finite&2\\ \hline
    $\rho_\Sigma(\vk,\omega)$& hypercubic &$d$&infinite&2\\ \hline
     $\rho_\Sigma(\omega)$& hypercubic& infinite&infinite&7\\ \hline
  $\rho_\Sigma(\omega)$& Bethe& infinite&infinite&7\\ \hline
   $\sigma(\omega)$ &hypercubic & $d$&infinite&2 \\ \hline
     $\sigma(\omega)$ &hypercubic & $d$&finite&0 \\ \hline
     $\sigma(\omega)$ &hypercubic & infinite&infinite&8 \\ \hline
    \end{tabular}
    \caption{The moments calculated in the present work for various models and quantities.\label{table_summary}}
    \end{table}
 
\subsection{High-temperature transport and optical conductivity for $U=\infty$: general results.}

In Sec.~\ref{infiniteU} we show, using an inspection of the general formalism and some simplifications applying at $U=\infty$, 
that the high-temperature expansion for the optical conductivity 
takes in this case the following form: 
\beq
\frac{\si(\omega)}{\sigmir} = (1-n) \frac{t}{T} \sigma^{(1)}(\frac{\omega}{t}) + (1-n)\frac{t^3}{T^3} \sigma^{(3)}(\frac{\omega}{t}) + \ldots
\label{eq:intro_opticalcond}
\eeq
In this expression, $\sigmir$ can be taken to be of order 
$\sigmir = a^{2-d}\,e^2/\hbar$, with $a$ a lattice spacing (corresponding in a quasi two-dimensional system 
to a sheet resistance of one quantum per plaquette). In the proximity
of a Mott insulator (i.e., for small doping), it becomes equivalent to
the Mott-Ioffe-Regel (MIR) value of the conductivity. In the rest of
the paper, we work in a system of units in which $e=a=1$, i.e. we normalize the conductivity to $\sigma_0$.

This expression calls for the following remarks:
\begin{itemize}
\item In the $U=\infty$ limit the general expansion (\ref{eq:general_beta_series_2pcf}) simplifies and only odd powers of $t/T$ remain.
\item The optical conductivity in the high-$T$ incoherent regime at $U=\infty$ is a smooth featureless function of frequency 
involving only the scale $t$. 
\item The functions $\sigma^{(i)}$ are dimensionless and depend only on density $n$. We provide in Sec. \ref{ancond} analytical expressions 
of the zeroth moments of $\sigma^{(1)}$ and $\sigma^{(3)}$, as well as the second moment of $\sigma^{(1)}$, on a $d$-dimensional cubic lattice.
Due to the $f$-sum rule~\cite{maldague_fsumrule,Bari_fsumrule, Sadakata_fsumrule,prelovsek_fsumrule,shastry_fsumrule} the zeroth moments are simply related to the corresponding 
high-$T$ expansion of the kinetic energy~\cite{calandra03epl}.
\item The functions $\sigma^{(i)}$ have non-singular behavior (i.e. do not vanish) in the $n\rightarrow 1$ limit of a half-filled band. 
Expression (\ref{eq:intro_opticalcond}) is written in a way that captures the dominant singularities in $1-n$ as the Mott 
insulator at $n=1$ is approached. In this limit, the conductivity vanishes, as expected. 
\item In the $n\rightarrow 0$ (empty band) limit, we find that the zeroth moment of the optical conductivity vanishes linearly in $n$, 
while the higher order moments vanish quadratically in $n$. This has consequences for dc-transport that will be summarized below. 
\end{itemize}

Hence, the dc-resistivity has the general expansion for $t \ll T \ll U=\infty$: 
\beq
\frac{\rho}{\rhomir}=\frac{T}{(1-n)t}\left[\tilde{c}_1(n)+\left(\frac{t}{T}\right)^2 \tilde{c}_3(n) + \ldots\right],
\label{eq:resist_intro}
\eeq
{\color{black} where $\rho_0\equiv\frac{1}{\si_0}$, and} the $\tilde{c}_i$'s are dimensionless coefficients depending on density, but which are non-singular in the $n\rightarrow 1$ limit, and in the case of $\tilde{c}_1$ in the $n\rightarrow0$ limit as well.  
Based on these general expressions, we can draw the following physical
conclusions:
\begin{itemize}
\item The resistivity at high-$T$ has a linear dependence on temperature, with a slope that diverges as the 
Mott insulator is approached, $n\rightarrow 1$.
\item In contrast, the slope of the $T$-linear dc resistivity reaches a finite value as $n\to0$ (i.e. at a fixed $T$, the resistivity saturates in the low-density limit). 
We show that this suprising result can be interpreted as a compensation between the density-dependence of the effective number of carriers and that of the 
scattering time. Furthermore, we find that to a very good approximation (see Figs. \ref{dccondnself} and \ref{dccondnselfBethe}).
\beq
\rho(n) \approx \frac{\rho(n=0)}{1-n},
\eeq
thus the saturated $n\to0$ resistivity sets the overall size of the
diverging $n\to1$ resistivity. This is the case not only in
the asymptotic $T \gg W$ limit ($W$ being the bare bandwidth), but even in the experimentally
relevant temperature range $T \lesssim W$. 
\item The first deviations from linearity, as $T$ is reduced, occur for $T^*\sim t \sqrt{|\tilde{c}_3/\tilde{c}_1|}$, of order $t$. 
At that scale, the resistivity is of order $\rhomir/(1-n)$. Hence, close to the Mott insulator, 
the first deviations from linearity occur at a scale where the resistivity is still much larger than the Mott-Ioffe-Regel limit, i.e. 
well into the `bad-metal' regime.  
\item This is consistent with the observation made in \refdisp{Badmetal}, that the scale at which the MIR limit is reached 
is the Brinkman-Rice scale $T_{\mathrm{BR}}\sim (1-n)T^* \ll T^*$.   
\end{itemize}

These properties of the resistivity can be rationalized by noting that the dominant $T$-dependence at high temperature is entirely controlled 
by that of the effective carrier number. To see this, we can use either of the following expressions for the conductivity:  
\begin{eqnarray}
\sigma\,&=&\, e^2 \kappa\, {\cal D}\,\,\,,\,\,\,\kappa \equiv \frac{\partial n}{\partial \mu}\\
\sigma\,&=&\,\frac{\omega_p^2}{4\pi}\,\tau_{\mathrm{tr}}\,\,\,,\,\,\,\omega_p^2\,=\,4\,\int_{-\infty}^\infty \si(\omega)  d\omega 
= 4\pi\,\frac{\sigma_0}{d}\,\frac{(-E_K)}{\hbar}
\label{eq:cond_pheno}
\end{eqnarray}
In the first expression, $\kappa=\partial n/\partial\mu$ is the electronic compressibility and $\mathcal{D}$ the diffusion constant 
defined from Fick's law $\vec{j}_n=- \mathcal{D} \nabla n$. Combining the latter with $\nabla n = \kappa\,\nabla\mu$ yields the 
above expression for $\sigma$, which expresses the Einstein relation \cite{Kokalj, Hartnoll}. 
The second expression, closer in spirit to the standard analysis of the conductivity~\cite{kohn_twist,shastry_sutherland} 
and the Drude formula for the complex conductivity {\color{black} $\sigma(\omega) = \omega_p^2/4\pi \times(-i\omega+1/\tau_{\mathrm{tr}})^{-1}$ }, 
relies on the fact that the integrated spectral weight of the optical conductivity is proportional to the kinetic energy $E_K$ 
(f-sum rule~\cite{maldague_fsumrule,Bari_fsumrule, Sadakata_fsumrule,prelovsek_fsumrule,shastry_fsumrule}). 
%
%
Note that for  $U=\infty$, all the spectral weight is contained in an energy range of order the bandwidth. 
Hence, the (absolute value of) the kinetic energy can be interpreted as setting the effective number of carriers $n_{\mathrm{eff}}$, 
which vanishes in both the low-density $n\rightarrow 0$ and the Mott insulating $n\rightarrow 1$ limits. 
In this view, the transport scattering time can be {\it defined} as: 
\begin{equation}
\tau_{\mathrm{tr}}\,=\,\frac{\sigma}{\sigma_0}\,\frac{\hbar}{(-E_K/d)}
\end{equation}
The compressibility and kinetic energy have, up to prefactors, the same high-$T$ expansion: 
\beq
\kappa \,=\,\frac{n(1-n)}{T}+\cdots\,\,\,,\,\,\, E_K\,=\,-\,n(1-n)
\frac{2dt^2}{T}+\cdots
\label{eq:kappa_kinetic}
\eeq  
Hence, the $T$-linear behaviour of the resistivity at high-$T$ is
directly related to the vanishing of the
\rzjm{compressibility}\cite{Kokalj}, or of the 
kinetic energy (effective carrier number) as $1/T$, as pointed out by Calandra and Gunnarsson~\cite{calandra03epl}.
In contrast, the diffusion constant and the transport scattering rate both reach a finite limit as $T\rightarrow \infty$. 
It is in that sense that one can talk of `saturation' at high
\rzjm{temperature}\cite{Kokalj}.
Furthermore, the divergence of the slope of the $T$-linear resistivity upon approaching the Mott insulator at $n=1$ is also captured by the 
vanishing of $\kappa$ or $E_K$. Note that, in contrast, in the low-density limit both the diffusion constant and the scattering time diverge as 
$1/n$ in order to insure that this slope approaches a finite value. Putting things together, we find that $\si(n,T) \simeq \si(0,T)(1-n)$, 
where $\si(0,T)/\sigma_0\propto t/T$ at high temperatures. This formula is surprising and works for all densities between $n=0$ and $n=1$. This is a characteristic of 
the high-temperature (bad-metal) regime $T \gtrsim T_{BR}$.

In Sec.~\ref{sec:thermal} we address  the high-$T$ expansion of the thermal conductivity and thermoelectric Seebeck coefficient.
The latter reaches at high-$T$ the value given by Heikes formula. We show furthermore that the Wiedemann-Franz law does not apply 
at high-temperature, and that the Lorenz number vanishes as $1/T^2$. 

\subsection{High-temperature transport and optical conductivity for $U=\infty$ in large dimensions, and DMFT}

In Sec. \ref{dinfinity}, we consider the infinite-$U$ Hubbard model in the limit of large dimensions. 
In this limit, the self-energy becomes momentum-independent:
\beq
\rho_\Sigma(-\mu + \domega) = D \sum_{i=0}^\infty  \left(\frac{D}{T}\right)^i   h^{(i)}(\frac{\domega}{D}), 
\eeq
where the half-bandwidth $D\propto t\sqrt{d}$ is kept finite. The functions $h^{(i)}(\frac{\domega}{D})$ are shown to be even (odd) in $\delta\omega$ for $i$ even (odd). 
In addition, all moments of  $\rho_\Sigma(-\mu + \domega)$ w.r.t to $\delta\omega$ vanish linearly in $n$ as $n\to0$. However, only the odd moments vanish linearly in $(1-n)$ as $n\to1$. In the finite-dimensional case, all moments continue to vanish linearly in $n$ as $n\to0$, while none of the moments vanish as $n\to1$. 

In this limit, we are able to calculate a larger number of moments (see table \ref{table_summary}) for both the optical conductivity 
and the self-energy. This allows us to reconstruct explicitly these correlation functions using two complementary reconstruction methods, 
the maximum entropy method (MEM) and Mori's relaxation function approach for the optical conductivity. 
The latter can also be obtained from the reconstructed self-energy, using the fact that vertex corrections vanish in infinite dimensions~\cite{khurana_vertex} 
so that only a bubble graph involving the convolution of two one-particle Green's functions has to be evaluated (see Eq.(\ref{dcbubble})). 

Furthermore, we obtained full numerical calculations of real-frequency correlation functions (self-energy and optical conductivity) 
by solving the DMFT equations using a numerical renormalization-group
algorithm (NRG), and compare the results to the analytical high-$T$ expansion in Sec.\ref{comparetoDMFT}. 
In order to reach this goal, we had to take special care in adapting the current NRG codes, 
using a very narrow kernel for the broadening of raw spectral data into a continuous spectral
function on the real-frequency axis. This leads to severe underbroadening of
spectral function and oscillatory artifacts, however intergrated
quantities such as optical and dc conductivity converge with the
decreasing kernel width to the exact high-temperature results.

\begin{figure}
\begin{center}
\includegraphics[width=0.4\columnwidth,keepaspectratio,angle=-90]{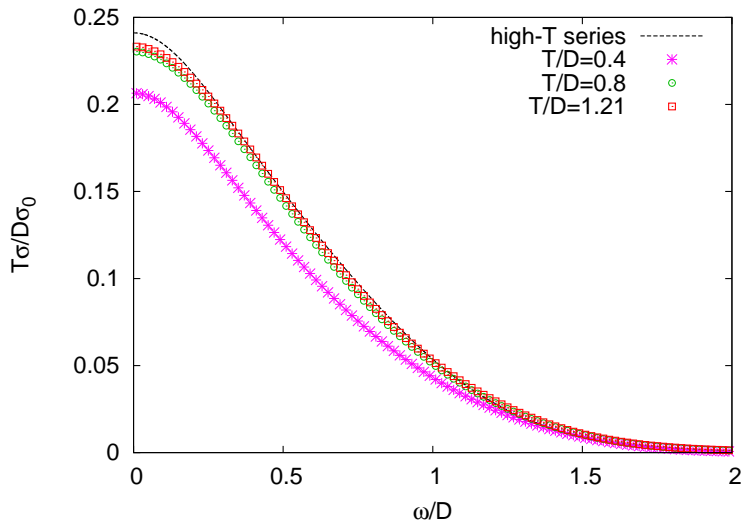}
\caption{The optical conductivity for the Bethe lattice at $U=\infty$ as calculated within NRG-DMFT for $n=.9$ at several high temperatures. 
When multiplied by $\frac{T}{D}$ (with $D$ the half-bandwidth), the curves collapse onto each other, in accordance with the first term in \disp{eq:intro_opticalcond}. 
Moreover, the resulting scaling function is in good agreement with that obtained using the high-temperature series.}
\label{compareNRGhighToptical}
\end{center}
\end{figure}

An example of such a comparison for the optical conductivity is displayed in Fig.~\ref{compareNRGhighToptical}. The shape of the optical 
conductivity displayed there confirms the qualitative points made above, and the agreement between the DMFT-NRG results and the high-$T$ expansion 
is seen to be excellent. 

\begin{figure}
\begin{center}
\includegraphics[width=0.4\columnwidth,keepaspectratio,angle=-90]{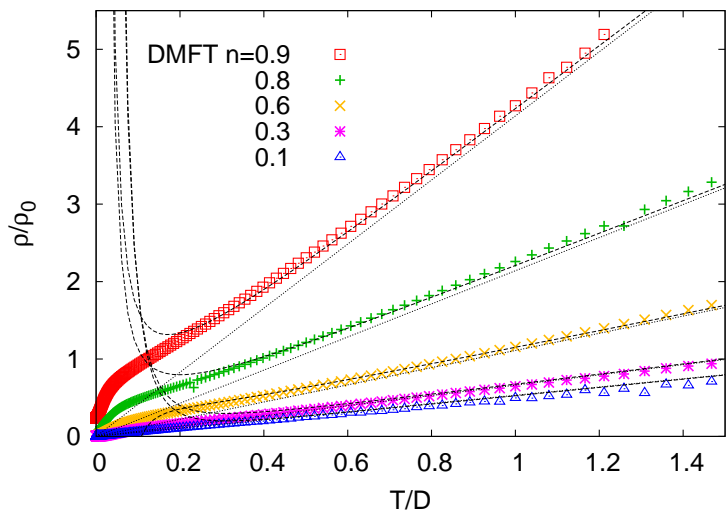}
\caption{Resistivity vs. temperature for the infinite-connectivity Bethe lattice at $U=\infty$, for several densities as calculated within NRG-DMFT (symbols). 
The results are compared to the ones from the high temperature series, both to leading order, $\rho \propto T/D$  (straight dotted lines), and including 
the first two sub-leading corrections, $D/T$ and $(D/T)^3$ (curved dashed lines). 
\label{compareNRGhighT}}
\end{center}
\end{figure}

Using moment reconstruction methods, we were also able to calculate both the leading and sub-leading coefficients $\tilde{c}_1(n)$, $\tilde{c}_3(n)$, and $\tilde{c}_5(n)$ in the 
high-temperature expansion \disp{eq:resist_intro} of the dc-resistivity. The resulting high-$T$ approximation to the resistivity is compared to the 
DMFT-NRG results in Fig.~\ref{compareNRGhighT}, for several densities.   
Remarkably, the series reproduces the NRG curves for $\frac{T}{D}\ge .3$, i.e. essentially throughout the bad-metal regime, well below its a priori range of applicability. 
The NRG curves confirm our finding that the slope of the resistivity in the linear high-$T$ regime saturates as $n\to0$ and diverges like $1/(1-n)$ as $n\to1$.  
In Sec.\ref{sec:resist_comp}, we furthermore present a physical interpretation of the `knee-like' feature displayed by the NRG resistivity curves at 
lower temperature, and previously noted in e.g.
\refdisp{Badmetal} (\rzjm{see also \cite{jaklic_prelovsek_prb_1994,jaklic_prelovsek_prb_1995}}). We show that above the knee, the resistivity is mostly controlled by the temperature 
dependence of the effective carrier number (as given by the
compressibility or kinetic energy according to Eq.~\eqref{eq:kappa_kinetic}) while below this scale it is mostly controlled 
by the $T$-dependence of the scattering rate or diffusion constant. 
\ag{See Sec.~\ref{sec:mechanisms_transport} and Fig.~\ref{fig:diffusion} therein, as well as the concluding section, 
for a discussion of the transport mechanism in the different regimes.}
%

In the very high-temperature regime, the DMFT results are most
reliably obtained using the interaction expansion continuous-time Quantum Monte-Carlo~\cite{gull_rmp_2011} 
rather than the NRG. This is also the regime in which the series works the best. On
the level of the imaginary-time Green's function, we find excellent
agreement between the QMC results and the series. Performing the
analytic continuation to obtain the self-energy on the real frequency
axis is somewhat problematic using the QMC data. Nonetheless, we find
good agreement between the two methods even for this quantity (see sec. \ref{comparespectra}).

\subsection{High-temperature transport and optical conductivity for finite $U$}

For the finite-$U$ Hubbard model, the spectral function has two peaks in the
high-temperature limit. In the parameter range $t\ll T$ with finite $U/T$, the lower and upper Hubbard bands can be individually 
expanded in the form of \disp{eq:general_beta_series_1pcf}:

\beq
\rho_{G,L}(\vk,-\mu+\delta\omega) = \frac{1}{t} \sum_{i=0}^\infty \left(\frac {t}{T}\right)^i g_L^{(i)}(\vk,\frac{\delta\omega}{t});\;\;\;\;\rho_{G,U}(\vk,-\mu + U+\delta\omega) = \frac{1}{t} \sum_{i=0}^\infty \left(\frac {t}{T}\right)^i g_U^{(i)}(\vk,\frac{\delta\omega}{t}).
\label{GLGUintro}
\eeq
The moments $\int dx x^m g^{(i)}_{L(U)}(\vk,x)$ are themselves a series in the expansion parameter $t/U$. The coefficients of this series depend on the density $n$ and $e^{-\beta U}$. Upon setting $U\rightarrow\infty$, the lower Hubbard band  becomes equal to the spectral function of the infinite-$U$ Hubbard model. 

In sec. \ref{momentsLHBUHB}, we compute the zeroth through second
moments of the lower and upper Hubbard bands in the case of a
$d$-dimensional hypercubic lattice, for arbitrary filling $n$. In sec. \ref{finiteUcond}, we use the results of sec. \ref{momentsLHBUHB} along with the ``bubble" formula, to address the temperature-dependence of the {\color{black}dc resistivity}.  
Finally, in Sec.~\ref{sec:Tinf_dinf}, we address the question to what
extent the DMFT becomes exact in the high-temperature limit
and in what sense is the physics `local' at high temperature.
We find that while the thermodynamic potential becomes exact, the same can only be said for the zeroth moment of the local self-energy. 
Moreover, the self-energy has non-local contributions which survive the high-$T$ limit, but are not captured by DMFT. 
DMFT becomes accurate however when both a high-temperature and a high-frequency expansion are performed.

\section{Behavior of 2-particle correlation functions in the high-T limit.\label{2particle}}
\label{sec:2pcf}
\subsection{Lehmann Representation}

Consider the two-particle correlator 
\beq
\chi_{O.O}(\vk,\tau) = - < T_\tau \ \hat{O}_k(\tau) \  \hat{O}_{-k}>,
\label{chitime}
\eeq
where $\hat{O}_k = \frac{1}{\sqrt{N_s}} \sum_j e^{- i \vk \cdot \vR_j}\hat{O}_j$, and $\hat{O}_j = \hat{n}_j - <\hat{n}_j>$, $\hat{J}_{j,\alpha}$, or $\hat{S}_{j,\alpha}$. These are the particle, current, and spin operators respectively, and $\alpha$ represents a spatial direction. We also consider the Fourier transform of $\chi_{O.O}(\vk,\tau)$, defined as

\beq
\chi_{O.O}(\vk,i\Omega_n) = \int_0^\beta d\tau  \ \chi_{O.O}(\vk,\tau) e^{i\Omega_n\tau}.
\label{chimomentum}
\eeq
Here, $\Omega_n = 2\pi n T$. We can also write $\chi(\vk,i\Omega_n)$ in terms of its spectral representation

\beq
\chi_{O.O}(\vk,i\Omega_n) =   \int d\nu \  \frac{\chi_{O.O}^{''}(\vk,\nu)}{i\Omega_n-\nu},
\label{spectral}
\eeq
 where $\chi_{O.O}^{''}$ is the spectral function corresponding to $\chi_{O.O}$.

We show quite generally, independent of the specific form of the Hamiltonian, that $\chi_{O.O}^{''}(\vk,\omega)$ has the following form 
when the temperature $T$ is the highest energy scale in the problem:  
\beq 
\frac{1}{\omega}\chi_{O.O}^{''}(\vk,\omega) = E^{\gamma(O)-1}\sum_{i=1}^\infty (\beta E)^i f^{(i)}(\vk,\frac{\omega}{E}).
\label{betaseries}
\eeq
Here, the $f^{(i)}(\vk,\frac{\omega}{E})$ are even functions of the frequency\cite{KadanoffMartin}, independent of the temperature, and hence this is an expansion in $\frac{1}{T}$. $E$ can be any unit of energy, and is most conveniently taken to be a characteristic energy unit of the Hamiltonian. For example, it can be taken to be the hopping. If $\hat{O}_j = \hat{J}_{j,\alpha}$, $\gamma(O) = 1$, while in the case that $\hat{O}_j = \hat{n}_j - <\hat{n}_j>$, or $\hat{O}_j = \hat{S}_{j,\alpha}$, $\gamma(O) = -1$. 

In the Lehmann representation, $\chi_{O.O}^{''}(\vk,\omega)$ is written as \cite{FW}

\beq
\chi_{O.O}^{''}(\vk,\omega) = \frac{-1}{Z}\sum_{m,n} e^{-\beta \epsilon_m} |\langle m | \hat{O}_k | n \rangle |^2 \delta(\omega + \epsilon_m-\epsilon_n)\left[e^{\beta (\epsilon_m-\epsilon_n)} - 1\right].
\label{Lehmannchi}
\eeq

Here, $Z= Tr(e^{-\beta H})$ is the partition function, and $m$ and $n$ are indices that run over all of the eigenstates. We assume the system to have inversion symmetry, and therefore $\chi_{O.O}^{''}(\vk,\omega) = \chi_{O.O}^{''}(-\vk,\omega)$.  We now expand the term in the brackets on the RHS of \disp{Lehmannchi} in powers of $\beta$, the inverse temperature. After some simplifications, this yields 

\beq
\frac{\chi_{O.O}^{''}(\vk,\omega)}{\omega} = \frac{\beta E^{\gamma^{(0)}}}{\sum_s e^{-\beta E \tilde{\epsilon}_s}} \sum_{m,n} e^{-\beta E\tilde{\epsilon}_m} |\langle m | \tilde{\hat{O}}_k | n \rangle |^2 \delta(\frac{\omega}{E} + \tilde{\epsilon}_m-\tilde{\epsilon}_n)\sum_{r=0}^\infty \frac{(\beta E)^r}{(r+1)!}(\tilde{\epsilon}_m-\tilde{\epsilon}_n)^r.
\label{Lehmannchiexp}
\eeq
Here, $\tilde{\epsilon}_m\equiv\frac{\epsilon_m}{E}$, and $\tilde{\hat{O}}_k = E^{-\left(\frac{\gamma^{(0)}+1}{2}\right)}\hat{O}_k$,  where both are dimensionless. Upon expanding all of the exponentials $e^{-\beta E \tilde{\epsilon}}$ in powers of $\beta$, the RHS of \disp{Lehmannchiexp} will have the form of the RHS of \disp{betaseries}. In particular, we find that 
\beq
f^{(1)}(\vk,\frac{\omega}{E}) = \frac{1}{D_{\cal{H}}} \sum_{m,n}  |\langle m | \tilde{\hat{O}}_k | n \rangle |^2 \delta(\frac{\omega}{E} + \tilde{\epsilon}_m-\tilde{\epsilon}_n),
\eeq
where $D_{\cal{H}}$ is the dimension of the Hilbert space.

\subsection{Short-time Expansion}

In this section, we derive a closed form expression for the moments of the functions $f^{(i)}(\vk,\frac{\omega}{E})$ appearing in \disp{betaseries}. We begin by writing the Hamiltonian in the following way

\beq
\hat{H} = \hat{H}_1 - \mu \hat{N}.
\label{Hamiltonian}
\eeq
Here, $\hat{H}_1$ is independent of the temperature, and the only part of the Hamiltonian which depends on the temperature is the chemical potential $\mu$. Also, we will assume that $\hat{H}_1$ conserves particle-number, therefore $[\hat{H}_1,\hat{N}]=0$. In order to be able to control the density through the chemical potential, it must be the case that as $T\to\infty$, $\mu$ is proportional to $T$. Therefore, we can write $\mu$ as a series in the inverse-temperature

\beq
\mu \equiv T\bar{\mu} = T\sum_{i=0}^\infty (\beta E)^i \bar{\mu}^{(i)}.
\label{muexp}
\eeq

Plugging \disp{Hamiltonian} into \disp{chitime} and choosing $\tau>0$, we obtain 
\beq
\chi_{O.O}(\vk,\tau) = - \langle e^{(\tau-\beta) \hat{H}_1} \hat{O}_{k} e^{-\tau \hat{H}_1}\hat{O}_{-k}\rangle_{0,c}
\eeq
where $\langle\hat{A}\rangle_0 \equiv \frac{Tr(e^{\beta\mu\hat{N}}\hat{A})}{Z_0}$, and $Z_0=Tr(e^{\beta\mu\hat{N}})$. Here, the partition function has been eliminated, and only ``connected" diagrams have been retained in the expectation values. The meaning of connected diagram in this context is that we keep only terms which are intensive, i.e. do not scale with the size of the lattice (see \refdisp{Edward}). Expanding the time-dependent exponentials, we obtain

\beq 
\chi_{O.O}(\vk,\tau) = - \sum_{a=0,b=0}^\infty \frac{(\tau-\beta)^a}{a!} \frac{\tau^b(-1)^b}{b!} \langle \hat{H}^a_1 \hat{O}_{k}  \hat{H}^b_1\hat{O}_{-k}\rangle_{0,c}.
\label{chiab}
\eeq
Inversion symmetry implies that $\langle \hat{H}^a_1 \hat{O}_{k}  \hat{H}^b_1\hat{O}_{-k}\rangle_{0,c} = \langle \hat{H}^b_1 \hat{O}_{k}  \hat{H}^a_1\hat{O}_{-k}\rangle_{0,c}$. Therefore, plugging \disp{chiab} into \disp{chimomentum}, and rearranging the sum yields
\beq
\chi_{O.O}(\vk,i\Omega_n) = -  \sum_{c=0,d=0}^\infty \frac{(-1)^c}{c!(c+d)!} \langle \hat{H}^{c+d}_1 \hat{O}_{k}  \hat{H}^c_1\hat{O}_{-k}\rangle_{0,c}\int_0^\beta d\tau e^{i\Omega_n\tau} (\tau-\beta)^c\tau^c\frac{(\tau-\beta)^d+(-1)^d\tau^d }{1+\delta_{d,0}}.
\label{chicd}
\eeq
The RHS of \disp{chicd} can now be expanded in powers of $\frac{1}{i\Omega_n}$. This expansion, performed in Appendix \ref{chi}, will allow us to derive a closed-form expression for the moments of $f^{(i)}(\vk,y)$. First, we derive some additional useful formulae.
\beq
\chi_{O.O}(\vk,i\Omega_n) =\sum_{m=0}^\infty\int d\omega \frac{\chi_{O.O}^{''}(\vk,\omega)}{\omega} \frac{\omega^{m+1}}{(i\Omega_n)^{m+1}} \;\;\;\;\;(n\neq0).
\label{highfreq}
\eeq
\beq
\chi_{O.O}(\vk,i\Omega_0) = -\int d\omega \frac{\chi_{O.O}^{''}(\vk,\omega)}{\omega} .
\label{zerofreq}
\eeq
\disp{highfreq} is the high-frequency expansion of \disp{spectral}, while \disp{zerofreq} is \disp{spectral} evaluated at $n=0$.
Plugging \disp{betaseries} into Eqs. (\ref{highfreq}) and  (\ref{zerofreq}) and comparing  with \disp{chicd4} yields the moments of $f^{(i)}(\vk,y)$. For $r$ even,
\beq
\int dy f^{(i)}(\vk,y)y^r = \sum_{\substack{c=0,d=0\\2c+d=r+i-1}}^\infty \frac{(-1)^{c+d}}{(c+d)!}\frac{2}{1+\delta_{d,0}} \langle \tilde{\hat{H}}^{c+d}_1 \tilde{\hat{O}}_{k}  \tilde{\hat{H}}^c_1\tilde{\hat{O}}_{-k}\rangle_{0,c}
\sum_{j=0}^{min[c,i-1]} \frac{(-1)^j}{j!(c-j)!} \frac{(r-1+i-j)!}{(i-j)!},
\label{momentsfst}
\eeq
while for $r$ odd, $\int dy f^{(i)}(\vk,y)y^r=0$.

\section{Behavior of 1-particle correlation functions in the high-T limit.\label{1particle}}

\subsection{Lehmann Representation}

Consider the single-particle Green's function 
\beq
G(\vk,\tau) = -\langle T_\tau c_{\vk\si}(\tau) c^\dagger_{\vk\si}\rangle.
\label{G}
\eeq
Its Fourier transform, defined as $G(\vk,i\omega_n)\equiv \int_0^\beta
d\tau e^{i\omega_n\tau} G(\vk,\tau)$ where $\omega_n = (2n+1)\pi T$, can be written in terms of the Dyson(-Mori) self-energy $\Sigma(\vk,i\omega_n)$:
\beq
G(\vk,i\omega_n) = \frac{a_G}{i\omega_n+\mu-\epsilon'_k - \Sigma(\vk,i\omega_n)}.
\label{Dyson}
\eeq
Here, $a_G$ is obtained from the high-frequency limit of the Green's
function, while $\epsilon'_k$ is some unspecified dispersion. In the case of the finite-$U$ Hubbard model $a_G=1$, and $\epsilon'_k=\epsilon_k$ (Dyson self-energy), where $\epsilon_k$ is the dispersion of the lattice. In the case of the infinite-$U$ Hubbard model $a_G=1-\frac{n}{2}$ and $\epsilon'_k = \epsilon_k(1-\frac{n}{2})$ (Dyson-Mori self-energy). Both the $G(\vk,i\omega_n)$ and $\Sigma(\vk,i\omega_n)$ can be written in terms of their respective spectral densities:
\beq
G(\vk,i\omega_n) = \int d\nu \frac{\rho_G(\vk,\nu)}{i\omega_n-\nu},
\label{spectralG}
\eeq
\beq
\Sigma(\vk,i\omega_n) = \Sigma_\infty(\vk) + \int d\nu \frac{\rho_\Sigma(\vk,\nu)}{i\omega_n-\nu}.
\label{spectralSig}
\eeq
We will show that when $T$ is the highest energy scale in the problem,
$\rho_G(\vk,\omega)$ and $\rho_\Sigma(\vk,\omega)$ can be expanded in the following series in $\frac{1}{T}$:
\beq
\rho_G(\vk,-\mu + x) = \frac{1}{E} \sum_{i=0}^\infty (\beta E)^i
g^{(i)}(\vk,\frac{x}{E}),
\label{betaseriesG}
\eeq
\beq
\rho_\Sigma(\vk,-\mu + x) = E \sum_{i=0}^\infty (\beta E)^i h^{(i)}(\vk,\frac{x}{E}).
\label{betaseriesSig}
\eeq
Note that in order to achieve this expansion, the frequency must be re-centered around $\omega=-\mu$. The new frequency variable $x$ measures the displacement of $\omega$ from $\omega=-\mu$. 

The Lehmann representation for the spectral density $\rho_G(\vk,\omega)$ is \cite{FW}
\beq
\rho_G(\vk,\omega) = \frac{1}{\sum_s e^{-\beta \epsilon_s}}\sum_{m,n} e^{-\beta \epsilon_m} |\langle m|c_{\vk\si}|n\rangle|^2[e^{\beta ( \epsilon_m- \epsilon_n)}+1]\delta(\omega+\epsilon_m-\epsilon_n).
\eeq 
We now define $\epsilon_m-\epsilon_n \equiv \mu + \Delta_{mn}\equiv T\bar{\mu} + E\tilde{\Delta}_{mn}$, where $\bar{\mu}=\frac{\mu}{T}$ and $\tilde{\Delta}_{mn}=\frac{\Delta_{mn}}{E}$ are dimensionless. Plugging these in, we find that
\beq
\rho_G(\vk,-\mu+x) = \frac{1}{\sum_s e^{-\beta E \tilde{\epsilon}_s}}\sum_{m,n} e^{-\beta E \tilde{\epsilon}_m} |\langle m|c_{\vk\si}|n\rangle|^2(e^{\bar{\mu}}e^{\beta E\tilde{\Delta}_{mn}}+1)\frac{1}{E}\delta(\frac{x}{E} + \tilde{\Delta}_{mn}).
\eeq
Expanding all exponentials in powers of $(\beta E)$, we find that $\rho_G(\vk,-\mu+x)$ has the form of \disp{betaseriesG}. For example, 
\beq
g^{(0)}(\vk,\frac{x}{E}) = \frac{1}{D_{\cal{H}}}\sum_{m,n} |\langle m|c_{\vk\si}|n\rangle|^2(e^{\bar{\mu}}+1)\delta(\frac{x}{E} + \tilde{\Delta}_{mn}).
\eeq

We will now use \disp{betaseriesG} to derive \disp{betaseriesSig}. 
Using Eqs. (\ref{spectralG}) and (\ref{spectralSig}), we find that 
\beq
\rho_\Sigma(\vk,-\mu+x) = \frac{a_G  \ \rho_G(\vk,-\mu+x)}{[\Re e G(\vk,-\mu+x)]^2 + [\pi\rho_G(\vk,-\mu+x)]^2}
\label{rhoSigrhoG}
\eeq
Plugging \disp{betaseriesG} into \disp{spectralG}, we obtain 
\beq
\Re e G (\vk,-\mu+x) = \ \frac{1}{E} \sum_{i=0}^\infty (\beta E)^i \bar{g}^{(i)}(\vk,\frac{x}{E});\;\;\;\; \bar{g}^{(i)}(\vk,\frac{x}{E})=\int dy \ \frac{g^{(i)}(\vk,y)}{\frac{x}{E}-y}.
\label{betaseriesReG}
\eeq
Finally, plugging Eqs. (\ref{betaseriesG}) and  (\ref{betaseriesReG}) into \disp{rhoSigrhoG}, we find that $\rho_\Sigma(\vk,-\mu + x)$ satisfies the form \disp{betaseriesSig}. For example,
\beq
h^{(0)}(\vk,\frac{x}{E}) = \frac{a_G g^{(0)}(\vk,\frac{x}{E})}{[\bar{g}^{(0)}(\vk,\frac{x}{E})]^2 + \pi^2 [g^{(0)}(\vk,\frac{x}{E})]^2}.
\eeq

\subsection{Short-time expansion}

We now use the short-time expansion of $G(\vk,\tau)$ to derive the moments of the functions $g^{(i)}(\vk,\frac{x}{E})$ appearing in \disp{betaseriesG}. For $\tau>0$, \disp{G} becomes
\beq
G(\vk,\tau) = -e^{\mu \tau} \langle e^{(\tau-\beta)H_1} c_{k\si} e^{-\tau H_1} c^\dagger_{k\si}\rangle_{0,c}
\label{Gvktau}
\eeq
Just as was the case for $\chi_{O.O}(\vk,i\Omega_n)$, $G(\vk,i\omega_n)$ can be expanded in power of $\frac{1}{i\omega_n+\mu}$. This expansion is derived starting from \disp{Gvktau} in Appendix \ref{Gapp}. Its final form is given in \disp{Gk3}.
Making the substitution $\nu = x-\mu$ in \disp{spectralG}, performing
a high-frequency expansion, plugging in \disp{betaseriesG}, and comparing with  \disp{Gk3} yields the moments of the $g^{(i)}(\vk,\frac{x}{E}) $:
\barray
 \int dy  \  g^{(i)}(\vk,y) y^{m}= &&\sum_{\substack{a=0,b=0\\a+b=m+i}}^\infty \frac{1}{b!} \langle \tilde{\hat{H}}_1^a c_{k\si} \tilde{\hat{H}}_1^b c^\dagger_{k\si}\rangle_{0,c} \nn\\
&&\sum_{j=0}^{min(i,a)} \frac{1}{j!(a-j)!}  (-1)^{j+m-b}\frac{(m+i-j)!}{(i-j)!}(e^{\bar{\mu}}+\delta_{j,i}).\nn\\
\label{momentsg}
\earray
The moments of $h^{(i)}(\vk,\frac{x}{E}) $ can be obtained by plugging the high frequency expansion for $G(\vk,i\omega_n)$ into Dyson's equation (\disp{Dyson}), comparing with the high-frequency expansion of \disp{spectralSig}, and using \disp{betaseriesSig}.

\section{Infinite-$U$ Hubbard model\label{infiniteU}}

The infinite-$U$ Hubbard model Hamiltonian is 

\beq
\hat{H} = -\sum_{ij\si} t_{ij} X_i^{\si0}X_j^{0\si} - \mu \sum_{i\si}X_i^{\si\si}.
\eeq
{\color{black} The Hubbard operator $X^{ab} \equiv |a\rangle\langle b|$ projects the state $|b\rangle$ onto the state $|a\rangle$, where $|a\rangle$ and $|b\rangle$ can be any of the three allowed states $|\uparrow\rangle$,$|\downarrow\rangle$, or $|0\rangle$. $X_i^{ab}$ acts only on the site $i$.} We can write this Hamiltonian as $\hat{H} = \hat{T} - \mu \hat{N}$, where $\hat{T}$ is the hopping term in the Hamiltonian. In the notation of sec.(\ref{2particle}), $E=t$, and $\tilde{\hat{H_1}}=\frac{1}{t}\hat{T}$. A special feature of this model is that 
\beq
\langle \tilde{\hat{H}}^{a}_1 \hat{O}_{k}  \tilde{\hat{H}}^b_1\hat{O}_{-k}\rangle_{0} \propto \delta_{0,p(a+b)},
\label{oddhops} 
\eeq
where $p(x)=1$ for $x$ odd, and $p(x)=0$ for $x$ even. This is due to
the fact that an odd number of hops cannot return the system back to its initial configuration. Moreover, the two identical $\hat{O}$ operators must contribute an even number of hops. Furthermore, the real space expectation value
\beq
\langle \tilde{\hat{H}}^{a}_1 c_i  \tilde{\hat{H}}^b_1c_j^\dagger \rangle_{0} \propto \delta_{p(|i-j|),p(a+b)},
\label{chain} 
\eeq
where $|i-j|$ is the Manhattan distance between site $i$ and site $j$.
This is the case since the number of hops required to get from $i$ to
$j$ is equal to the separation between them, while the number of
remaining hops must be even for the system to return to its initial configuration.

\subsection{General form of the conductivity\label{conductivitygeneral}}
Using \disp{oddhops}, we see that the RHS of \disp{momentsfst} must vanish for $i$ even. Therefore, in the infinite-U Hubbard model, \disp{betaseries} acquires the specific form 
\beq 
\frac{1}{\omega}\chi_{O.O}^{''}(\vk,\omega) = \frac{t^{\gamma(O)}}{T} f^{(1)}(\vk,\frac{\omega}{t}) + \frac{t^{\gamma(O) + 2}}{T^3} f^{(3)}(\vk,\frac{\omega}{t}) + \ldots
\label{betaseriesinfU}
\eeq

To discuss the conductivity, we choose $\hat{O}=\hat{J}_\alpha$, in which case $\frac{\chi_{J_{\alpha}.J_{\alpha}}^{''}(\vk,\omega)}{\omega}\equiv \frac{\si(\vk,\omega)}{\pi\si_0}$. Applying \disp{betaseriesinfU} with $\gamma(\hat{J})=1$ yields 
\beq 
\frac{\si(\vk,\omega)}{\pi\si_0} = \frac{t}{T } f^{(1)}(\vk,\frac{\omega}{t}) + \frac{t^3}{T^3 } f^{(3)}(\vk,\frac{\omega}{t}) + \ldots
\label{betaseriesinfUcond}
\eeq
We will see below that the moments of $f^{(i)}(\vk,\frac{\omega}{t})$ are each proportional to exactly one power of $(1-n)$. Explicitly pulling the factor $(1-n) $ out of the functions $f^{(i)}(\vk,\frac{\omega}{t})$, and absorbing the factor of $\pi$, we write
\beq 
\frac{\si(\vk,\omega)}{\sigma_0} = (1-n) \frac{t}{T} \si^{(1)}(\vk,\frac{\omega}{t}) + (1-n)\frac{t^3}{T^3} \si^{(3)}(\vk,\frac{\omega}{t}) + \ldots
\eeq
Specializing to the case of the optical-conductivity $\sigma(\omega)=\lim_{k\to0}\sigma(\vk,\omega)$, we write 
\beq
\frac{\si(\omega)}{\sigma_0} = (1-n) \frac{t}{T} \sigma^{(1)}(\frac{\omega}{t}) + (1-n)\frac{t^3}{T^3} \sigma^{(3)}(\frac{\omega}{t}) + \ldots
\label{opticalcond}
\eeq
Here, $\sigma^{(i)}(\frac{\omega}{t})=\lim_{k\to0}\si^{(i)}(\vk,\frac{\omega}{t})$. The dc conductivity $\sigma_{dc} = \lim_{\omega\to0}\sigma(\omega)$ can be obtained by taking the $\omega\to0$ limit of \disp{opticalcond}, producing the series 
\beq
\frac{\si_{dc}}{\sigma_0} =  (1-n) \frac{t}{T}d_1 + (1-n)\frac{t^3}{T^3} d_3+ \ldots
\label{dccond}
\eeq
Here, $d_i = \lim_{\omega\to0}\sigma^{(i)}(\frac{\omega}{t}) $ and is non-singular in the $n\to 1$ limit. This form implies that $\rho_{dc}$ is linear in $T$ for $T\geq t$. Deviations set in for $T\sim t$. At that scale, $\rho_{dc}\sim\frac{\rho_0}{\delta}$, where $\delta=1-n$ is the doping away from half-filling.

\subsection{Analytically calculated moments of the conductivity.\label{ancond}}

The current density on a hypercubic lattice is given by
\beq
J_{i,\alpha} = \frac{i}{2}\sum_{j\si} t_{ij}(\vR_j-\vR_i)_\alpha (X_j^{\si0}X_i^{0\si}-X_i^{\si0}X_j^{0\si}).
\eeq
Plugging this into \disp{momentsfst}, and taking the $k\to0$ limit, we calculate the zeroth and second moments of  $\si^{(1)}(\frac{\omega}{t})$, and the zeroth moment of $\si^{(3)}(\frac{\omega}{t})$ in $d$ dimensions. Note that a direct application of \disp{momentsfst} yields the moments in terms of $\mu$. To obtain them in terms of the density $n$, we must also carry out an expansion of the chemical potential \cite{Edward}:

\beq
\mu= T \left\{\log\left[\frac{n}{2(1-n)}\right] + (\beta t)^2  d (2n-1)+\ldots\right\}.
\label{expmu}
\eeq

This yields the moments:

\beq
\int d(\frac{\omega}{t}) \sigma^{(1)}(\frac{\omega}{t}) &=& 2\pi n, \nn\\
\int  (\frac{\omega}{t})^2 d(\frac{\omega}{t}) \sigma^{(1)}(\frac{\omega}{t})  &=& \pi n^2(4-n)(2d-2), \nn\\
\int d(\frac{\omega}{t}) \sigma^{(3)}(\frac{\omega}{t}) &=&\frac{\pi}{6}  n\{ n[d (13 n-16)-25 n+28]-6\}.
\eeq
 
The above moments of $\sigma^{(1)}(\frac{\omega}{t})$ for the square lattice were computed in \refdisp{jaklic_prelovsek_prb_1995}. Using expression (\ref{eq:kappa_kinetic}) for the kinetic energy, we see that the zeroth-order moment above is consistent 
with the $f$-sumrule~\cite{maldague_fsumrule,Bari_fsumrule, Sadakata_fsumrule,prelovsek_fsumrule,shastry_fsumrule} $\frac{1}{\pi}\int d\omega \sigma_{\alpha\alpha} = - e^2 E_K /d$.
We note that both the zeroth and second moments of $\sigma(\omega)$ vanish linearly in $(1-n)$ in the limit $n\to1$. However, in the limit $n\to 0$, the zeroth moment vanishes linearly in $n$ while the second moment vanishes quadratically in $n$. As will be seen in sec \ref{dinfinity}, in the $d\to\infty$ limit, the higher order moments continue to vanish quadratically in $n$. This will affect both the shape of the optical conductivity, and the value of the $dc$ conductivity, which saturates as $n\to0$.
\subsection{Analytically calculated moments of the Green's function and the self-energy. \label{angrself}}

The first few moments of the Green's function and self-energy have been calculated for a $d$-dimensional hypercubic lattice in \refdisp{Edward}. We reproduce the results 
in Tables.~\ref{table_mom_g} and \ref{table_mom_h}.
We note that the moments of $\rho_\Sigma(\vk,-\mu+x)$ and $\Sigma_\infty(\vk)$ vanish linearly in $n$ as $n\to0$. On the other hand, only $\Sigma_\infty(\vk)$  vanishes as $n\to1$.

\begin{table}
    \begin{tabular} { | p{2.0cm}  | p{10cm} |}
    \hline
    
    $m_0[g^{(0)}(\vk,y)]$&$1-\frac{n}{2}$  \\ \hline
    $m_1[g^{(0)}(\vk,y)]$ &$(1-\frac{n}{2})^2\tilde{\epsilon}_k$ \\ \hline
    $m_1[g^{(1)}(\vk,y)]$ &$d (1-n) n$ \\ \hline
    $m_1[g^{(2)}(\vk,y)]$&$-\frac{1}{4} (1-n)^2 n^2  \tilde{\epsilon}_k$ \\ \hline
    $m_1[g^{(3)}(\vk,y)]$& $\frac{1}{12} d (1-n) n \{n [d (13 n-16)-25 n+28]-6\}$\\ \hline
    $m_2[g^{(0)}(\vk,y)]$ &$2 d (1-\frac{n}{4}) (1-\frac{n}{2}) n + (1-\frac{n}{2})^3 \tilde{\epsilon}_k^2$ \\ \hline
     $m_2[g^{(1)}(\vk,y)]$&  $(2 d-1) (1-\frac{n}{2}) (1-n) n  \tilde{\epsilon}_k$\\ \hline
  $m_2[g^{(2)}(\vk,y)]$& $-\frac{1}{4} (1-n) n^2  \left\{(2-n) (1-n) \tilde{\epsilon}_k^2-d  [6 d-2 (3-n) n+1]\right\}$\\ \hline
   $m_3[g^{(0)}(\vk,y)]$ & $(1-\frac{n}{2})^2 \tilde{\epsilon}_k \left[(4 d-1) (1-\frac{n}{4}) n +(1-\frac{n}{2})^2 \tilde{\epsilon}_k^2\right]$ \\ \hline
    $m_3[g^{(1)}(\vk,y)]$ &$-\frac{1}{4} (1-n) n  \left\{2 d  [d (3 n (n+2)-8)-n (2 n+1)]-
    4(3 d-2) (1-\frac{n}{2})^2 \tilde{\epsilon}_k^2\right\}$ \\ \hline
 $m_4[g^{(0)}(\vk,y)]$ &$\frac{1}{2} (3 d-1) (4-n) n (1-\frac{n}{2})^3 \tilde{\epsilon}_k^2-
d n (1-\frac{n}{2})  \{2 d (7 n-10)+n [(1-\frac{n}{8})n-9]+10\}+
 (1-\frac{n}{2})^5\tilde{\epsilon}_k^4$ \\ \hline     
    \end{tabular}
    \caption{Moments of $g^{(i)}(\vk,y)$ for the infinite-$U$ Hubbard
    model on a hypercubic lattice of dimension $d$, where $\rho_G(\vk,-\mu + x) = \frac{1}{t} \sum_{i=0}^\infty (\beta t)^i g^{(i)}(\vk,\frac{x}{t})$. Here, $m_n[g^{(i)}(\vk,y)]\equiv \int dy  \  g^{(i)}(\vk,y)y^n $. All moments displayed for $n+i\le 4$, except for those which vanish. We have defined $\tilde{\epsilon_k} \equiv \frac{1}{t} \epsilon_k$.
    \label{table_mom_g}}
    \end{table} 

\begin{table}
    \begin{tabular} { | p{2.0cm}  | p{6cm} |}
    \hline   
    $m_0[h^{(0)}(\vk,y)]$&$2 d (1-\frac{n}{4}) n$  \\ \hline
    $m_0[h^{(1)}(\vk,y)]$ &$-(1-n) n  \tilde{\epsilon} _k$ \\ \hline
    $m_0[h^{(2)}(\vk,y)]$ &$-\frac{d (n-1) n^2 \{2 d (n+2)+2(1-\frac{n}{2}) [2 (n-3) n+1]\}}{8 (1-\frac{n}{2})^2}$ \\ \hline
    $m_1[h^{(0)}(\vk,y)]$&$- (1-\frac{n}{4}) (1-\frac{n}{2}) n \tilde{\epsilon} _k$ \\ \hline
    $m_1[h^{(1)}(\vk,y)]$& $-\frac{d (1-n) n \{d [n (n+14)-8]-n (2 n+1)\}}{2(1-\frac{n}{2})}$\\ \hline
    $m_2[h^{(0)}(\vk,y)]$ &$ \frac{1}{8} d (2 d-1) n \{n [(1-\frac{n}{8}) n-9]+10\}$ \\ \hline
  $\Sigma^{(1)}_\infty(\vk)$& $\frac{2d (1-n) n }{1-\frac{n}{2}}$\\ \hline
   $\Sigma^{(2)}_\infty(\vk)$ & $-\frac{(1-n)^2 n^2 \tilde{\epsilon _k}}{4 (1-\frac{n}{2})}$ \\ \hline
    $\Sigma^{(3)}_\infty(\vk)$ &$\frac{d (1-n) n  \{n [d (13 n-16)-25 n+28]-6\}}{12 (1-\frac{n}{2})}$ \\ \hline
    \end{tabular}
    \caption{Moments of $h^{(i)}(\vk,y)$ for the infinite-$U$ Hubbard
    model on a hypercubic lattice of dimension $d$, where $\rho_\Sigma(\vk,-\mu + x) = t \sum_{i=0}^\infty (\beta t)^i h^{(i)}(\vk,\frac{x}{t})$ and $\Sigma(\vk,i\omega_n) = \Sigma_\infty(\vk) + \int d\nu \frac{\rho_\Sigma(\vk,\nu)}{i\omega_n-\nu}$. Here, $m_n[h^{(i)}(\vk,y)]\equiv \int dy  \  h^{(i)}(\vk,y)y^n $. All moments displayed for $n+i\le 2$, except for those which vanish. We have expanded $\Sigma_\infty(\vk)$ as $\Sigma_\infty(\vk) = t\sum_{i=0}^\infty (\beta t)^i \Sigma^{(i)}_\infty(\vk)$.
    \label{table_mom_h}}
    \end{table}

\section{Finite-$U$ Hubbard Model.\label{finiteU}}

The Hamiltonian for the Hubbard model is 

\beq
\hat{H} = -\sum_{ij\si}t_{ij} c_{i\si}^\dagger c_{j\si} +U\sum_i n_{i\uparrow}n_{i\downarrow}- \mu\sum_{i\si}n_{i\si}. 
\eeq

Suppose that $T>t$. Then, we can perform an expansion \cite{Metzner} in $(\beta t)$ and $\frac{t}{U}$, while holding $(\beta U)$ constant. Let us consider the resulting Green's function, which has the following form.

\beq
G(\vk,i\omega_n)= \sum_{j=0,r=0}^\infty\sum_{s=0}^r \frac{t^r (\beta t)^j l^{(j,r,s)}[\vk,n,e^{-\beta U}]}{U^s (i\omega_n+\mu)^{r+1-s}} + \sum_{j=0,r=0}^\infty\sum_{s=-\infty}^r\frac{t^r (\beta t)^j u^{(j,r,s)}[\vk,n,e^{-\beta U}]}{U^s(i\omega_n+\mu-U)^{r+1-s}},
\label{GLHBUHB}
\eeq
where $l^{(j,r,s)}[\vk,n,e^{-\beta U}]$ and $u^{(j,r,s)}[\vk,n,e^{-\beta U}]$ are coefficients which are a function of $\vk$, $n$ and $e^{-\beta U}$. For future reference, we also write down the real-space version of \disp{GLHBUHB} (see \disp{rs}).
\beq
G_{i,m}(i\omega_n)= \sum_{\substack{j=0,r=0\\p(j+r)=p(|i-m|)}}^\infty \sum_{s=0}^r \frac{t^r (\beta t)^j l^{(j,r,s)}[|i-m|,n,e^{-\beta U}]}{U^s(i\omega_n+\mu)^{r+1-s}} + \sum_{\substack{j=0,r=0\\p(j+r)=p(|i-m|)}}^\infty \sum_{s=-\infty}^r \frac{t^r (\beta t)^j u^{(j,r,s)}[|i-m|,n,e^{-\beta U}]}{U^s(i\omega_n+\mu-U)^{r+1-s}},\nn\\
\label{GLHBUHBrs}
\eeq
Here, the restriction that the parity of $(j+r)$ equal the parity of $|i-m|$ stems from \disp{chain} with $\hat{H}_1$ equal to the hopping term in the Hubbard Hamiltonian. In the second term on the RHS of Eqs. (\ref{GLHBUHB}) and (\ref{GLHBUHBrs}), the negative values of $s$ stem from terms proportional to $\frac{1}{i\omega_n + p U}$, where p is an integer, in which the denominator is expanded around $i\omega_n +U$. We now write the spectral function as the sum of two parts,
\beq
\rho_G(\vk,\nu)=\rho_{G,L}(\vk,\nu)+\rho_{G,U}(\vk,\nu),
\label{LHBUHB}
\eeq 
where $\rho_{G,L}(\vk,\nu)$ refers to the lower Hubbard band (LHB),
and $\rho_{G,U}(\vk,\nu)$ refers to the upper Hubbard band (UHB).
We expand the spectral functions of the lower and upper Hubbard bands in analogy with \disp{betaseriesG}.
\beq
\rho_{G,L}(\vk,-\mu+x) &=& \frac{1}{t} \sum_{j=0}^\infty (\beta t)^j g_L^{(j)}(\vk,\frac{x}{t}), \nn\\
\rho_{G,U}(\vk,-\mu + U+x) &=& \frac{1}{t} \sum_{j=0}^\infty (\beta t)^j g_U^{(j)}(\vk,\frac{x}{t}),
\label{GLGU}
\eeq
where both $g_L^{(j)}(\vk,\frac{x}{t})$ and $g_U^{(j)}(\vk,\frac{x}{t})$ are centered on $x=0$. Their moments are given by
\beq
\int g_{L}^{(j)}(\vk,y) y^r dy &=& \sum_{s=0}^\infty  \left(\frac{t}{U}\right)^sl^{(j,r+s,s)}[\vk,n,e^{-\beta U}],\nn\\
\int g_{U}^{(j)}(\vk,y) y^r dy &=& \sum_{s=-r}^\infty  \left(\frac{t}{U}\right)^su^{(j,r+s,s)}[\vk,n,e^{-\beta U}].
\label{gLHBUHBmoments}
\eeq
Therefore, in the case of the finite-$U$ Hubbard model, the moments of $g_{L}^{(j)}(\vk,y)$ and $g_{U}^{(j)}(\vk,y)$ are themselves an infinite series in $\frac{t}{U}$. In the following section, we evaluate some of these moments to low orders.

\subsection{Analytically calculated moments of the upper and lower Hubbard bands. \label{momentsLHBUHB}}
We perform the expansion from \refdisp{Metzner} through second order in $(\beta t)$ and $\frac{t}{U}$. This yields
\beq
G^{(0)}(\vk,i\omega_n) &=& \frac{1-\frac{n}{2}}{i\omega_n+\mu} + \frac{\frac{n}{2}}{i\omega_n+\mu-U},\nn\\
G^{(1)}(\vk,i\omega_n) &=& \epsilon_k [G^{(0)}(\vk,i\omega_n)]^2,\nn\\
G^{(2)}(\vk,i\omega_n) &=& G^{(2)}_{loc}(i\omega_n) - 2 d t^2 [G^{(0)}(\vk,i\omega_n)]^3 + \epsilon_k^2 [G^{(0)}(\vk,i\omega_n)]^3,
\label{G2ndorderfiniteU}
\eeq
where 
\beq
G^{(2)}_{loc}(i\omega_n)&=& \frac{2 c  t^2 \beta }{U \left( i \omega _n+\mu\right)}-\frac{2 c  t^2 \beta }{U \left(i \omega
   _n+\mu-U \right)}+\frac{c  t^2 \beta }{\left(i \omega _n+\mu-U
   \right){}^2}\nn\\
   &&+\frac{c  t^2 \beta }{\left(i \omega _n+\mu \right){}^2}+\frac{d n
   t^2}{\left(i \omega _n+\mu-U \right){}^3}-\frac{d (n-2) t^2}{\left(i \omega _n+\mu
   \right){}^3},
\label{G2locfiniteU}   
\eeq
where $c=\frac{d (n-2) (n-1) n}{\sqrt{(n-1)^2-(n-2) n e^{-U \beta
}}+1}$. Note that this gives the correct result in the case of half-filling (\refdisp{Pairault}). Using Eqs. (\ref{GLHBUHB}) and (\ref{gLHBUHBmoments}), we compute the moments of $g_{loc,L}^{(j)}(\vk,\frac{x}{t})$ and $g_{loc,U}^{(j)}(\vk,\frac{x}{t})$, where the subscript $loc$ denotes local. The results are given in table.~\ref{table_mom_g_finiteU}.

\begin{table}
    \begin{tabular} { | p{2.0cm}  | p{10cm} |}
    \hline   
    $m_0[g_{loc,L}^{(0)}(y)]$&$1-\frac{n}{2}+O[\left(\frac{t}{U}\right)^4]$  \\ \hline
    $m_1[g_{loc,L}^{(0)}(y)]$ &$0+O[\left(\frac{t}{U}\right)^3]$ \\ \hline
    $m_2[g_{loc,L}^{(0)}(y)]$ &$(2-n)d+O[\left(\frac{t}{U}\right)^2]$ \\ \hline
    $m_0[g_{loc,L}^{(1)}(y)]$&$2 c \left(\frac{t}{U}\right) +O[\left(\frac{t}{U}\right)^3]$ \\ \hline
    $m_1[g_{loc,L}^{(1)}(y)]$& $c+O[\left(\frac{t}{U}\right)^2]$\\ \hline
    $m_0[g_{loc,L}^{(2)}(y)]$ &$0+O[\left(\frac{t}{U}\right)^2]$ \\ \hline
    $m_0[g_{loc,U}^{(0)}(y)]$&$\frac{n}{2}+O[\left(\frac{t}{U}\right)^4]$  \\ \hline
    $m_1[g_{loc,U}^{(0)}(y)]$ &$0+O[\left(\frac{t}{U}\right)^3]$ \\ \hline
    $m_2[g_{loc,U}^{(0)}(y)]$ &$d n+O[\left(\frac{t}{U}\right)^2]$ \\ \hline
    $m_0[g_{loc,U}^{(1)}(y)]$&$-2 c \left(\frac{t}{U}\right) +O[\left(\frac{t}{U}\right)^3]$ \\ \hline
    $m_1[g_{loc,U}^{(1)}(y)]$& $c+O[\left(\frac{t}{U}\right)^2]$\\ \hline
    $m_0[g_{loc,U}^{(2)}(y)]$ &$0+O[\left(\frac{t}{U}\right)^2]$ \\ \hline
    \end{tabular}
     \caption{Moments of $g_{loc,L}^{(i)}(y)$ and $g_{loc,U}^{(i)}(y)$ for the finite-$U$ Hubbard model on a hypercubic or Bethe lattice of dimension $d$, where $\rho_G(\vk,-\mu+x)=\frac{1}{t} \sum_{i=0}^\infty (\beta t)^i \left[g_L^{(i)}(\vk,\frac{x}{t})+g_U^{(i)}(\vk,\frac{x-U}{t})\right]$. Here, $m_n[g_{loc,L(U)}^{(i)}(y)]\equiv \int dy  \  g_{loc,L(U)}^{(i)}(y)y^n $, and $c=\frac{d (n-2) (n-1) n}{\sqrt{(n-1)^2-(n-2) n e^{-U \beta }}+1}$. All moments displayed for $n+i\le 2$. Upon taking the $U\to\infty$ limit, we recover the results of Table~\ref{table_mom_g}.
\label{table_mom_g_finiteU}}
    \end{table}

\subsection{Analytically calculated moments of the optical conductivity \ag{and high-temperature transport}  for $d\to\infty$. \label{finiteUcond}}

\ag{In this section, we present an analysis of transport at high-temperature in the finite-$U$ case. 
This analysis uses the simplifications associated with the $d=\infty$ limit. In contrast to the more 
rigorous derivation at $U=\infty$, it is not based on the calculation of a large number of moments but 
rather on a qualitative analysis of the optical conductivity in the lower frequency range based on 
just the dominant term in the high-$T$ series.}
{\color{black} The series considered is an expansion in both ($\beta D$)
  and $\frac{D}{U}$, where $D$ is the half-bandwidth. Therefore, we
  only consider values of $U> D$, in which case the density of states
  (DOS) consists of the two well separated Hubbard bands.}
The optical conductivity reflects optical transitions that either
occur within a single Hubbard band, or transitions that occur between
the two Hubbard bands. The former contribute to $\sigma(\omega)$ in
the vicinity of $\omega \sim 0$, while the latter contribute in the
vicinity of $\omega\sim U$. To determine the $dc$ conductivity, we
compute the moments of the part of the $\sigma(\omega)$ curve in the
vicinity of $\omega\sim0$. To simplify the calculation, we shall
assume that $d\to\infty$, which allows us to neglect vertex
corrections~\cite{khurana_vertex} in computing $\sigma(\omega)$. Since we are in the
infinite-dimensional limit, the conductivity along a single spatial
direction vanishes like $\frac{1}{d}$, therefore we redefine
$\sigma(\omega)$ as $\sigma(\omega)=\sum_\alpha
\sigma_{\alpha\alpha}(\omega)$. Recalling that $\lim_{k\to0}\frac{\chi_{J_{\alpha}.J_{\alpha}}^{''}(\vk,\omega)}{\omega}\equiv \frac{\si_{\alpha\alpha}(\omega)}{\pi\si_0}$, we make use of the ``bubble" formula for a hypercubic lattice  in the infinite-dimensional limit:
\beq
\sum_\alpha\lim_{k\to0}\chi_{J_\alpha.J_\alpha}(\vk,\tau) = 2\sum_{\vq} D^2 \sin^2 q_x \ G(\vq,\tau)G(\vq,-\tau).
\label{bubble}
\eeq
Taking the Fourier transform of the top line in \disp{G2ndorderfiniteU}, we find that
\beq
G^{(0)}(\vk,\tau) &=& -(1-\frac{n}{2}) e^{\mu^{(0)}\tau}\left[\frac{1}{1+e^{\beta\mu^{(0)}}}\Theta(\tau)-\frac{1}{1+e^{-\beta\mu^{(0)}}}\Theta(-\tau)\right]\nn\\
&&-\frac{n}{2}e^{(\mu^{(0)}-U)\tau}\left[\frac{1}{1+e^{\beta(\mu^{(0)}-U)}}\Theta(\tau)-\frac{1}{1+e^{-\beta(\mu^{(0)}-U)}}\Theta(-\tau)\right],
\label{G0taufiniteU}
\eeq
where $\mu^{(0)}$ is the chemical potential in the atomic limit, and therefore $e^{\beta\mu^{(0)}}=\frac{n-1+\sqrt{(1-n)^2+2n(1-\frac{n}{2})e^{-\beta U}}}{2(1-\frac{n}{2})e^{-\beta U}}$.  Plugging \disp{G0taufiniteU} into \disp{bubble}, using \disp{zerofreq}, and making the substitutions $(1-n)\to\delta$ and $e^{-\beta U}\to y$, we find that
\beq
\int d\omega \frac{\tilde{\si}(\omega)}{\pi\si_0} = \frac{\beta D^2  \left[y \left(\sqrt{-y \delta ^2+y+\delta ^2}+\delta^2-2\right)+\sqrt{-y \delta ^2+y+\delta ^2}-\delta ^2\right]}{2
   (y-1)^2} + O(\beta^2D^3).
\label{fsumrulefiniteU}
\eeq
\ag{In this expression, $\tilde{\si}(\omega)$ designates} the part of the $\si(\omega)$ curve in the vicinity of $\omega\sim0$, as opposed to the part in the vicinity of $\omega\sim U$. Therefore, in writing \disp{fsumrulefiniteU}, we have only considered optical transitions which occur within the two Hubbard bands and neglected those which occur between the bands. {\color{black} At this point, we make the approximation that all higher order moments of $\tilde{\si}(\omega)$ have the same dependence on $U$ and $\delta$ as does the zeroth moment. We shall later justify this approximation in all parameter ranges in which it is applied. Then, $\si_{dc}$ is equal to the RHS of \disp{fsumrulefiniteU} up to some scale factor, i.e. $\frac{\si_{dc}}{\si_0} \propto \frac{1}{D}\int d\omega \frac{\tilde{\si}(\omega)}{\pi\si_0}$. In the infinite temperature limit, i.e. $T\gg U$, we set $y\to 1$ in \disp{fsumrulefiniteU} to find

\beq
\frac{\rho_{dc}}{\rho_0} \propto \frac{8T}{D(1-\delta^4)}.
\label{reshighTfiniteU}
\eeq
The resistivity is linear in the temperature, as we have shown must be the case on general grounds.} 
\ag{Note that, in this regime, the slope tends to a constant value at low-doping, 
 in contrast to the $U=\infty$ case where it diverges.}

\begin{figure}
\begin{center}
\includegraphics[width=.45 \textwidth ]{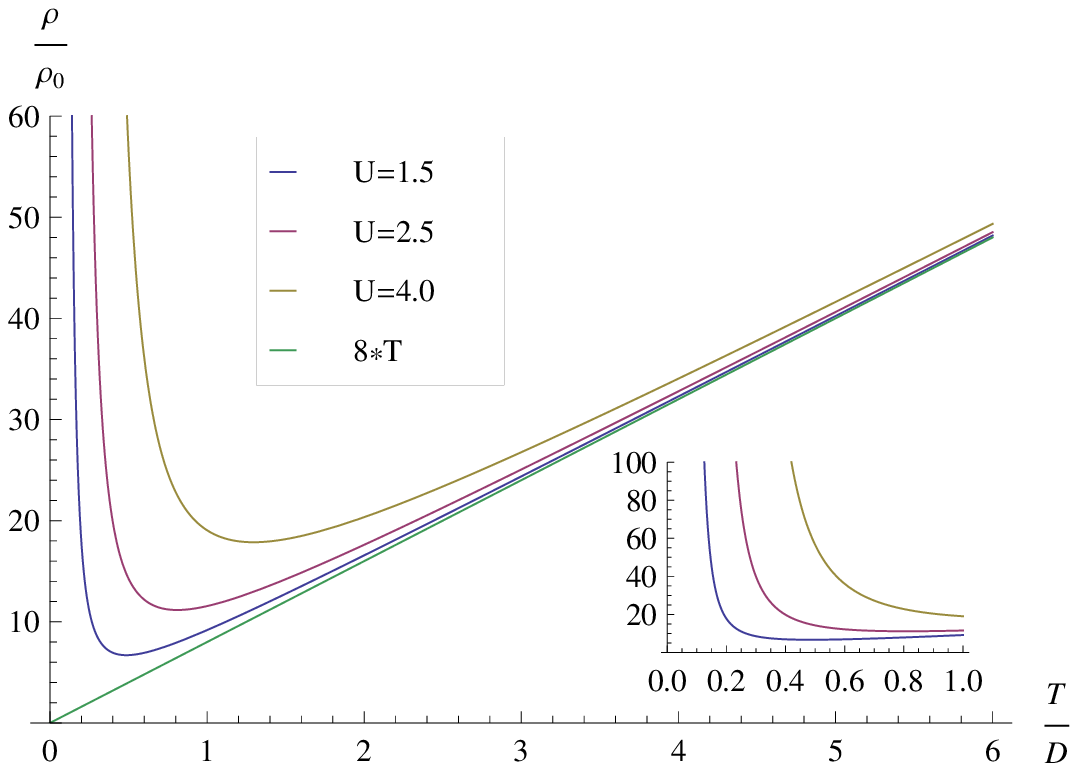}
\includegraphics[width=.45 \textwidth ]{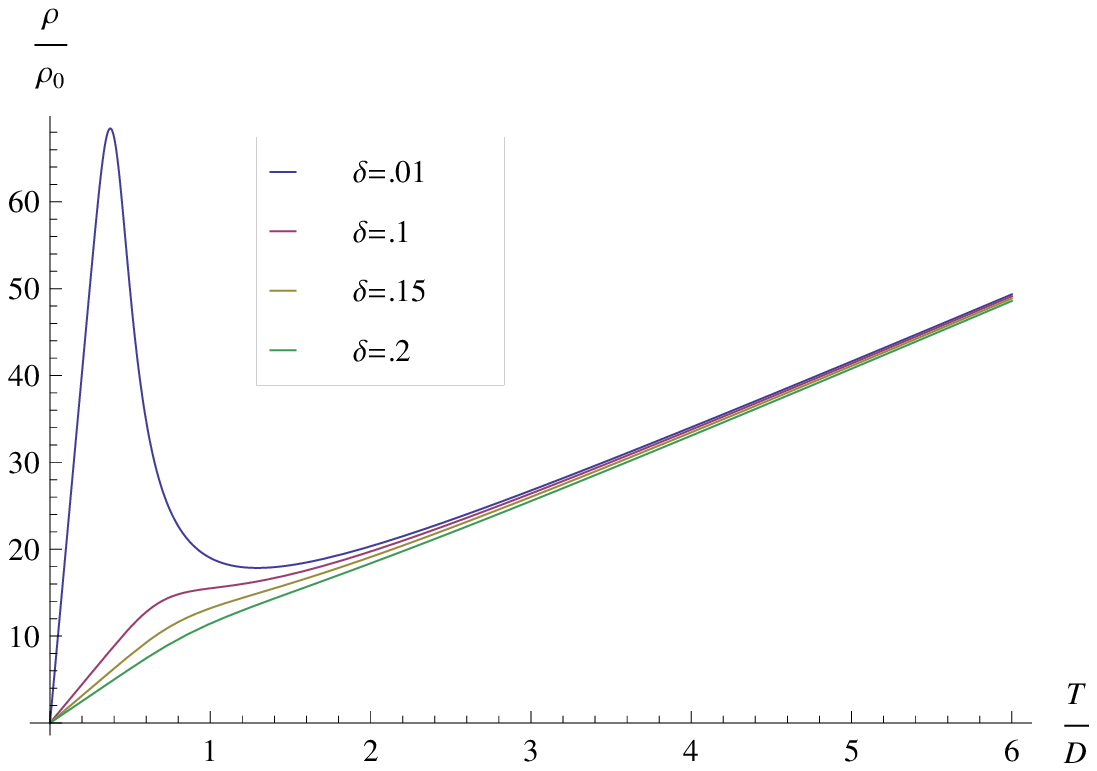}
\caption{ \rzjm{Leading order series result for the temperature
dependence of the dc resistivity for the finite-$U$ Hubbard model on
an infinite-dimensional hypercubic lattice. Left panel: half-filled
case for various values of $U$. The resistivity first decays
exponentially with the temperature followed by a linear regime with a
slope which asymptotically approaches $8$, independent of the value of
$U$. The inset is a close-up to the range
$\frac{T}{D}\le1$.  Right panel: the case of $U=4$ for various values
of the doping $\delta$. The resistivity curve consists of two linear
regimes connected by a regime of exponential decay, which is both
wider and more pronounced for smaller $\delta$. } }
\label{res_finiteU_hf}
\end{center}
\end{figure}

{\color{black} We now examine the dc resistivity in the half-filled
Hubbard model. Setting $\delta\to0$ in \disp{fsumrulefiniteU}, we obtain
\beq
\frac{\rho_{dc}}{\rho_0}\propto\frac{2 T }{D}\left(1+e^{\frac{U}{2T}}\right)\left(1+e^{-\frac{U}{2T}}\right).
\label{res_eq_finiteU_hf}
\eeq
We plot this resistivity vs. temperature curve in \figdisp{res_finiteU_hf} for $U=1.5$, $U=2.5$, and $U=4.0$. For all values of $U$, the resistivity decays exponentially with increasing temperature starting from $T=0$ followed by a crossover to the high-$T$ regime, in which all curves converge to a linear resistivity 
\ag{$\sim 8T/D$, in accordance with \disp{reshighTfiniteU} with $\delta=0$.}

DMFT results for the half-filled case can be found in 
\ag{refs. [\onlinecite{limelette_prl_2003}], [\onlinecite{pakhira15}] and [\onlinecite{Kurosakietal}].
}
For $U>U_c$, the system is a Mott insulator, and the resistivity decays exponentially with increasing temperature starting from $T=0$. 
\ag{There is a characteristic value $U^*$ such that} for $U^*<U<U_c$, starting from $T=0$, the
resistivity increases rapidly with increasing temperature reaching a maximum, after which it decays exponentially with increasing
temperature. Finally, for $U<U^*$,  the resistivity \rzjm{increases} monotonically as a function of the temperature for all values of the temperature. In all cases, the DMFT calculations have not been done to sufficiently high temperatures to display the high-T ($T\gg U$) linear regime, although we know that it must be there. In addition, the slope must be independent of the value of $U$ (\disp{reshighTfiniteU}), and hence all of these resistivity curves must converge at high enough temperatures.
 
For the case of the Mott insulator ($U>U_c$), the leading order series result accurately describes the DMFT result for all values of the temperature. For the metallic state with $U^*<U<U_c$, the leading order series correctly captures the exponential decay and subsequent linear growth of the resistivity at high-$T$, but \ag{obviously} not the initial increase of the resistivity with increasing temperature in the vicinity of $T=0$, \ag{which has to do with the Fermi liquid regime of the metal}. For the metallic state with $U<U^*$, the leading order series captures only the asymptotic high-$T$ linear regime. It is an open question whether or not the higher order terms in the series can at least partially correct for the shortcomings of the leading order term in the metallic state.

Also in \figdisp{res_finiteU_hf}, we plot the leading order series result for the doped Mott insulator, corresponding to $U=4$ and $\delta=.01$, $\delta=.1$, $\delta=.15$, and $\delta=.2$. The resulting resistivity vs. temperature curves all have three distinct regimes. In the high-$T$ ($T\gg U$) regime, each resistivity curve is linear in temperature with slope given by \disp{reshighTfiniteU}. As the temperature is lowered, we reach a regime in which $U\gg T$, but $\delta < e^{-\frac{U}{2T}}$. Asymptotically, this corresponds to taking the $\delta \to 0$ limit followed by the $y\to 0$ limit of \disp{fsumrulefiniteU}, and gives $\frac{\rho_{dc}}{\rho_0} \propto \frac{2 T }{D}  e^{\frac{U}{2T}}$, as can be seen from \disp{res_eq_finiteU_hf}. For finite $\delta$, increasing $\delta$ lowers the peak height from this asymptotic result. Finally, as the temperature is lowered even further, we reach the regime in which $U\gg T$ and $\delta > e^{-\frac{U}{2T}}$. Asymptotically, this corresponds to taking the $y\to0$ limit of \disp{fsumrulefiniteU}, and hence we recover the infinite-$U$ result $\frac{\rho_{dc}}{\rho_0} \propto \frac{2 T }{D \delta (1-\delta)}$ (\disp{f1zerothmomentinfU}). This yields a linear resistivity whose slope diverges as $\delta\to0$. We expect that the higher order terms in the series will correct this leading order result and reproduce the DMFT resistivity \cite{Badmetal} in the bad-metal regime ($\frac{\rho_{dc}}{\rho_0}>1$), as we have shown to be the case for the infinite-$U$ Hubbard model (see sec.~\ref{sec:resist_comp}).

We now go back and comment on our approximation that the higher order
moments of $\tilde{\si}(\omega)$ have the same functional dependence
on $U$ and $\delta$ as the zeroth moment. This will generally be true
if $\tilde{\si}(\omega)$ is a smooth featureless function that does
not have any sharp peaks. To illustrate this, we consider the
infinite-$U$ Hubbard model. In \figdisp{f1}, we see that in the
$\delta\to0$ limit, $\sigma(\omega)$ (which in this case is the same
as $\tilde{\si}(\omega)$) is a featureless function. However in the
$\delta\to1$ limit, it contains a sharp peak followed by a long tail.
Accordingly, in the $\delta\to0$ limit, all higher order moments have
the same functional dependence on $\delta$ as does the zeroth moment,
while in the $\delta\to1$ limit, this is not the case. Therefore, it
turns out that the estimate for $\rho_{dc}$ given above on the basis
of the zeroth moment, i.e. $\frac{\rho_{dc}}{\rho_0} \propto \frac{2 T
}{D \delta (1-\delta)}$ is only correct in the $\delta\to0$ limit, but
not in the $\delta\to1$ limit. Returning to the finite-$U$ Hubbard
model, we know that at low-$U$ and low-$T$, $\tilde{\si}(\omega)$
contains a sharp Drude peak. Therefore, we would not expect the zeroth
moment to provide a good estimate for the dc resistivity in this case.
However, here we have considered the limit where both $U$ and $T$ are
large. In this parameter range, the Drude peak is either greatly diminished or disappears altogether, and $\tilde{\si}(\omega)$ becomes a smooth featureless function \cite{Badmetal,PCJ, JFP, Merino_2008}. 
}

\section{The infinite-U Hubbard model in the $d\to\infty$ limit\label{dinfinity}}

In this section, we consider the $U=\infty$ Hubbard model in the limit
of large dimensionality, $d\to\infty$, where the formalism simplifies
further. A quantitative comparison to DMFT numerical 
solutions will be made. We will see in particular that the
high-temperature series allows for an essentially 
analytical calculation of the optical conductivity and dc-resistivity, which compares very well to numerical 
solutions down to a surprisingly low temperature. This allows one to better understand qualitatively the transport 
mechanisms in the Hubbard model at high and intermediate temperatures.

We shall consider two cases:
\begin{itemize}
\item The $d=\infty$ cubic lattice (referred to as 'hypercubic' lattice). In this case, the non-interacting 
density of states (DOS) is a Gaussian: 
\beq
D(\epsilon) = \frac{1}{D\sqrt{\pi}} e^{-\frac{\epsilon^2}{D^2}}
\eeq
with $D^2=4 d t^2$, which fulfills the relation $\int d\epsilon D(\epsilon) \epsilon^2 = 2 d t^2$. 
\item The infinite-connectivity Bethe lattice, which has a semi-circular DOS:  
\beq
D(\epsilon) = \frac{2}{\pi D^2} \sqrt{D^2-\epsilon^2} \ \Theta(D-|\epsilon|)
\eeq
with $D^2=8 d t^2$, which fulfills the relation $\int d\epsilon D(\epsilon) \epsilon^2 = 2 d t^2$.
\end{itemize}

Thus far, we have mostly considered quantities in momentum space, but
we can equally well consider their real-space versions. The relationship between the two is given by the relation
\beq
Q_{i,j} = \frac{1}{N_s} \sum_{\vk} e^{i \vk \cdot (\vR_i-\vR_j)} Q(\vk),
\label{rs}
\eeq
where $Q$ can stand for any quantity such as $G$, $\Sigma$, $\chi$, $f$, $g$, or $h$. Furthermore, all of the formulas derived thus far (with the exception of Dyson's equation \disp{Dyson}) continue to hold if we make the substitutions $Q(\vk)\to Q_{i,j}$, $c_{k\si}\to c_{i\si}$, $c^\dagger_{k\si}\to c^\dagger_{j\si}$, $\hat{O}_{k} \to \hat{O}_i$, and $\hat{O}_{-k} \to \hat{O}_j$. Now, consider the moments of the real-space density of states $\rho_{G,ij}(\omega)$. Then, the real space version of \disp{momentsg} together with \disp{chain} tell us that 
\beq
\int dy  \  g_{i,j}^{(i)}(y) y^{m} \propto \delta_{p(|i-j|),p(m+i)}.
\label{ijeqmi}
\eeq
In particular, if we consider the local density of states $\rho_{G,jj}(\omega)$, this tells us that  
$g_{j,j}^{(i)}(\vk,y)$ is an even function of $y\equiv (\omega+\mu)/t$ 
for $i$ even, and an odd function for $i$ odd. Hence, for example, the
leading high-$T$ contribution 
to the local spectral density is an even function centred at $\mu=T\overline{\mu}$

In the $d\to\infty$ limit, the self-energy becomes local \cite{Kuramoto,Muller-Hartmann,Metzner}, i.e. $\Sigma(\vk,i\omega_n)\to \Sigma(i\omega_n)$. Therefore, in the infinite-$U$ Hubbard model in the $d\to\infty$ limit, Dyson's equation (\disp{Dyson}) takes on the particular form
\beq
G(\vk,i\omega_n) = \frac{1-\frac{n}{2}}{i\omega_n+\mu-(1-\frac{n}{2})t\tilde{\epsilon_k} - \Sigma(i\omega_n)}.
\label{DysoninfU}
\eeq
Furthermore, due to \disp{chain}, the series for $G_{i,j}(i\omega_n)$ contains only powers of $t$ of the same parity as $|i-j|$. Expanding the RHS of \disp{DysoninfU} in powers of $t$ then implies that the series for $\Sigma(i\omega_n)$  contains only even powers of $t$. Finally, performing the high-frequency expansion of \disp{spectralSig} and plugging in \disp{betaseriesSig}
yields
\beq
\Sigma(i\omega_n) = t\sum_{i=0}^\infty (\beta t)^i \Sigma_\infty^{(i)}+ \sum_{m=1}^\infty \frac{t^{m+1}}{(i\omega_n+\mu)^{m}}\sum_{i=0}^\infty (\beta t)^i \int dy  \  h^{(i)}(y) y^{m-1} .
\label{spectralSigmuinfd}
\eeq
Putting this all together, we find that 
\beq
\int dy  \  h^{(i)}(y) y^{m} \propto \delta_{0,p(m+i)};\;\;\;\;\;\; \Sigma_\infty = t\sum_{\substack{i=0\\odd}}^\infty (\beta t)^i \Sigma^{(i)}_\infty.
\eeq
For $i$ even (odd), $h^{(i)}(y)$ is an even (odd) function. 

\subsection{Moments and reconstruction of the self-energy\label{momentsSEinfd}}


Using the method from \refdisp{Metzner}, we have evaluated the zeroth
through sixth moments of the self-energy in the $d\to\infty$ limit. Changing the
characteristic energy from $t$ to $D$, and thereby redefining the
constants $\Sigma_\infty^{(i)}$ and the functions $h^{(i)}(y)$, we define
\begin{equation}
\begin{split}
\rho_\Sigma(-\mu + x) &= D
\sum_{i=0}^\infty (\beta D)^i h^{(i)}(\frac{x}{D}), \\
\Sigma(i\omega_n) &= \Sigma_\infty+ \int d\nu
\frac{\rho_\Sigma(\nu)}{i\omega_n-\nu}.
\end{split}
\label{betaseriesSiginfd}
\end{equation}
We first consider the case of a hypercubic lattice. We
give the leading contributions to the moments of $h^{(i)}(y)$ in
Table~\ref{table_mom_h_dinf}, 
with $m_n[h^{(i)}(y)]\equiv \int dy \ h^{(i)}(y)y^n $. We have
expanded $\Sigma_\infty$ as $\Sigma_\infty = D\sum_{i=0}^\infty(\beta
D)^i \Sigma^{(i)}_\infty$. We also calculated higher order expressions
but they are are too lengthy to write here \cite{momentsweb}. They will be used to reconstruct the
spectral function of $\Sigma$, $\rho_\Sigma$. Note that these
expressions can be obtained from those in section \ref{angrself} by
taking the leading order term in $d$ and then setting
$d\to\frac{1}{4}$ (reflecting the use of $D$ instead of $t$ as our
basic energy unit in this section). 
Once again we find that both the moments of $\rho_\Sigma(-x+\mu)$ and $\Sigma_\infty$ vanish linearly as $n\to0$, and that $\Sigma_\infty$ vanishes linearly as $n\to1$. In addition, in contrast to the finite-dimensional case, we find that the odd moments of $\rho_\Sigma(-x+\mu)$ vanish linearly as $n\to1$.

\begin{table}
    \begin{tabular} { | p{2.0cm}  | p{4cm} |}
    \hline    
    $m_0[h^{(0)}(y)]$&$-\frac{1}{8} (n-4) n$  \\ \hline
    $m_0[h^{(2)}(y)]$ &$-\frac{n^2 \left(n^2+n-2\right)}{16 (n-2)^2}$ \\ \hline
    $m_1[h^{(1)}(y)]$& $-\frac{(n-1) n (n (n+14)-8)}{16 (n-2)}$\\ \hline
    $m_2[h^{(0)}(y)]$ &$-\frac{1}{64} n (n ((n-8) n+72)-80)$ \\ \hline
  $\Sigma^{(1)}_\infty$& $\frac{(n-1) n}{2 (n-2)}$\\ \hline
    $\Sigma^{(3)}_\infty$ &$\frac{(n-1) n^2 (13 n-16)}{96 (n-2)}$ \\ \hline
    \end{tabular}
\caption{Moments of $h^{(i)}(y)$ for the infinite-$U$ Hubbard model on
a hypercubic lattice for $d=\infty$, where $m_n[h^{(i)}(y)]\equiv \int dy \ h^{(i)}(y)y^n $. All moments displayed for $n+i\le 2$,
except for those which vanish.
\label{table_mom_h_dinf} 
}
    \end{table} 


In Table~\ref{table_mom_h_Bethe}, we display the moments for the infinite-connectivity Bethe lattice. 
Once again, the higher order expressions will not be written
explicitly \cite{momentsweb}, but will be used to reconstruct the
spectral function $\rho_\Sigma$. The $n\to0$ and $n\to1$ limits behave in the same way as for the hypercubic case above.

\begin{table}
    \begin{tabular} { | p{2.0cm}  | p{4cm} |}
    \hline    
    $m_0[h^{(0)}(y)]$&$ -\frac{1}{16} (n-4) n$  \\ \hline
    $m_0[h^{(2)}(y)]$ &$ -\frac{n^2 \left(n^2+n-2\right)}{64 (n-2)^2}$ \\ \hline
    $m_1[h^{(1)}(y)]$& $-\frac{(n-1) n (n (n+14)-8)}{64 (n-2)}$\\ \hline
    $m_2[h^{(0)}(y)]$ &$\frac{1}{64} (16-13 n) n$ \\ \hline
  $\Sigma^{(1)}_\infty$& $\frac{(n-1) n}{4 (n-2)}$\\ \hline
    $\Sigma^{(3)}_\infty$ &$-\frac{(n-1) n}{96 (n-2)}$ \\ \hline 
    \end{tabular}
\caption{Moments of $h^{(i)}(y)$ for the infinite-$U$ Hubbard model on
a Bethe lattice for $d=\infty$, where $m_n[h^{(i)}(y)]\equiv \int dy \ h^{(i)}(y)y^n $.
All moments displayed for $n+i\le 2$,
except for those which vanish.
\label{table_mom_h_Bethe}
    }
    \end{table} 
    
We now consider the reconstruction of the asymptotic high-temperature self-energy, $h^{(0}(x)$, from the calculated moments. 
Note that, in this limit, $h^{(0}(x)$ is an even function. In order to reconstruct this function from the moments calculated 
above, we use the maximum entropy method (MEM), which can be summarized as follows. 
Suppose that one has an even distribution $P(x)$, whose first $k +1$ even moments are known. The MEM estimate for $P(x)$ is 
\beq
P(x) = \exp[{-\sum_{n=0}^k \lambda_{2n} x^{2n}}].
\eeq
The $\lambda_{2n}$ are then chosen such that
\beq
\int dx P(x) x^{2n}=m_{2n} \;\;\;\;\;\;(n=0\ldots k),
\eeq
where the $m_n$ are the known moments of $P(x)$. Details of the MEM are provided in Appendix~\ref{MEM}. 
%

\begin{figure}
\begin{center}
\includegraphics[width = .44 \textwidth ]{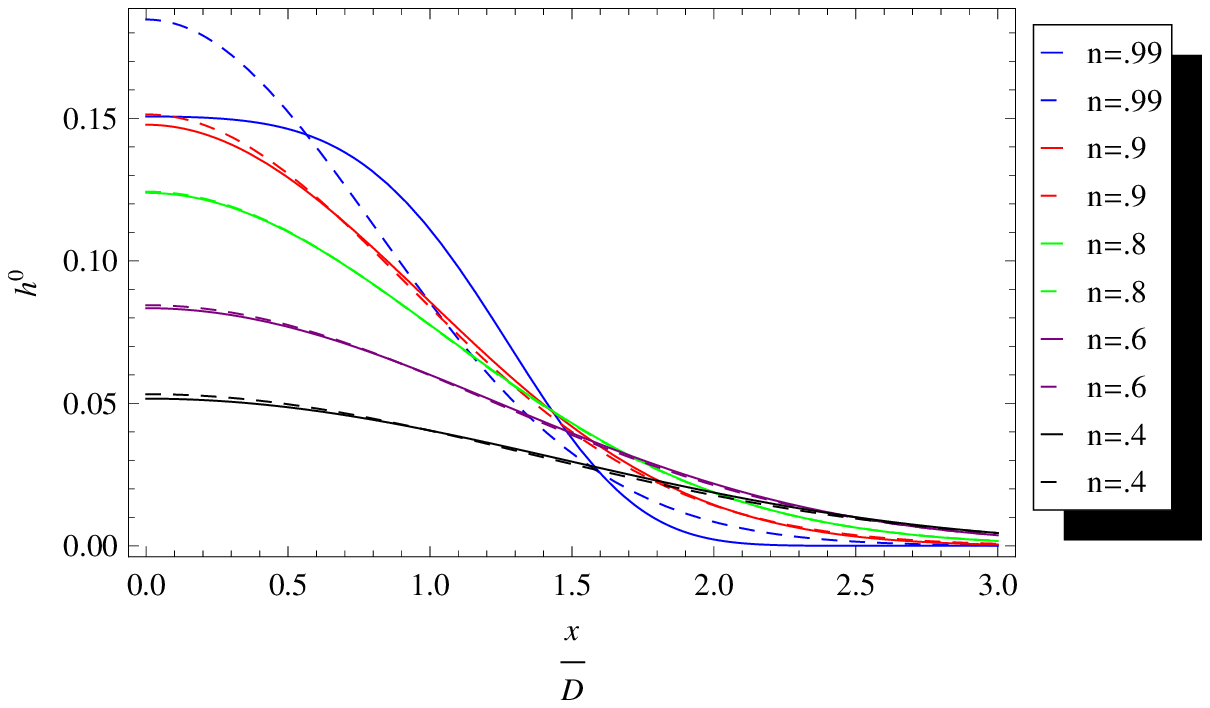}
\includegraphics[width = .45 \textwidth ]{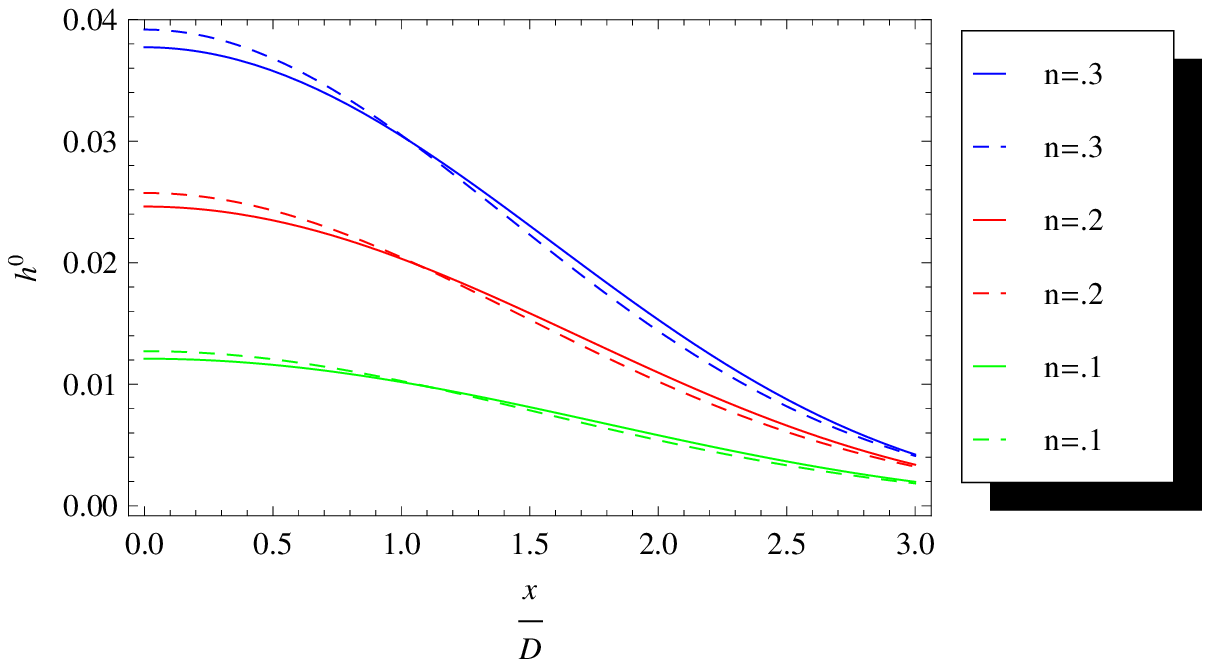}
\caption{
The function $h^{(0)}(\frac{x}{D})$ associated with the self-energy spectral density: 
$\rho_\Sigma(-\mu + x) = D h^{(0)}(\frac{x}{D}) + \ldots$, for the infinite-$U$ Hubbard model on
a hypercubic lattice for $d=\infty$. This function is even and displayed here for $x>0$ only.
The solid curves are obtained using the MEM with the zeroth through eighth moments, while the dashed curves 
are the Gaussian approximation obtained using the MEM with just the zeroth and second moments. 
The latter agrees well with the former at all but the highest density.}
\label{h0}
\end{center}
\end{figure}

\begin{figure}
\begin{center}
\includegraphics[width = .43 \textwidth ]{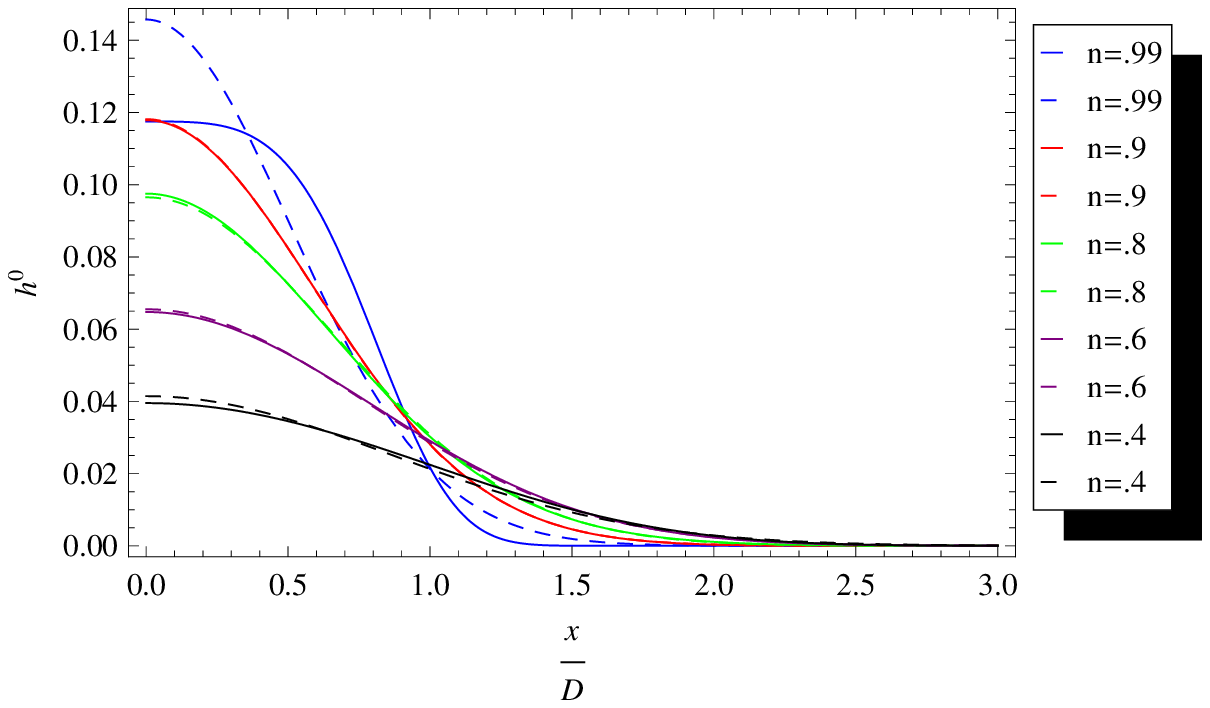}
\includegraphics[width = .45 \textwidth ]{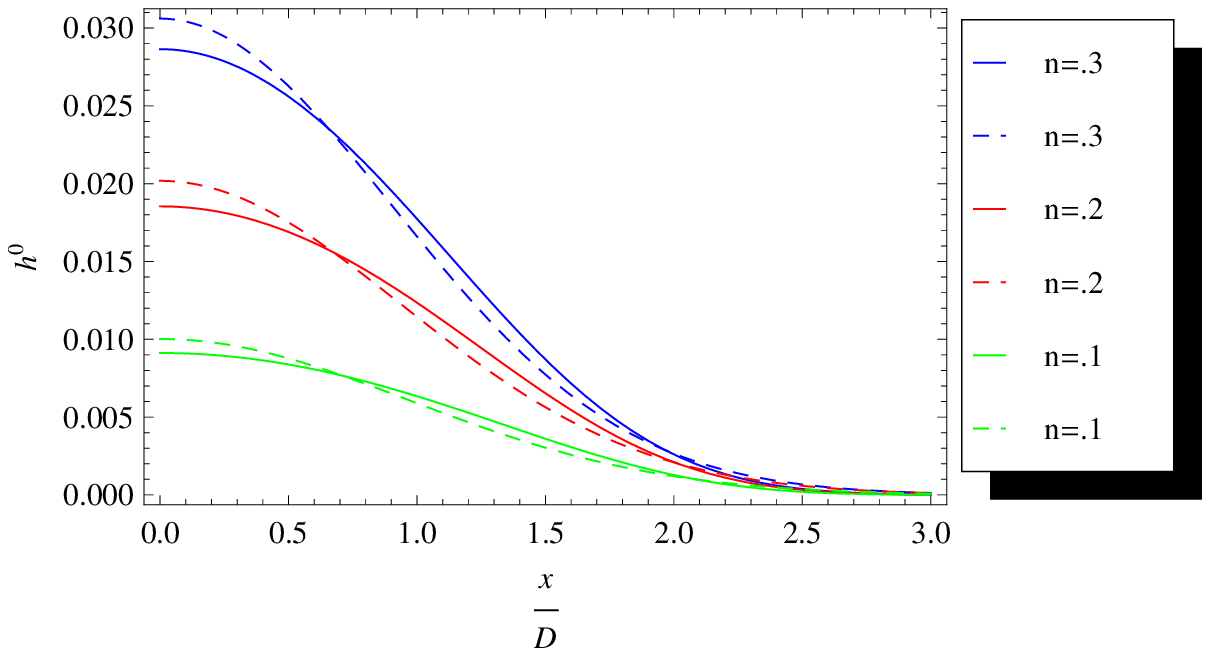}
\caption{
The function $h^{(0)}(\frac{x}{D})$ associated for the infinite-$U$ Hubbard model on the infinite-connectivity 
Bethe lattice.
}
\label{h0Bethe}
\end{center}
\end{figure}

The results for the MEM-reconstructed $h^{(0)}$ are displayed in Figs.~\ref{h0} and \ref{h0Bethe} for the hypercubic and Bethe lattice, respectively.  
We see that this function turns out to be quite close to a Gaussian form, at all but the highest values of density. 
As noted above, the function $h^{(0)}(\frac{x}{D})$ vanishes linearly with $n$ as $n\to0$.

\subsection{Moments and reconstruction of the optical conductivity}
\label{reconstructionoptical}

We now consider the reconstruction of the optical conductivity for the infinite-dimensional hypercube based on the moments of 
the current-current correlation function. 
The high-$T$ expansion of the $\vec{q}=\vec{0}$ conductivity takes the form (\disp{betaseriesinfUcond}): 
\beq 
\frac{\si(\omega)}{\pi\si_0} = \frac{D}{T} f^{(1)}(\frac{\omega}{D}) + \frac{ D^3}{T^3} f^{(3)}(\frac{\omega}{D}) + \ldots,
\label{betaseriesinfUcondinfd}
\eeq
where $f^{(i)}(\frac{\omega}{D}) \equiv \lim_{k\to0}f^{(i)}(\vk,\frac{\omega}{D})$. Note that while in \disp{betaseriesinfUcond} the conductivity is calculated along a single direction, in \disp{betaseriesinfUcondinfd} it is summed over all spatial directions. 
Due to the relation $dt^2=\frac{D^2}{4}$, there is a factor of $\frac{1}{4}$ which has been absorbed into the functions $f^{(i)}(\frac{\omega}{D})$ in \disp{betaseriesinfUcondinfd} relative to those in \disp{betaseriesinfUcond}. 
On the hypercubic lattice, we have calculated the moments $m_n[f^{(i)}(y)]\equiv\ \int dy  \  f^{(i)}(y)y^n $ for all $n+i\le8$.
\beq
\int dy f^{(1)}(y) = - (n - 1) \frac{n}{2}.
\label{f1zerothmomentinfU}
\eeq
\beq
\int dy f^{(3)}(y) = -\frac {1} {96} (n - 1) n^2 (13 n - 16)
\eeq
\beq
\int dy f^{(1)}(y) y^2 = \frac {1} {8} (n - 4) (n - 1) n^2.
\eeq
The higher order terms are too lengthy to be written here \cite{momentsweb} but shall be used in the calculation of the optical conductivity below. These expressions can be obtained from those in section \ref{ancond} by taking the leading order term in $d$, setting $d\to\frac{1}{4}$, and multiplying by the factor $\frac{(1-n)}{4\pi}$. 

We use two methods to reconstruct the functions $f^{(i)}(\frac{\omega}{D})$ from their moments. 
The first one is the maximum-entropy method (MEM) briefly explained above, and further detailed in Appendix~\ref{MEM}. 
The second one (see Appendix~\ref{momentsrec} for more details) is in the context of Mori's approach to transport, in which the infinite sequence of relaxation functions is truncated at some order. 
The relaxation function of that order is then assumed to have a specific form determined by physical principles. 
In the Mori approach to transport\cite{Mori,Mori2, Lovesey}, the relaxation function $R(\vk,t)$ is given by the Fourier Transform of $\frac{\sigma(\vk,\omega)}{\pi\si_0}$. 
Its normalized Laplace transform is related to $\frac{\si(\vk,\omega)}{\pi\si_0}$ by the formula
\beq
\frac{R(\vk,s)}{m_0(\vk)} = \int d\omega \frac{\sigma(\vk,\omega)}{\pi\si_0} \frac{1}{m_0(\vk)} \frac{1}{s-i\omega},
\label{Rsig2mt} 
\eeq
where $m_n(\vk)=\int d\omega \frac{\sigma(\vk,\omega)}{\pi\si_0} \omega^n$. $\frac{\sigma(\vk,\omega)}{\pi\si_0}$ can be recovered from $\frac{R(\vk,s)}{m_0(\vk)} $ via the formula
\beq
\frac{\sigma(\vk,\omega)}{\pi\si_0} = \frac{m_0(\vk)}{\pi} \ \Re e \left[\frac{R(\vk,i\omega + \eta)}{m_0(\vk)} \right].
\label{Rsig3mt}
\eeq
To reconstruct $\frac{\sigma(\vk,\omega)}{\pi\si_0}$ from its first $r+1$ even moments, we make the following approximation for $\frac{R(\vk,s)}{m_0(\vk)}$,
\beq
\frac{R(\vk,s)}{m_0(\vk)} = \frac{1}{s+}  \;\; \frac{\delta_1(\vk)}{s +}\ldots\frac{\delta_{r-2}(\vk)}{s+\frac{\delta_{r-1}(\vk)} {s+\sqrt{\delta_r(\vk)}}}.
\label{finalRmt}
\eeq 
in which we have used standard notations for continued-fraction expansions. 
The $\delta_n$ are given in terms of the $m_n$ as
\beq
\delta_1=\tilde{m}_2;\;\;\;\;\delta_2=\frac{\tilde{m}_4}{\tilde{m}_2}-\tilde{m}_2;\;\;\;\;\;\delta_3=\frac{\tilde{m}_6\tilde{m}_2-\tilde{m}_4^2}{\tilde{m}_2(\tilde{m}_4-\tilde{m}_2^2)}\;\ldots,
\label{deltamt}
\eeq
where $\tilde{m}_n=\frac{m_n}{m_0}$. When performing the reconstruction, we use $r=3$. This leads to the following form for $\frac{\sigma(\omega)}{\pi\si_0}$.
\beq
\frac{\sigma(\omega)}{\pi\si_0} = \frac{m_0}{\pi}\frac{\tau \delta_1\delta_2}{[\tau\omega(\omega^2-\delta_1-\delta_2)]^2+(\omega^2-\delta_1)^2},
\label{functional form}
\eeq
where $m_i\equiv\lim_{k\to0}m_i(\vk)$, $\delta_i\equiv \lim_{k\to0}\delta_i(\vk)$, and $\tau = \frac{1}{\sqrt{\delta_3}}$. This function reproduces the moments $m_0$ through $m_5$ exactly, while $m_6$ onwards diverge due to the $\omega^6$ power law decay. However, we are interested in the low-frequency behavior of the optical conductivity, which this functional form is expected to capture correctly. Note that \disp{functional form} can be directly applied to compute $f^{(1)}(y)$ in terms of its moments w.r.t $y$. 

\begin{figure}
\begin{center}
\includegraphics[width=.44 \textwidth]{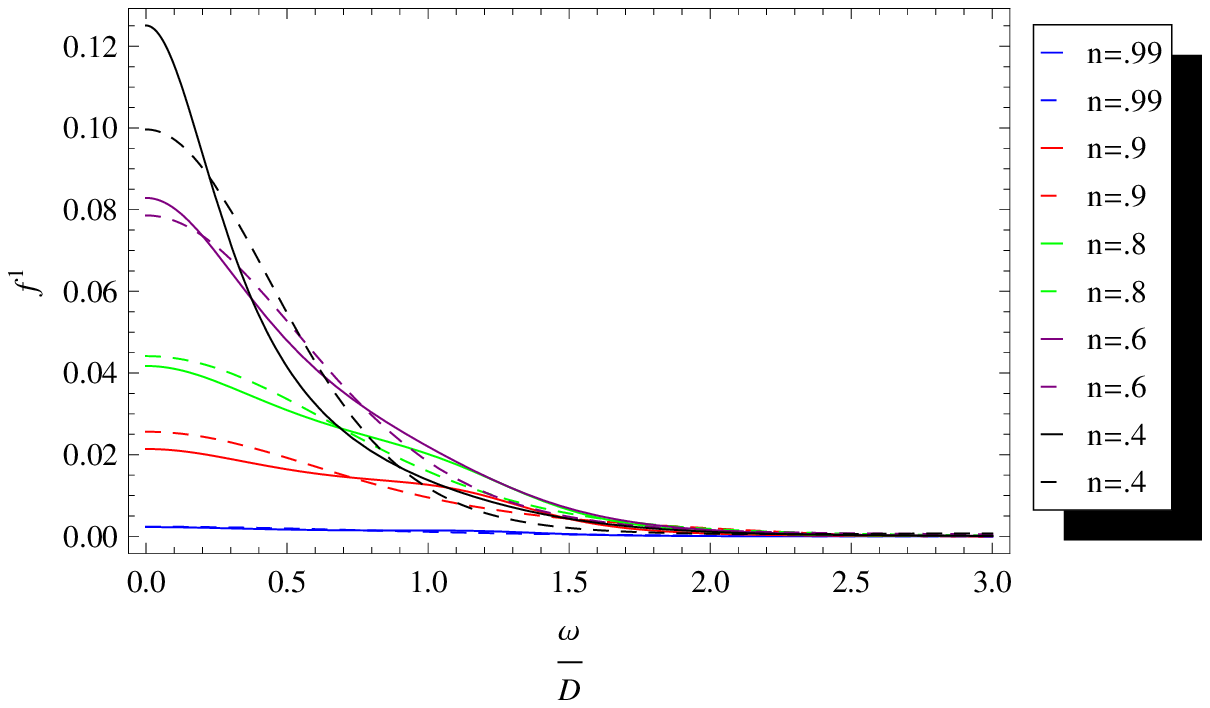}
\includegraphics[width=.5 \textwidth]{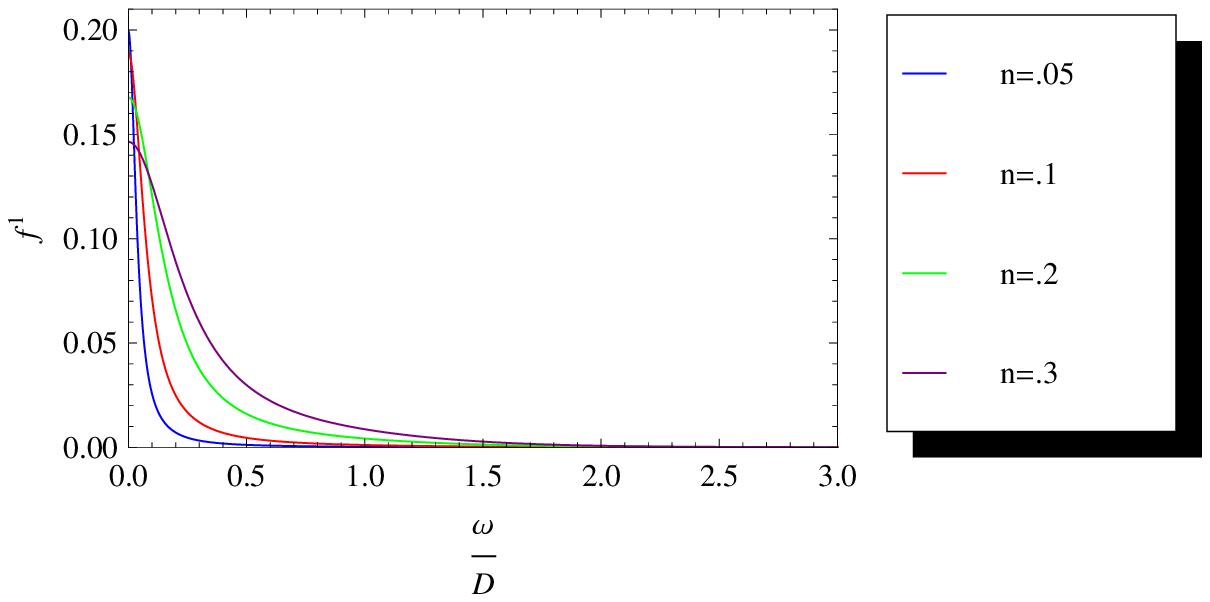}
\caption{The function $f^{(1)}(\frac{\omega}{D})$ associated with the optical conductivity  
$\frac{\si(\omega)}{\pi\si_0} = \frac{D}{T} f^{(1)}(\frac{\omega}{D}) + \ldots$ for the infinite-$U$ Hubbard
model on a $d=\infty$ hypercubic lattice. The dashed curves are obtained using the MEM with the zeroth through eighth moments, while the solid curves are obtained using \disp{functional form}. As the density is lowered, the MEM becomes less reliable, and is therefore not displayed for $n\le .3$.}
\label{f1}
\end{center}
\end{figure}

The function $f^{(1)}(\frac{\omega}{D})$ obtained using both methods is displayed in \figdisp{f1} for several densities. 
We observe that, in this high-temperature incoherent regime at $U=\infty$, the optical conductivity for a generic density 
is a featureless function of frequency, with a width of order $D$. The interesting limiting case of low densities is discussed in more details below. 

\begin{figure}
\begin{center}
\includegraphics{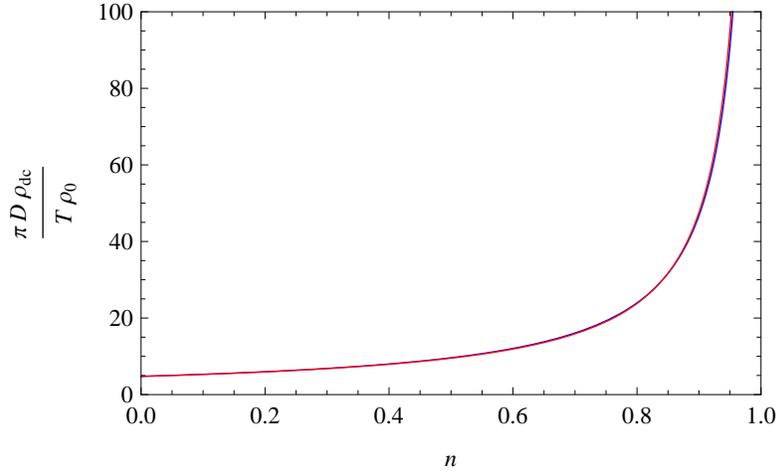}
\caption{Slope of the $T$-linear dc-resistivity for the infinite-$U$ Hubbard model on a hypercubic lattice for $d=\infty$, 
in units of $\frac{\rho_0}{\pi D}$ plotted as a function of $n$. The blue curve is plotted using \disp{dccn}, while the red curve is plotted using the approximation $\rho_{dc}\approx \rho_{dc,0}/(1-n)$, where $\rho_{dc,0}/\rho_0=4.76064\frac{T}{\pi D}\simeq 1.51 \frac{T}{D} $ is the $n=0$ value of the dc resistivity.
\label{dccondn}}
\end{center}
\end{figure}

Using \disp{functional form}, we obtain the dominant linear high-$T$ behaviour of the $dc$-resistivity on the hypercubic lattice as: 
\begin{eqnarray}\label{dccn}
\frac{\rho_{dc}}{\rho_0}\,&=&\,c_1(n)\,\frac{T}{D}+\cdots\\
c_1(n)\,&=&\,\frac{(n-4)^2}{12 \sqrt{3} (1-n) (8-5 n)} 
\sqrt{\frac{7936-n (n (n (n((n-16) n+124)-136)-1752)+8896)}{(n-4) (5 n-8)}}\,\simeq\,\frac{1.51}{1-n} 
\nonumber
\end{eqnarray}
The slope of the dc resistivity is plotted as a function of $n$ in \figdisp{dccondn}. Except for the factor of $1/(1-n)$, it very
weakly depends on $n$, and the slope is very well approximated by $c_1(n)\simeq c_1(0)/(1-n)\simeq 1.51/(1-n)$. 

\paragraph{Behaviour near the Mott insulator $n\rightarrow 1$.}

In the $n\to1$ limit, all of the moments of $f^{(1)}$ vanish linearly in $(1-n)$. 
Therefore, $f^{(1)}(\frac{\omega}{D}) = (1-n)\tilde{f}^{(1)}(\frac{\omega}{D})$, 
where $\tilde{f}^{(1)}(\frac{\omega}{D})$ is independent of $n$. 
This reflects the fact that the spectral weight of the optical conductivity is proportional to the 
kinetic energy (f-sum rule) which vanishes proportionally to $1-n$ as the Mott insulator is reached for 
$U=\infty$. 
Note that the width of the optical conductivity remains of order $D$ in this limit, while its height 
at $\omega=0$ vanishes as $1-n$.  
Correspondingly, the slope of the $dc$-resistivity diverges as $1/(1-n)$ as described by Eq.~\ref{dccn}. 

\paragraph{Behaviour in the low-density limit $n\rightarrow 0$.}

The low-density limit proves to be especially interesting. The zeroth moment of $f^{(1)}(\frac{\omega}{D})$ vanishes linearly as $n\to0$, 
while {\it all moments} higher than the zeroth moment vanish proportionally to $n^2$ as $n\to0$, as exemplified by the second moment given above. 
Therefore,  $f^{(1)}(\frac{\omega}{D})$  can be decomposed in the following way at small $n$:
\beq
f^{(1)}(\frac{\omega}{D}) = f_p^{(1)}(\frac{\omega}{D n}) + n^2 f_t^{(1)}(\frac{\omega}{D}),
\label{nto0limit}
\eeq
where $f_p^{(1)}(\frac{\omega}{D n})$ describes the peak of $f^{(1)}(\frac{\omega}{D})$, while $f_t^{(1)}(\frac{\omega}{D})$ describes its tail. The leading order $n$-dependence has been displayed explicitly in \disp{nto0limit}, and therefore $f_p^{(1)}(y)$ and $f_t^{(1)}(y)$ are $n$-independent in this limit. 
Therefore, the low-density limit involves two distinct scales: $D$ and $Dn$.  
The Mori-based expression \disp{functional form} respects the decomposition in \disp{nto0limit}, and is therefore well suited to describe the 
low-density limit. Note that $\delta_1\propto n$, while $\delta_2,\delta_3\propto n^0$. Plugging $\omega\sim n$ into \disp{functional form}, it becomes $f^{(1)}(\frac{\omega}{D})=\frac{A}{1+B\frac{\omega^2}{n^2}}$, where $A$ and $B$ are both $n$-independent constants. Therefore, in this range of $\omega$, $f^{(1)}(\frac{\omega}{D})=f_p^{(1)}(\frac{\omega}{D n})$. Plugging $\omega\gg n$ into the RHS of \disp{functional form}, it becomes $f^{(1)}(\frac{\omega}{D})=n^2 f_t^{(1)}(\frac{\omega}{D})$, with $ f_t^{(1)}(\frac{\omega}{D})$ an $n$-independent function. 

One can observe the decomposition \disp{nto0limit} as it applies to \disp{functional form} in the right panel of \figdisp{f1}. As $n\to0$, the peak height saturates, while the peak width and the tail height vanish linearly and quadratically in $n$, respectively. The MEM on the other hand, does not easily capture the decomposition \disp{nto0limit}, and is therefore not suitable for addressing the $n\to0$ limit. We can already see a fairly large discrepancy between the two methods at $n=.4$ in \figdisp{f1}, and it becomes worse as $n$ is lowered. At $n=.1$, the MEM can no longer be made to correctly reproduce the moments of $f^{(1)}(\frac{\omega}{D})$. 

At very low density, $n\to0$, the tail in \disp{nto0limit} can be neglected, and this decomposition implies that the width of the optical 
conductivity vanishes proportionally to $n$ in this limit, while the $T$-linear slope of the $dc$-resistivity {\it saturates in this limit}. 
This may appear rather surprising at first. This is actually due to the fact that the effective carrier number 
vanishes as $n/T$ (as the kinetic energy), while the scattering time diverges as $1/n$. Indeed, the spectral density $\rho_\Sigma$ 
was shown in the previous section to vanish as $n$. Hence the density drops out of the conductivity, which is the product of the effective carrier 
number by the scattering time. 

\subsection{Optical conductivity obtained from the self-energy.}

\begin{figure}
\begin{center}
\includegraphics[width=.45 \textwidth]{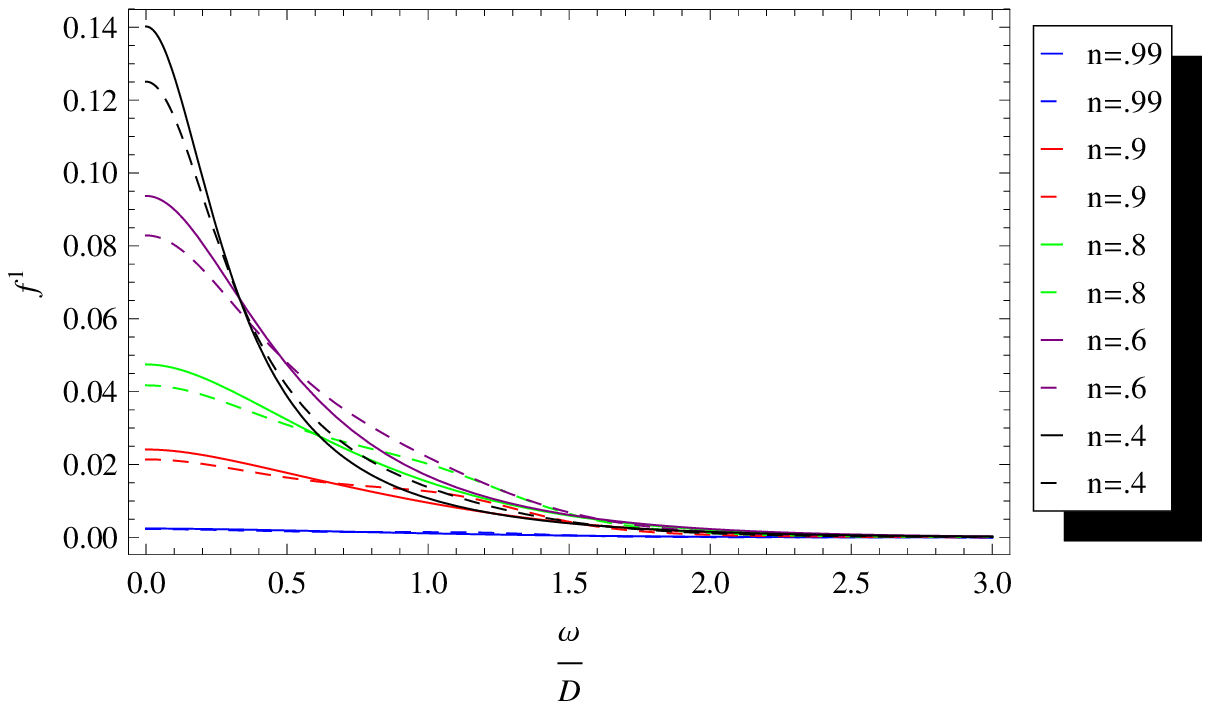}
\includegraphics[width=.48 \textwidth]{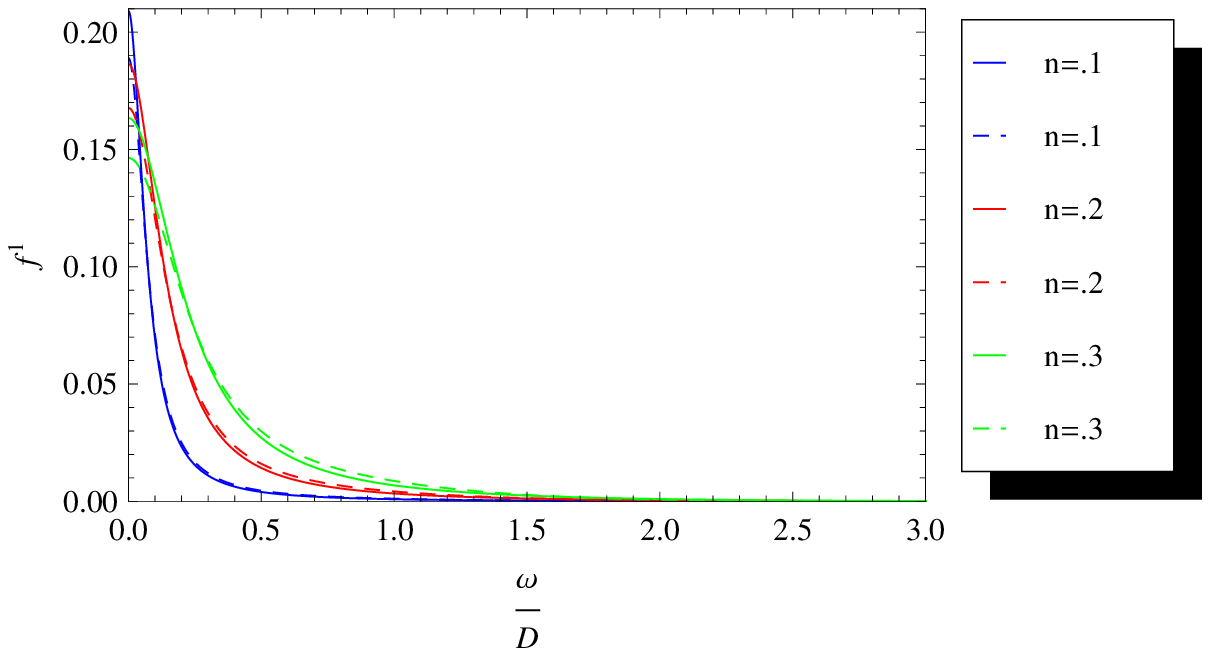}
\caption{High-temperature optical conductivity $\frac{\si(\omega)}{\pi\si_0} = \frac{D}{T} f^{(1)}(\frac{\omega}{D}) + \ldots$ for the infinite-$U$ Hubbard
model on a $d=\infty$ hypercubic lattice. The dashed curves are obtained using the Mori reconstruction \disp{functional form}, while the solid curves are obtained 
from the single-particle self-energy using the `bubble formula' \disp{f1bubblehypercube}.}
\label{f1fromselflargen}
\end{center}
\end{figure}

In the $d\to\infty$ limit, vertex corrections drop out~\cite{khurana_vertex} and the conductivity can be obtained from the single particle 
Green's function via the bubble graph \disp{bubble}. 
{High-temperature dc transport in large dimensions was first discussed in this perspective 
in Refs. (\onlinecite{palsson_phd}) and (\onlinecite{palsson_thermo_1998_prl}).}
Performing the analytical continuation of expression \disp{bubble} yields: 
\beq
\frac{\si(\omega)}{\pi\si_0} = \frac{2}{\omega}\int d\epsilon \  \Phi(\epsilon) \int dx \ \rho_G(-\mu+x,\epsilon) \rho_G(-\mu+\omega+x,\epsilon) (1-e^{-\beta\omega})f(x-\mu)\bar{f}(x-\mu+\omega),
\label{opticalbubble}  
\eeq
where $f(\nu)=\frac{1}{e^{\beta\nu}+1}$, $\bar{f}(\nu)=1-f(\nu)$, and $\Phi(\epsilon)$ is the transport function, which reads, for the hypercubic lattice:
\beq
\Phi(\epsilon) = \sum_{\vq,\alpha} \left(\frac{\partial \epsilon_{\vq}}{\partial_{q_\alpha}}\right)^2 \delta(\epsilon-\epsilon_{\vq})\rightarrow_{d=\infty} \frac{D}{2\sqrt{\pi}}e^{-\frac{\epsilon^2}{D^2}}
\eeq
In (\ref{opticalbubble}), $\rho_G$ is the one-particle spectral function:
\beq
\rho_G(-\mu+x,\epsilon)=\frac{(1-\frac{n}{2})\rho_\Sigma(-\mu+x)}{[x-(1-\frac{n}{2})\epsilon-\Re e\Sigma(-\mu+x)]^2+\pi^2[\rho_\Sigma(-\mu+x)]^2}.
\label{Fdef}
\eeq
Taking the $T\to\infty$ limit and switching to dimensionless variables ($x\to x D$, $\epsilon\to \epsilon D$), we find that
\beq
f^{(1)}(\frac{\omega}{D}) = \frac{n(1-n)}{(1-\frac{n}{2})^2 D}\int d\epsilon \ \Phi(\epsilon D)\int dx  \ \rho_G^{(0)}(-\mu+x,\epsilon) \rho_G^{(0)}(-\mu+\frac{\omega}{D}+x,\epsilon),
\label{f1bubblehypercube}
\eeq
where
\beq
\rho_G^{(0)}(-\mu+x,\epsilon)=\frac{(1-\frac{n}{2})h^{(0)}(x)}{[x-(1-\frac{n}{2})\epsilon-\bar{h}^{(0)}(x)]^2+\pi^2[h^{(0)}(x)]^2},
\eeq
where $\bar{h}^{(0)}(x)= P\int\frac{h^{(0)}(x')}{x-x'}dx'$, and we have made use of  Eqs. (\ref{expmu}) and (\ref{betaseriesinfUcondinfd}). 
Hence, from these expressions, one can calculate $f^{(1)}(\frac{\omega}{D})$ from $h^{(0)}(\frac{x}{D})$. 
Plugging the $h^{(0)}(x)$ obtained in section \ref{momentsSEinfd} into \disp{f1bubblehypercube}, we display in \figdisp{f1fromselflargen} the resulting function $f^{(1)}(\frac{\omega}{D})$ 
for the hypercubic lattice and compare with the one obtained by the ``Mori reconstruction method" in section \ref{reconstructionoptical} (\disp{functional form}).

\begin{figure}
\begin{center}
\includegraphics[width=0.8\textwidth]{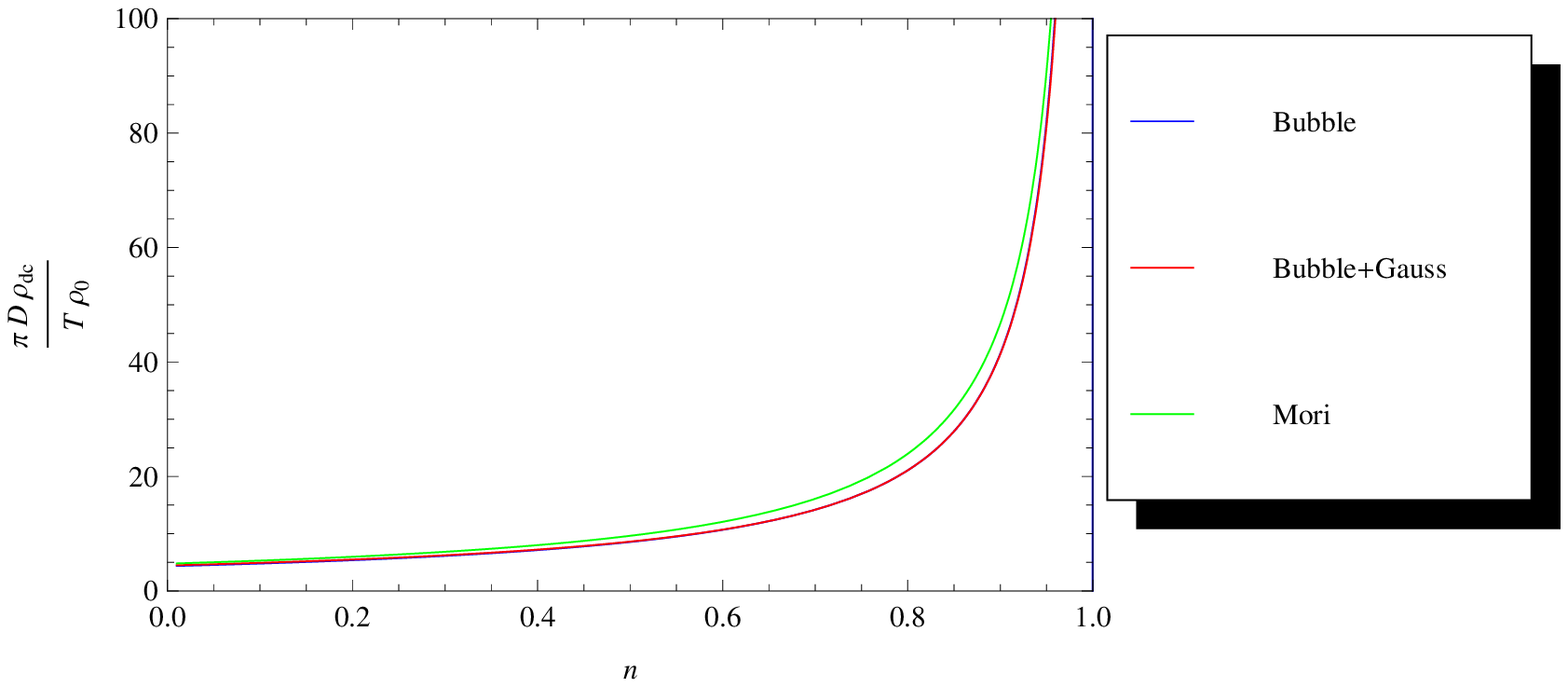}
\caption{The slope of the dc resistivity for the infinite-$U$ Hubbard model on a
hypercubic lattice for $d=\infty$ in units of $\frac{\rho_0}{\pi D}$
plotted as a function of $n$. The blue and red curves are plotted using \disp{f1bubblehypercube}, the former with  $h^{(0)}(\frac{x}{D})$ reconstructed using the MEM, the latter by assuming $h^{(0)}(\frac{x}{D})$ to be Gaussian,
which is an excellent approximation. The green curve is plotted using the ``Mori reconstruction method" (\disp{dccn}). We find reasonable agreement between these two complementary methods. }
\label{dccondnself}
\end{center}
\end{figure}

Setting $\omega=0$ in \disp{f1bubblehypercube}, we plot the resulting dc resistivity vs. density curve in \figdisp{dccondnself}, and compare it to the one obtained from \disp{dccn}. We find reasonable agreement between the two complementary methods. We also compare it to the curve obtained from \disp{f1bubblehypercube} assuming $h^{(0)}(x)$ to be a Gaussian. 
This Gaussian form is an excellent approximation and will be used to extend our results for the conductivity to a broader range of temperatures.

The $n\rightarrow 1$ and low-density $n\rightarrow 0$ limits can also be discussed from this perspective, using \disp{f1bubblehypercube}. Since  $h^{(0)}(x)$ has non-singular behavior as $n\to1$, $\si_{dc}$ vanishes linearly in $(1-n)$ due to the pre-factor.  As $n\to0$, $h^{(0)}(x)$ vanishes linearly in $n$. Therefore, $\rho_G^{(0)}(-\mu+x,\epsilon)$ becomes asymptotically a Lorentzian of width $n$ and height $\frac{1}{n}$. The integral on the r.h.s. of \disp{f1bubblehypercube} is therefore of order $\frac{1}{n}$. This is cancelled by the $n$ in the pre-factor, and therefore the dc conductivity saturates as $n\to0$. These results are the same ones that we found using the Mori method in section \ref{reconstructionoptical}. It is reassuring that these two complementary methods lead us to the same conclusions. Finally, comparing \disp{f1bubblehypercube} with Eqs. (\ref{eq:cond_pheno}) and (\ref{eq:kappa_kinetic}), we see that the pre-factor in \disp{f1bubblehypercube} can be associated with $E_K$, while the integral can be associated with $\tau_{\mathrm{tr}}$.

\begin{figure}
\begin{center}
\includegraphics[width=.45 \textwidth]{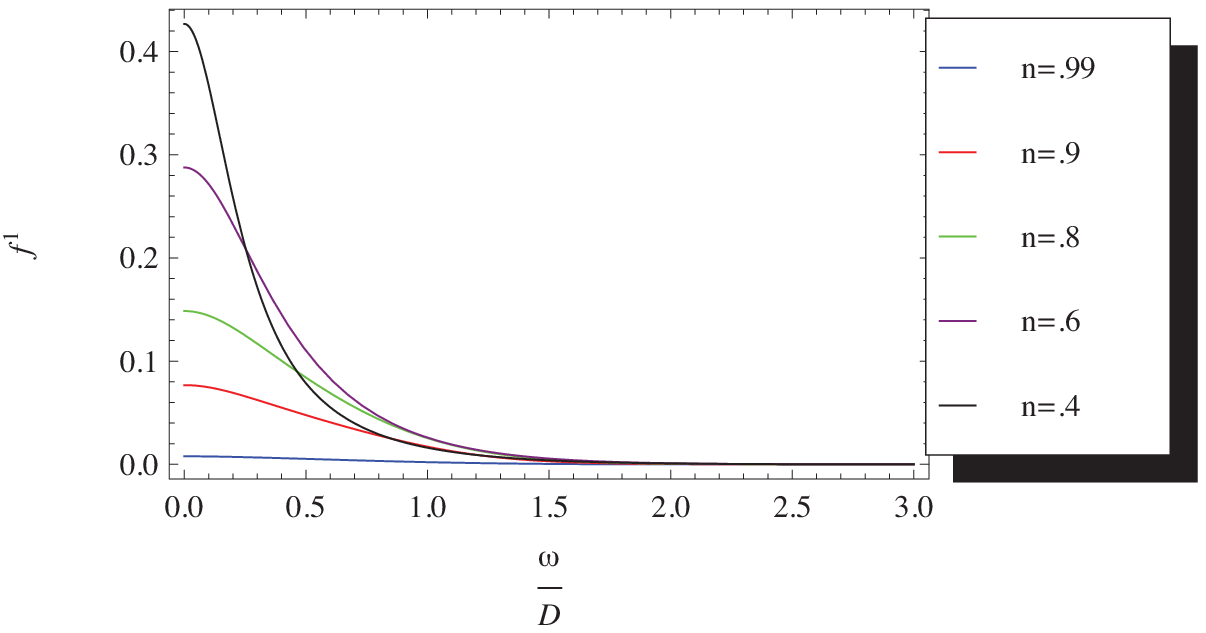}
\includegraphics[width=.5 \textwidth]{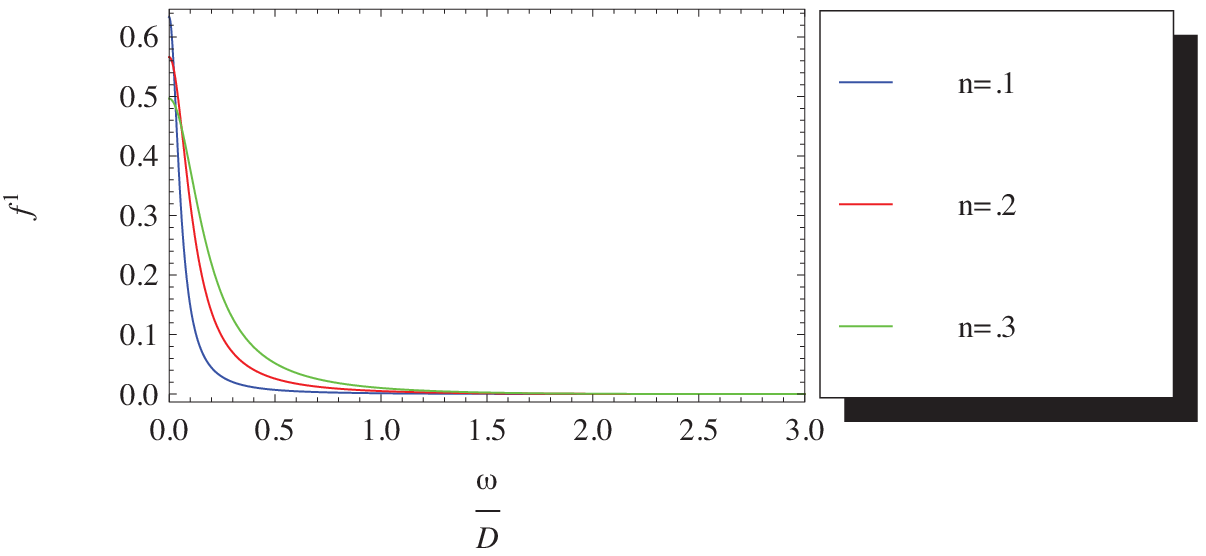}
\caption{$f^{(1)}(\frac{\omega}{D})$ for the infinite-$U$ Hubbard model on a Bethe lattice for $d=\infty$, where $\frac{\si(\omega)}{\pi\si_0} = \frac{D}{T} f^{(1)}(\frac{\omega}{D}) + \ldots$. 
The curves are obtained using \disp{f1bubblehypercube}.}
\label{f1fromselflargenBethe}
\end{center}
\end{figure}

For the infinite-connectivity Bethe lattice, it is customary to choose the transport function in the form\cite{Badmetal}:
\beq
\Phi(\epsilon) = D\left[1-\left(\frac{\epsilon}{D}\right)^2\right]^{\frac{3}{2}}
\eeq
which is such that the f-sum rule keeps its standard form in which the integral of the optical conductivity is proportional to the kinetic energy.  
We plot the resulting $f^{(1)}(\frac{\omega}{D})$ for the Bethe lattice in \figdisp{f1fromselflargenBethe}. 

\begin{figure}
\begin{center}
\includegraphics{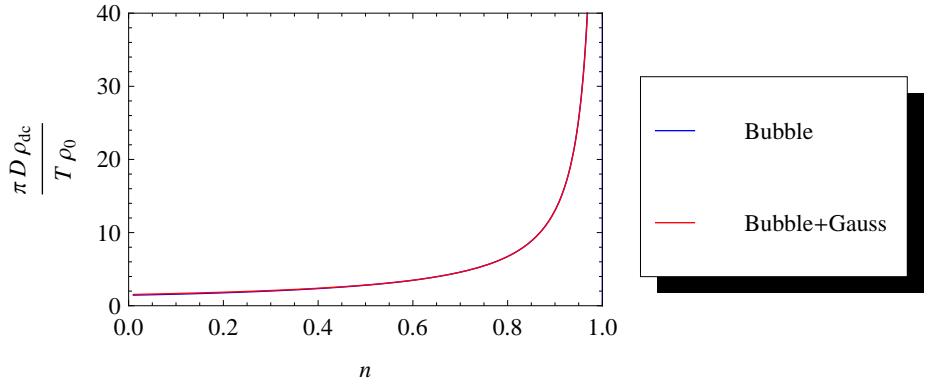}
\caption{The slope of the dc resistivity for the infinite-$U$ Hubbard model on a
Bethe lattice for $d=\infty$ in units of $\frac{ \rho_{0}}{\pi D}$
plotted as a function of $n$. The blue and red curves are plotted
using \disp{f1bubblehypercube}, the former with  $h^{(0)}(\frac{x}{D})$
reconstructed using the MEM, the latter by assuming
$h^{(0)}(\frac{x}{D})$ to be Gaussian. The Gaussian form is an
excellent approximation.}
\label{dccondnselfBethe}
\end{center}
\end{figure}

We plot the dc resistivity vs. density curve for the Bethe lattice in
\figdisp{dccondnselfBethe}.  Again, the Gaussian approximation for
$h^{(0)}(x)$ is excellent, and will be used to extend our results for the conductivity to a broader range of temperatures.

\subsection{DC resistivity: corrections beyond the dominant $T$-linear behaviour \label{corrections}}

To obtain the dc conductivity, we take the $\omega\to0$ limit of \disp{opticalbubble}. 
\beq
\frac{\si_{dc}}{\pi\si_0} = \frac{2}{T}\int d\epsilon \ \Phi(\epsilon) \int dx [\rho_G(-\mu+x,\epsilon)]^2  f(x-\mu)\bar{f}(x-\mu).
\label{dcbubble}  
\eeq
We use in this expression the Gaussian approximation of $\frac{1}{D}\rho_\Sigma(-\mu+Dx)$, which was shown above to be quite accurate:
\beq
\frac{1}{D}\rho_\Sigma(-\mu+Dx) = \frac{m_0}{\sqrt{2\pi(\tilde{m}_{2}-\tilde{m}_{1}^2)}} \exp\left[-\frac{(x-\tilde{m}_{1})^2}{2(\tilde{m}_{2}-\tilde{m}_{1}^2)}\right],
\label{hgaussian}
\eeq
where $m_n\equiv \int dx \frac{1}{D}\rho_\Sigma(-\mu+Dx) x^n$, and $\tilde{m}_n\equiv\frac{m_n}{m_0}$. The $m_n$ can be expanded in powers of $(\beta D)$, i.e. $m_n=\sum_{i=0}^\infty(\beta D)^i m_n^{(i)}$. From \disp{betaseriesSiginfd}, $m_n^{(i)}=\int dx  \ h^{(i)}(x) x^n$. Plugging in the series for the chemical potential and the self-energy into Eqs. (\ref{dcbubble}) and (\ref{hgaussian}), 
and using $\Phi(\epsilon)$ appropriate for the hypercubic lattice, we calculate the series for $\rho_{dc}$ through fourth order in $\frac{D}{T}$:
\beq
\frac{\rho_{dc}}{\rho_0}= \frac{T}{D} \left[c_1 + c_3\left(\frac{D}{T}\right)^2 + c_5\left(\frac{D}{T}\right)^4 + \ldots \right].
\label{resseriesinfd}
\eeq 
Keeping only the coefficient $c_1$ leads to the red curve in \figdisp{dccondnself}. In Table~\ref{table_c_dinf}, we give the values of $c_1$, $c_3$, and $c_5$ for various densities.

\begin{table}
    \begin{tabular} { | p{2.0cm}  | p{2cm} | p{2cm} |p{2cm} |}
    \hline
    $n$&$c_1$&$c_3$&$c_5$  \\ \hline
    $.1$ &$1.56337 $&$-0.228875$&$0.0849724$ \\ \hline
    $.2$ &$1.74593$&$-0.108281$&$0.0880068$ \\ \hline
    $.3$ &$1.98192$&$0.00922823 $&$ 0.0652345$ \\ \hline
    $.4$ &$2.29802$&$ 0.1241$&$0.0287058$ \\ \hline
    $.6$ &$3.4083$&$0.345571$&$ -0.0321044$ \\ \hline
    $.8$ &$6.70832$&$0.536409$&$-0.000556513$ \\ \hline
    $.9$ &$13.1668$&$0.59144$&$0.0458907$ \\ \hline   
    \end{tabular}
    \caption{Values of $c_1$, $c_3$, and $c_5$  for the infinite-$U$
    Hubbard model on a hypercubic lattice for $d=\infty$, for various
    values of the density, where $\rho_{dc}$ is given in
    Eq.~\eqref{resseriesinfd}.
    \label{table_c_dinf}
    }
    \end{table}

To obtain analogous results for the Bethe lattice, we use the appropriate form of $\Phi(\epsilon)$ in \disp{dcbubble}. This leads to the series 
\beq
\frac{\rho_{dc}}{\rho_0}= \frac{T}{D} \left[c_{1,B} + c_{3,B}\left(\frac{D}{T}\right)^2 + c_{5,B}\left(\frac{D}{T}\right)^4 + \ldots \right],
\label{resseriesinfdBethe}
\eeq 
where the subscript B indicates the Bethe lattice. Retaining only the $c_{1,B}$ coefficient leads to the red curve in \figdisp{dccondnselfBethe}. In Table~\ref{table_c_Bethe}, we give the values of $c_{1,B}$, $c_{3,B}$, and $c_{5,B}$ for various densities. We will use these coefficients in sec. \ref{sec:resist_comp} to plot the $dc$ resistivity for $\frac{T}{D}<1$. {\color{black} We note that the discrepancy in the coefficients for the hypercubic and Bethe lattices is largely due to $\Phi(0)$, which sets the overall scale of the transport function.}

\begin{table}
    \begin{tabular} { | p{2.0cm}  | p{2cm} | p{2cm} |p{2cm} |}
    \hline
    $n$&$c_{1,B}$&$c_{3,B}$&$c_{5,B}$  \\ \hline
    $.1$ &$0.531979$&$-0.00873688$&$ 0.00193936$ \\ \hline
    $.2$ &$0.588682$&$ 0.00101664$&$0.00218447$ \\ \hline
    $.3$ &$0.662356$&$0.0107786 $&$0.00172337$ \\ \hline
    $.4$ &$0.761394$&$ 0.0207135$&$0.000838343$ \\ \hline
    $.6$ &$1.11009$&$0.0419418$&$ -0.0006315$ \\ \hline
    $.8$ &$2.14253$&$0.0680721$&$0.000236013$ \\ \hline
    $.9$ &$4.14565$&$0.0946119$&$ 0.000263639$ \\ \hline     
    \end{tabular}
    \caption{Values of $c_{1,B}$, $c_{3,B}$, and $c_{5,B}$  for the infinite-$U$ Hubbard model on the Bethe lattice for $d=\infty$, for various values of the density, where $\frac{\rho_{dc}}{\rho_0}= \frac{T}{D} \left[c_{1,B} + c_{3,B}\left(\frac{D}{T}\right)^2 + c_{5,B}\left(\frac{D}{T}\right)^4 + \ldots \right]$.
    \label{table_c_Bethe}
    }
    \end{table} 

\section{Comparing the high-$T$ expansion and DMFT numerical solutions\label{comparetoDMFT}}

\subsection{Comparison of moments for the single-particle Green's function}

We compare the high-temperature series with the DMFT for the infinite-connectivity Bethe lattice. 
In this subsection, we compare the moments of the spectral function in the parameter regime $D<T\ll U$. 
In this very high-temperature regime, the most reliable numerical solver for DMFT equations is the 
interaction-expansion continuous-time Monte Carlo algorithm (CT-INT QMC)\cite{gull_rmp_2011}.

Using Eqs. (\ref{G2ndorderfiniteU}) and (\ref{G2locfiniteU}), we find that 

\beq
G_{loc,L}^{(0)}(i\omega_n) &=& \frac{1-\frac{n}{2}}{i\omega_n+\mu},\nn\\
G_{loc,U}^{(0)}(i\omega_n) &=&  \frac{\frac{n}{2}}{i\omega_n+\mu-U},\nn\\
G_{loc,L}^{(2)}(i\omega_n)&=& \frac{2 c  D^2 \beta }{U \left( i \omega _n+\mu\right)} +\frac{c  D^2 \beta }{\left(i \omega _n+\mu \right){}^2}-\frac{ (n-2) D^2}{8\left(i \omega _n+\mu
   \right){}^3},
\label{G2ndorderfiniteULHBUHB}
\eeq
where the subscripts $L$ and $U$ refer to the lower and upper Hubbard bands, $dt^2=\frac{D^2}{8}$, and $c\to\frac{(1-n)n}{8}$, as $e^{-\beta U}\to0$.

Expanding $\mu$ in powers of $D$, i.e. $\mu = T\left(\bar{\mu}^{(0)} + (\beta D)^2\bar{\mu}^{(2)}+\ldots\right)$, and setting $e^{-\beta U}\to0$, we find that
$\bar{\mu}^{(0)}=\log \frac{n}{2(1-n)}$, and $\bar{\mu}^{(2)}= \frac{2n -1}{8} - \frac{n T}{4 U}$. Plugging this into \disp{G2ndorderfiniteULHBUHB} and going into the time-domain yields

\beq
G_{loc,L}^{(0)}(\bar{\tau}) &=& e^{(\bar{\mu}^{(0)}\bar{\tau})}(n-1),\nn\\
G_{loc,U}^{(0)}(\bar{\tau}) &=& e^{(\bar{\mu}^{(0)}-\beta U)\bar{\tau}}(-\frac{n}{2}) ,\nn\\
G_{loc,L}^{(2)}(\bar{\tau})&=& e^{(\bar{\mu}^{(0)}\bar{\tau})}\left[-\frac{\beta D^2 (n-1) \bar{\tau}  (2 n+U \beta )}{8 U}+\frac{D^2 (n-1) n \beta }{4
   U}+\frac{1}{8} \beta ^2 D^2 (n-1) \bar{\tau} ^2\right],
\label{G2ndorderfiniteULHBUHBmuexptime}
\eeq
where $\tau\equiv\beta\bar{\tau}$. Our objective is to compare $G_{loc}(\bar{\tau})$ obtained from DMFT data with the analytical expression \disp{G2ndorderfiniteULHBUHBmuexptime}. In the range of times $1-\bar{\tau}\ll1$, $G_{loc,U}^{(0)}(\bar{\tau})$ is of the same magnitude as $G_{loc,L}^{(2)}(\bar{\tau})$ due to the exponential factor $e^{-\beta U \bar{\tau}}$. Therefore,
\beq
G_{loc}(\bar{\tau}) +e^{\bar{\mu}^{(0)}\bar{\tau}}e^{-\beta U \bar{\tau}}\frac{n}{2}=G_{loc,L}(\bar{\tau}) + O\left[(\beta D)^3\right];\;\;\;\;\;1-\bar{\tau}\ll1.
\eeq

\begin{figure}
\begin{center}
\includegraphics{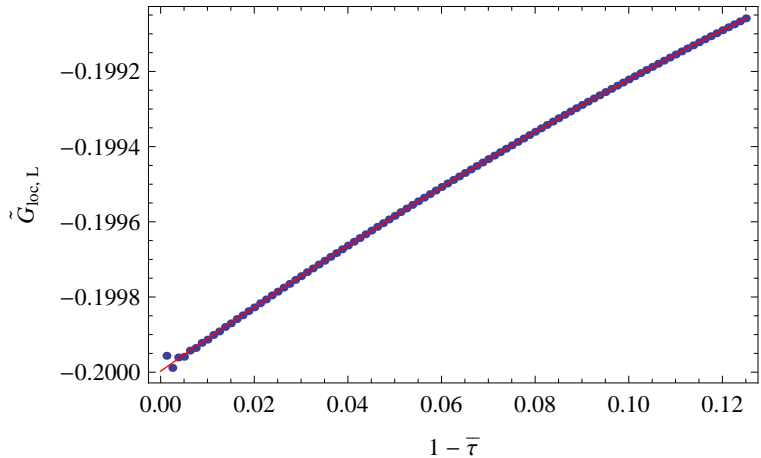}
\caption{$\tilde{G}_{loc,L}(\bar{\tau})\equiv G_{loc,L}(\bar{\tau}) e^{-\bar{\mu}^{(0)}\bar{\tau}}=G_{loc}(\bar{\tau}) e^{-\bar{\mu}^{(0)}\bar{\tau}}+e^{-\beta U \bar{\tau}}\frac{n}{2}$ plotted versus $(1-\bar{\tau})$, where $G_{loc}(\bar{\tau})$ is obtained from DMFT calculations (using CT-INT Quantum Monte-Carlo) for the Hubbard model on a Bethe lattice with $n=.8$, $U=20$, and $\frac{T}{D}=1.6$. The DMFT data, represented by the blue dots, is fit to the functional form $\left\{\gamma_0 + (\beta D)^2\left[\gamma_{22}(1-\bar{\tau})^2+\gamma_{21}(1-\bar{\tau})\right]\right\}$, with fit parameters $\gamma_0 =-0.199997$, $\gamma_{21}=0.02242$, and $\gamma_{22}=-0.0255256$.} 
\label{comparemoments}
\end{center}
\end{figure}

According to \disp{G2ndorderfiniteULHBUHBmuexptime}, 
\beq
G_{loc,L}(\bar{\tau})=e^{\bar{\mu}^{(0)}\bar{\tau}}\left[\alpha_0 + (\beta D)^2(\alpha_{22}\bar{\tau}^2+\alpha_{21}\bar{\tau}+\alpha_{20})+\ldots\right],
\label{expLHBbeta}
\eeq
where the $\alpha$ coefficients are functions of the density and the ratio $\frac{T}{U}$. They are given by the expressions
\beq
\alpha_0 &=& n-1, \nn\\
\alpha_{22}&=& \frac{1}{8}(n-1),\nn\\
\alpha_{21}&=& - \frac{1}{8}(n-1)(1+2n\frac{T}{U}),\nn\\
\alpha_{20}&=& \frac{(n-1)n}{4}\frac{T}{U}.
\label{alphaanalytical}
\eeq
\disp{expLHBbeta} can also be rewritten in powers of $(1-\bar{\tau})$ as
\beq
G_{loc,L}(\bar{\tau}) e^{-\bar{\mu}^{(0)}\bar{\tau}} =\left\{\gamma_0 + (\beta D)^2\left[\gamma_{22}(1-\bar{\tau})^2+\gamma_{21}(1-\bar{\tau})+\gamma_{20}\right]+\ldots\right\},
\label{expLHBbeta2}
\eeq
where $ \alpha_0=\gamma_0$, $\alpha_{22}=\gamma_{22} $, $\alpha_{21}=-2\gamma_{22} -\gamma_{21} $, and $\alpha_{20}=\gamma_{22} +\gamma_{21} +\gamma_{20}$. Finally, using the DMFT data, we fit the quantity $G_{loc}(\bar{\tau})e^{-\bar{\mu}^{(0)}\bar{\tau}} +e^{-\beta U \bar{\tau}}\frac{n}{2}$ to the RHS of \disp{expLHBbeta2} for $1-\bar{\tau}\ll1$. This yields the $\gamma$ and hence $\alpha$ coefficients. In \figdisp{comparemoments}, we show the result of the fit for the case of $n=.8$, $U=20$, and $\frac{T}{D}=1.6$. It yields the values $\alpha_{22}=-0.0255256$, $\alpha_{21}=0.0286311$, and $\alpha_{20}=-0.00309885$. Since the fit in \figdisp{comparemoments} does not distinguish between $\gamma_0$ and $\gamma_{20}$, we have assumed $\gamma_0=\alpha_0=n-1=-.2$. The analytical expressions in \disp{alphaanalytical} yield the values $\alpha_{22}=-0.025$, $\alpha_{21}=0.0282$, and $\alpha_{20}=-0.0032$. Therefore, we find excellent agreement between DMFT and the analytical calculations. 

\subsection{Comparison of self-energy\label{comparespectra}}

We next compare the self-energies obtained using the high-$T$ expansion and numerical solutions 
of the DMFT equations. In the intermediate temperature regime, we use the NRG method~\cite{bulla2008,zitko2009}, 
which provides direct real-frequency results, while at very high temperature we use the CT-INT QMC algorithm which requires analytical continuation. 
 
\begin{figure}
\begin{center}
\includegraphics{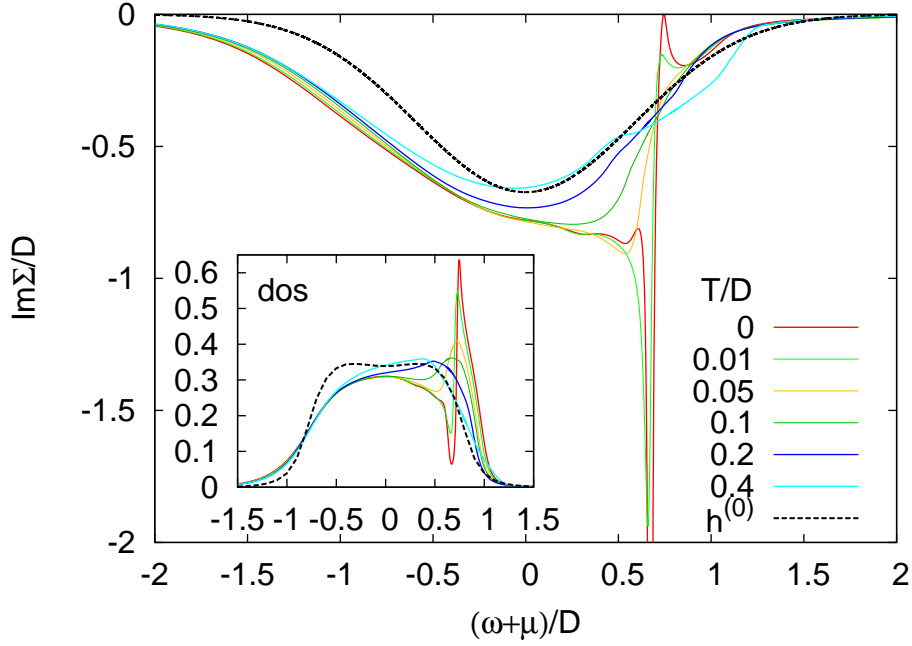}
\caption{$\frac{1}{D} Im \ \Sigma(\omega)$ vs. $\frac{\omega+\mu}{D}$
for the infinite-$U$ Hubbard model on the Bethe lattice for $n=.9$.
The different colored curves are the NRG results at different
temperatures, while the black dashed line is
$\frac{-\pi}{1-\frac{n}{2}} h^{(0)}(x)$, which is the asymptotic high-T
result. \rzjm{The inset shows the corresponding local density of states}
{\color {black} at each temperature.} At $\frac{T}{D}=.4$, the high-T result for $\Sigma$ is almost
in perfect agreement with the actual self-energy and remains a good
approximation down to $\frac{T}{D}\approx .2$. {\color{black} At the two
lowest temperatures displayed, the self-energy acquires a quasi-pole
followed by a sharp minimum. The latter corresponds to the
quasiparticle peak in the density of states, while the former
corresponds to the dip between the quasiparticle peak and the lower
Hubbard band \cite{DMFT-ECFL}. }}
\label{compareselfNRGhighT}
\end{center}
\end{figure}

In \figdisp{compareselfNRGhighT}, we plot $\frac{1}{D} Im \ 
\Sigma(\omega)$ vs. $\frac{\omega+\mu}{D}$ for $n=.9$. The different
colored curves are NRG results at different temperatures, while the
black dashed line is $\frac{-\pi}{1-\frac{n}{2}} h^{(0)}(x)$, which is
the asymptotic high-T result (we have now set $a_G=1$ in
\disp{Dyson}). As can be seen in \figdisp{compareselfNRGhighT}, at
$\frac{T}{D}=.4$, the high-T result is almost in perfect agreement
with the actual self-energy. We note that in this case a broader
kernel was used in the DMFT(NRG) procedure to find a smooth
representation for $\Sigma(\omega)$ without oscillatory artifacts; 
this also explains the disagreement at large negative frequencies
which are mainly due to over-broadening in the NRG. At lower
temperatures, deviations from the high-$T$ limit result begin to
occur, but it remains a good approximation down to $T\simeq 0.2 D$ . 

\begin{figure}
\begin{center}
\includegraphics[width=0.4\columnwidth,keepaspectratio,angle=-90]{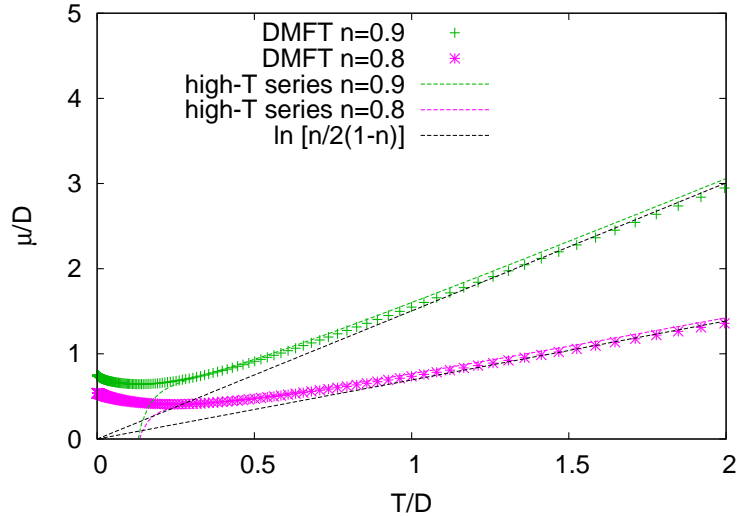}
\caption{The chemical potential $\mu$ vs. $\frac{T}{D}$ plotted for
$n=.8$ and $n=.9$ for the infinite-$U$ Hubbard model on the Bethe
lattice. For each density, the NRG result is compared to the high-$T$
series, $\mu(T)/D= \bar{\mu}^{(0)} \frac{T}{D}+ \bar{\mu}^{(2)}
\frac{D}{T} + \bar{\mu}^{(4)} (\frac{D}{T})^3$ (dashed curved lines),
and the asymptotic high-T result $\mu=T \log
[\frac{n}{2(1-n)}]$ (thin dashed straight lines). The series extends to a lower temperature for $n=.9$ since the effective Fermi temperature shrinks like $(1-n)$ with increasing density.} 
\label{comparemuNRGhighT}
\end{center}
\end{figure}

In \figdisp{comparemuNRGhighT}, we compare the chemical potential from
NRG to the  high-$T$ series, $\frac{\mu(T)}{D}= \bar{\mu}^{(0)} \frac{T}{D}+
\bar{\mu}^{(2)} \frac{D}{T} + \bar{\mu}^{(4)} (\frac{D}{T})^3$, for
$n=.8$ and $n=.9$. The agreement is excellent for $\frac{T}{D}>0.3$.
Both are also compared to the asymptotic high-T result
$\mu=\frac{T}{D}\log [\frac{n}{2(1-n)}]$. The actual behavior deviates
from the high-$T$ asymptote significantly for $\frac{T}{D}\le1$. 


\begin{figure}
\begin{center}
\includegraphics[width=0.6\columnwidth,keepaspectratio]{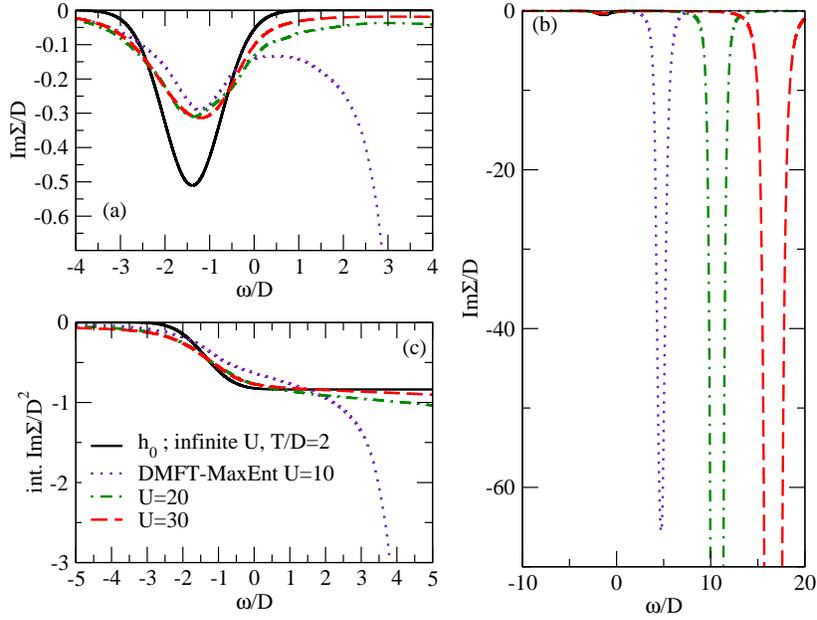}
\caption{(a,b) Imaginary part of the self energy for the Bethe lattice for $n=0.8,\frac{T}{D}=2$.  The analytically continued QMC-DMFT data are compared to the $U=\infty$ high-T result $\frac{-\pi h^{(0)}(\frac{\omega+\mu}{D})}{1-\frac{n}{2}}$. (c) $\int_{-\infty}^{\omega} \mathrm{Im} \Sigma(\omega) \ d\omega$. 
\label{compareimsigtoQMC}}
\end{center}
\end{figure}

At the very high temperature of $\frac{T}{D}=2$, the high-$T$ self-energy
is compared to the CT-INT QMC solution of the DMFT equations for several
interaction strengths in Fig.~\ref{compareimsigtoQMC}. The latter is obtained from the analytic
continuation of the interaction-expansion QMC data.  The right panel shows the self-energies on a bigger
scale, while the left panel displays a close-up on the lower Hubbard band. In the right panel the LHB can be barely seen as a
pronounced peak that separates the lower and
upper Hubbard band dominates the signal. Therefore, the agreement in the exact shape of the lower Hubbard band with the high-T result is only approximate. Nonetheless, the agreement in the integrated weight of the LHB is exact (panel c). 

\subsection{Comparison of dc-resistivity and physical mechanisms of transport \label{sec:resist_comp}}
\label{sec:mechanisms_transport}

\begin{figure}
\begin{center}
\includegraphics{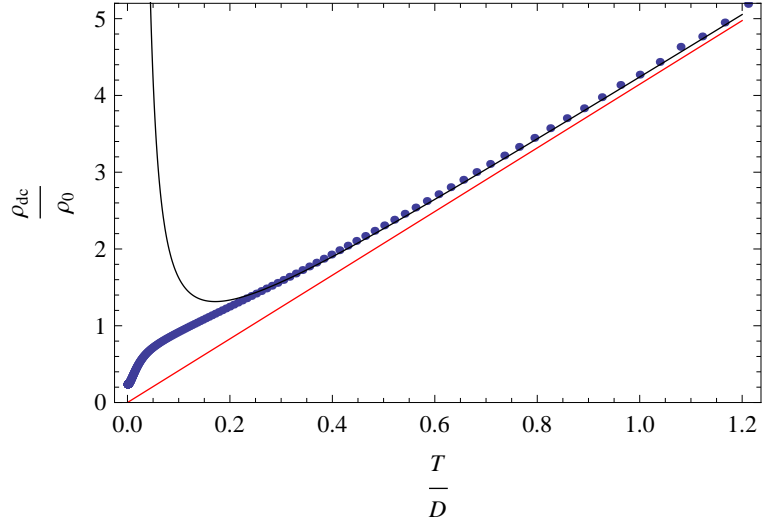}
\caption{Resistivity vs. temperature for the infinite-$U$ Hubbard model on the infinite-connectivity Bethe lattice for $n=.9$. 
The blue dots are the NRG result. 
The red line is the leading order high-$T$ result, while the black line contains the
first two sub-leading corrections as in Eq.~(\ref{resseriesinfdBethe}). 
The high-temperature series is in very good agreement with the NRG solution for $\frac{T}{D}\gtrsim 0.2$, throughout most of the 
`bad-metal' regime. } 
\label{compareNRGhighTn9}
\end{center}
\end{figure}

In \figdisp{compareNRGhighT}, we displayed the dc resistivity vs. temperature curve for various densities using both the NRG and the
high-temperature series (\disp{resseriesinfdBethe}). In Fig.~\ref{compareNRGhighTn9}, 
we display it again separately for the density $n=0.9$. 

The $T$-dependence of the resistivity displays several distinct regimes. At very high temperature the resistivity approaches the asymptotic high-$T$ linear behaviour with the 
slope $c_1(n)$ discussed above. As $T$ is reduced, non-linear corrections become visible, with the resistivity deviating upwards from the asymptotic linear behaviour, 
with correspondingly an apparent $T=0$ positive intercept.  
We observe that the high-temperature expansion at the order considered is in excellent agreement with the NRG data down to $T\simeq 0.2 D$ for $n=0.9$. 
This is almost comparable with the scale at which the resistivity reaches the Mott-Ioffe-Regel scale $\sim \rho_0$, which happens at $T\simeq 0.1 D \simeq (1-n)D$, 
the Brinkman-Rice scale\cite{Badmetal}. 
%
At very low-temperature $T\lesssim T_{\mathrm{FL}}\simeq 0.05 (1-n)D$, the resistivity obtained from NRG obeys Fermi liquid behaviour $\rho\propto T^2$. 
Revealing this behaviour would require displaying the data on a much lower scale - this is discussed in details in e.g. \refdisp{Badmetal} and we do not repeat this analysis 
in the present paper whose main emphasis is on the high-temperature regime. 
As $T$ is increased above $ T_{\mathrm{FL}}$, a more complex crossover with a `knee-like' feature connects to the bad-metal regime. 
This intermediate regime, in which the resistivity is smaller than the MIR value, extends 
over a decade or so in temperature, and is associated with the presence of `resilient' quasiparticle excitations as discussed in \refdisp{Badmetal}. 
These excitations are beyond the reach of the high-$T$ expansion. 
It is nonetheless remarkable that the high-$T$ expansion performed in the present work provides a good estimate of the resistivity essentially 
throughout the `bad-metal' regime where $\rho\gtrsim \rho_0$.   
%
%

As emphasized in the outline Sec.~\ref{sec:outline} at the beginning of this article, the high-temperature behaviour of the resistivity 
can be analyzed either as the product of compressibility and diffusion constant $\sigma=e^2\kappa {\cal D}$, or as the product 
of a scattering time by an effective carrier number proportional to the kinetic energy $\sigma/\sigma_0 = e^2 |E_K| \tau_{\mathrm{tr}}/\hbar$. 
In the high-$T$ limit, the diffusion constant and scattering time saturate. The non-saturating $T$-linear behaviour of the resistivity is 
entirely due to the fact that the carrier number (and the compressibility) decreases as $n(1-n)/T$ as $T$ increases. 

\begin{figure}
\begin{center}
\includegraphics[width=0.5\columnwidth,keepaspectratio,angle=-90]{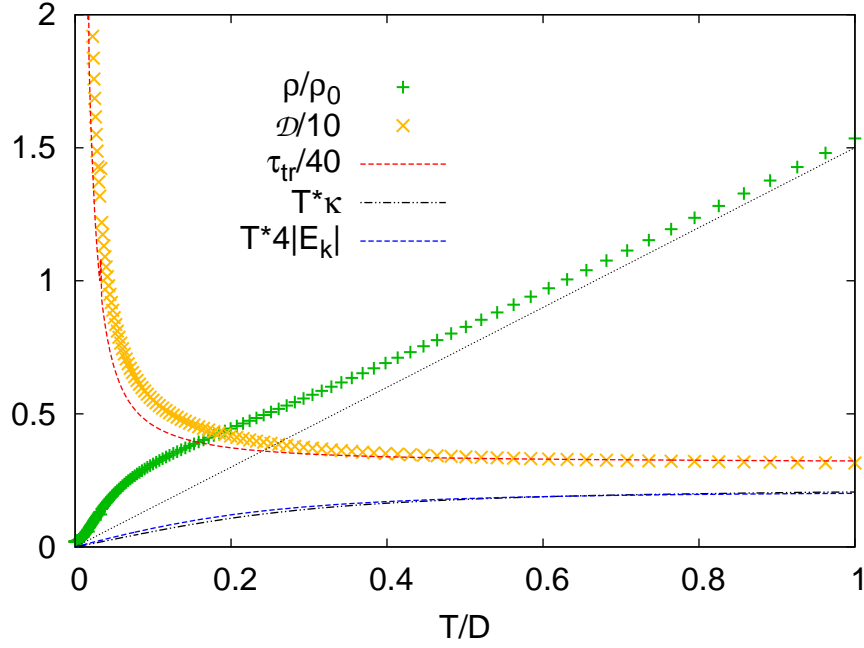}
\caption{Resistivity of the infinite-connectivity Bethe lattice at $U=\infty$ and $n=0.7$ as obtained from a NRG 
solution of the DMFT equations, plotted together with the absolute value of the kinetic energy $|E_K|$ and compressibility $\kappa$ (multiplied by $T$). 
The resulting diffusion constant and transport scattering time obtained as ${\cal D}=\sigma/\kappa$ and $\tau_{tr}=(\sigma/\sigma_0)/|E_K|$ are also 
displayed. For the plotted quantities, we measure energy in units of the half-bandwidth $D$.
}
\label{fig:diffusion}
\end{center}
\end{figure}

In order to test the validity of this analysis and its range of applicability, we display on the same plot in Fig.~\ref{fig:diffusion}, as a function of temperature, 
the $dc$-resistivity, the kinetic energy and compressibility, as well as 
the resulting diffusion constant and transport scattering time obtained as ${\cal D}=\sigma/\kappa$ and $\tau_{tr}=(\sigma/\sigma_0)/|E_K|$. 
This plot indeed reveals that the high-temperature (bad-metal) regime is characterized by a saturating transport time (and diffusion constant) 
and an effective carrier number (and compressibility) decreasing as $n(1-n)/T$. 
It also provides a physical interpretation of the `knee' feature of the resistivity (at $T/D\simeq 0.08$ for the density $n=0.7$ displayed in this plot). 
This feature was noted in previous work\cite{Badmetal}, but not understood in simple terms. Here we see that the `knee' is associated 
with the temperature below which the temperature-dependence of the
scattering time and the diffusion constant become significant: both
diverge at low $T$ as 
they should for inelastic scattering. Hence, it separates a high-$T$ regime in which the temperature dependence of the resistivity is 
dominated by that of the effective carrier number or compressibility, from a low-$T$ regime where it is dominated by that of the scattering time or diffusion constant. See also Refs.~\onlinecite{Kokalj,Hartnoll,pakhira15} for a discussion of the diffusion constant. 
%
%

\section{Thermoelectric transport coefficients and thermal conductivity in the high-$T$ limit.} 
\label{sec:thermal}

In this section, we describe the asymptotic high-$T$ behavior of thermoelectric transport coefficients in the infinite-$U$ Hubbard model. In addition to the conductivity, denoted by $\sigma$, already discussed above, we will also focus on the Seebeck coefficient, and the thermal conductivity, denoted by $\alpha$ and $\kappa$, respectively. We first define the particle, energy, and heat currents as \cite{shastry_electrothermal}

\beq
J_{\alpha}^N &=& i\sum_{im\si} t_{im}(\vR_i-\vR_m)_\alpha X_i^{\si0}X_m^{0\si},\nn\\
J_\alpha^E &=& - \frac{i}{2} \sum_{ijm\si\si'} (\vR_i-\vR_m)_\alpha t_{ij}t_{jm} X_i^{\si0}(\delta_{\si\si'} - \si\si' X_j^{\sib\sib'})X_m^{0\si'},\nn\\
J_\alpha^Q &=& J_\alpha^E - \mu J_\alpha^N.
\eeq

In terms of these, we define the generalized conductivities $\sigma_{A.B}(\omega)$ as

\beq
\frac{\sigma_{A.B}(\omega)}{\pi} = \frac{\chi_{A.B}''(\omega)}{\omega},
\eeq
where $A,B$ can be $N$, $E$, or $Q$. Here, the $\chi_{A.B}''(\omega)$ are related to correlation functions of the corresponding currents, $\chi_{A.B}(\tau) = - \langle T_\tau J_\alpha^A(\tau)J_\alpha^B\rangle$, through \disp{spectral}. The Onsager coefficients $L_{ij}$ are defined in terms of these as
\beq
L_{11} = \sigma_{N.N}(0);\;\;\;\;L_{12}=\frac{1}{ T}\sigma_{N.Q}(0);\;\;\;\;L_{21}=\frac{1}{ T}\sigma_{Q.N}(0);\;\;\;\;L_{22}=\frac{1}{T^2}\sigma_{Q.Q}(0).
\label{Onsager}
\eeq

By Onsager's reciprocity, $L_{12}=L_{21}$. 
The conductivity $\sigma$, the Seebeck coefficient $\alpha$, and the
thermal conductivity $\condth$ are defined in terms of the Onsager coefficients as
\beq
\si = e^2 L_{11};\;\;\;\; \alpha = - \frac{L_{12}}{e L_{11}};\;\;\;\; \condth = T\left(L_{22} - \frac{L_{12}L_{21}}{L_{11}}\right).
\label{transport}
\eeq

Plugging \disp{Onsager} into \disp{transport} and simplifying, we find that

\beq
\sigma &=& e^2 \sigma_{N.N}(0),\nn\\
\alpha &=& -\frac{1}{eT} \frac{\sigma_{E.N}(0)}{\sigma_{N.N}(0)} + \frac{\mu}{e T},\nn\\
\condth &=& \frac{1}{T} \sigma_{E.E}(0) - \frac{\sigma^2_{E.N}(0)}{T \sigma_{N.N}(0)}.
\label{transport2}
\eeq

We now take the $T\to\infty$ limit. To leading order, $\mu = T\bar{\mu}$ and $\sigma_{A.B}(\omega) = \frac{1}{T} \sigma^{(1)}_{A.B}(\omega)$, where $\bar{\mu}$ and $\sigma^{(1)}_{A.B}(\omega)$ are independent of the temperature. Furthermore, $\sigma^{(1)}_{N.N}(\omega)$ and $\sigma^{(1)}_{E.E}(\omega)$ are even in $\omega$, while $\sigma^{(1)}_{E.N}(\omega)$ is odd in $\omega$. This is true since the $m^{th}$ moment of $\sigma^{(1)}_{A.B}(\omega)$ is derived solely from the $m^{th}$ order term of the expansion of $\chi_{A.B}(\tau)$ around the atomic limit. The expansion of $\chi_{N.N}(\tau)$ and $\chi_{E.E}(\tau)$ contain only even orders, while the expansion of $\chi_{E.N}(\tau)$ contains only odd orders. Therefore, $\sigma^{(1)}_{E.N}(0)=0$, and in the $T\to\infty$ limit, \disp{transport2} simplifies to 
\beq
\sigma &=& \frac{e^2}{T} \sigma^{(1)}_{N.N}(0),\nn\\
\alpha &=&  \frac{\bar{\mu}}{e},\nn\\
\condth &=& \frac{1}{T^2} \sigma^{(1)}_{E.E}(0).
\label{transporthiT}
\eeq

The expression for $\alpha$ is the Heikes formula, often used in the
discussion of the high-temperature behavior of Seebeck coefficient~\cite{chaikin_beni_1976}.
It is interesting to look also at the Lorenz number ${\cal L}\equiv\frac{\condth}{T\sigma}$.
According to the Wiedemann-Franz law that holds for good (impurity
scattering dominated) metals, the Lorenz number is temperature independent quantity ${\cal L}={\cal L}_0 =\pi^2/3$.
From the equations above, we see that the Lorenz number is of
$O(1/T^2)$:  the Wiedemann-Franz law is violated at
high-temperatures. 

\section{How much of the $T\to\infty$ limit is captured by the $d\to\infty$ limit ?}
\label{sec:Tinf_dinf}

Finally, in this section, we comment on some formal issues regarding
the connection between the $T\to\infty$ limit and the $d\to\infty$
limit, focusing on the behaviour of the self-energy.  In particular,
we show how much of the former is captured by the latter.  Using
\disp{betaseriesSig}, the spectral density associated with the
self-energy for a given pair of sites $(ij)$ is given by, in the
$T\to\infty$ limit: \beq \rho_{\Sigma_{ij}}(-\mu + x) = t \
h_{ij}^{(0)}(\frac{x}{t}). \label{SiginfTrs} \eeq Therefore, using
\disp{spectralSig} \beq \tilde{\Sigma}_{ij}(i\omega_n) = t
\Sigma_{\infty,ij}^{(0)} + \sum_{m=0}^\infty
\frac{t^{m+2}}{(i\omega_n+\mu)^{m+1}} \int dy \ h_{ij}^{(0)}(y) y^{m},
\label{spectralSigmuinfTrs} \eeq where
$\tilde{\Sigma}_{ij}(i\omega_n)$ is the part of
$\Sigma_{ij}(i\omega_n)$ which comes form $h_{ij}^{(0)}(\frac{x}{t})$
and $\Sigma_{\infty,ij}^{(0)}$. We will show that the $d\to\infty$;
$T\to\infty$ limit captures only the high-frequency behavior of
$\tilde{\Sigma}_{ij}(i\omega_n)$. Equivalently, it captures only the
zeroth moment of $h_{ij}^{(0)}(y)$, but not the higher order moments.

In \refdisp{Metzner}, the author formulates a diagrammatic series to compute $\Sigma_{ij}(i\omega_n)$ in powers of $t$. The diagrams through $O(t^4)$ are given in Fig. 4 of \refdisp{Metzner}. Only those which can't be split into two by cutting a single line contribute to $\Sigma_{ij}(i\omega_n)$ (not including the atomic limit, which has zero lines). An immediate consequence is that $\Sigma_{ij}(i\omega_n)$ is $O(t^2)$, and therefore $\Sigma_{\infty,ij}^{(0)}=0$.

To power count the $d$-dependence of a diagram, we use the rule that a
path running from a fixed site $i$ to a fixed site $j$ contributes
$O\left(\left(\frac{1}{\sqrt{d}}\right)^{r_{ij}}\right)$. Hence,
polygons have no $d$ dependence. In the $d\to\infty$ limit, all 
$\Sigma_{ij}(i\omega_n)$ diagrams that vanish faster than
$O\left(\left(\frac{1}{\sqrt{d}} \right)^{r_{ij}}\right)$ must be
discarded. An inspection of the diagrams shows that for all orders
higher than second, and for all separations between $i$ and $j$, some
of the self-energy diagrams must be discarded in the $d\to\infty$
limit. On the other hand, they contribute to $h_{ij}^{(0)}(y)$ through
\disp{spectralSigmuinfTrs}, and hence survive the $T\to\infty$ limit.
Therefore, the $d\to\infty$; $T\to\infty$ limit captures only the
zeroth moment of $h_{ij}^{(0)}(y)$, but not the higher order moments.

We now consider the thermodynamic potential $\Omega$, whose diagrams are displayed in Fig. 1 of \refdisp{Metzner}. Once again, for all orders higher than second, some of the diagrams must be discarded in the $d\to\infty$ limit. However, since $\Omega$ is a static quantity, only the zeroth order contribution (the atomic limit) must be kept in the $T\to\infty$ limit. Therefore, the $d \to\infty$; $T\to\infty$ limit completely captures $\Omega$ in the $T\to\infty$ limit.

Finally, we address to what extent $h_{ij}^{(0)}(y)$ is local in the
$T\to\infty$ limit. By inspection, only diagrams of $O(t^{3
r_{ij}+2\delta_{ij}+2l})$, where $l$ is a non-negative integer,
contribute to $\Sigma_{ij}(i\omega_n)$. Therefore, using
\disp{spectralSigmuinfTrs}, $h_{ij}^{(0)}(y)$ has non-vanishing
moments of order $3 \ r_{ij} + 2\delta_{ij} - 2 + 2l$. Therefore, the
$T\to\infty$ limit kills off the first $3 \ r_{ij} + 2\delta_{ij} - 2$
moments of $\rho_{\Sigma_{ij}}(-\mu+x)$, but does not completely
eliminate non-local contributions to the self-energy.

The general conclusion is that the $d=\infty$ limit (DMFT)
captures the dominant term in correlation functions when both a high-temperature {\it and} a high-frequency expansion are performed, while  
it captures the dominant orders in $1/T$ in thermodynamic quantities. This clarifies which limits have to be taken so that a local approximation becomes accurate.

\section{Conclusion and Perspectives}
\label{sec:conclusion}

We have analyzed the transport properties of the paradigmatic model
for strong correlations between electrons residing on a lattice, the
Hubbard model. We used analytical series expansion in $1/T$ to high
order, and have applied standard as well as novel techniques developed here (e.g. Appendix \ref{momentsrec}) to reconstruct frequency-dependent response functions from their moments. The results have been  compared to the numerical calculation within the
framework of the DMFT (using NRG and interaction expansion CT-QMC
impurity solvers).  We have found an excellent overlap of the results
over a surprisingly wide range of temperatures, not only in the
asymptotic high-T limit but also well into the $T < D$ range, covering
all the bad-metal regime.  In fact, this significantly exceeds a
priori expectations on two counts: 1) the high-temperature expansion
is reliable down to surprisingly low (experimentally relevant)
temperatures, 2) the NRG works to surprisingly high temperatures as
an impurity solver in the DMFT, in particular when integrated
quantities such as resistivity are computed. This agreement in
the infinite-$d$ limit suggests that in the generic finite-$d$
situation the series expansion technique is likewise reliable in such
a wide temperature range.  The series expansion approach allows us to
formulate some very general statements about the transport properties
of strongly-correlated electron systems that we summarize below. Some
of these have been suggested before, but our work provides their
definitive proofs in the high-$T$ limit.

\ag{The strongly-correlated Hubbard model displays bad metal behaviour at high temperature with no 
resistivity saturation: the} resistivity smoothly crosses
the Mott-Ioffe-Regel limit where the mean-free path of quasiparticles
is reduced to the lattice spacing. The (resilient) quasiparticles
disappear at the Brinkman-Rice scale of order $(1-n)D$, above which
the transport is fully incoherent. The $T$-linear dependence of the
resistivity in this high-$T$ regime can be understood by factorizing
the conductivity $\sigma$ either into the diffusion constant and
charge compressibility, or into the transport scattering time and the
kinetic energy {(which can be interpreted as the effective carrier number)}. 
The diffusion constant $\mathcal{D}$ and the transport
scattering time $\tau_{tr}$ saturate at high temperatures as the
consequence of the presence of the lattice. The charge carriers at
high temperatures are essentially bare electrons obeying the
constraint of no double occupancy on lattice sites. 
The temperature
dependence is then entirely due to the static quantities,
compressibility or kinetic energy, which have asymptotically the same
behavior as functions of electron density and temperature, $\propto
n(1-n)/T$. From this perspective, it is seen that the violation of the
MIR limit is no surprise, since the Drude-Boltzmann picture only
applies to quasiparticle transport. For fully incoherent transport on
a lattice, the only relevant element is the exclusion on lattice sites
and the only parameter at play is the electron density $n$. Indeed, we
find that $\mathcal{D} \propto 1/n$ \ag{at low density}.  

Another general finding concerns the frequency dependence of response
functions in the high temperature regime. It is a priori unclear what
is expected. On one hand the atomic picture could be relevant,
suggesting sharp (essentially $\delta$-like) spectral features. On the
other hand, the strong incoherence might suggest rather flat
(essentially constant) spectral functions. Instead, we find an
intermediate situation with smooth functions that saturate in the
high-$T$ limit to (a sum of) peaks whose positions are determined by
atomic physics, while their widths are of the order hopping parameter
$t$. This again suggests the crucial role of the presence of the
lattice which sets this scale.

It is interesting to contrast these results to what one expect from
theories formulated in the continuum. There are two
key differences, a) the optical sum-rule in absence of the lattice is
simply a temperature independent quantity, $\frac{n e^2}{m}$, and b), 
the lower bound on the diffusion constant that is in the fully incoherent
regime on a lattice set by {$\mathcal{D}\mathrm{min} \sim a^2 t/n$} clearly cannot apply. 

\ag{Finally, we briefly comment on the experimental relevance of this work. 
Bad metal behaviour without resistivity saturation is indeed observed in many transition-metal oxides. 
Most of these materials are however multi-orbital systems which require an extension of our calculations 
beyond the single-band Hubbard model for a reliable comparison to be possible. 
Cuprates provide an example of materials better described by a single-band Hubbard model, and 
indeed some of the qualitative features observed in the high-temperature bad metallic transport of these materials 
in the underdoped regime are in qualitative agreement with our findings. 
For instance, the high-temperature resistivity in some cuprates is indeed
found to be proportional to $T/(1-n)$ \cite{barisic13,takagi_1992}, with a slope becoming larger 
at low doping levels, in agreement with Eq.~\eqref{eq:resist_intro}. 
Note however that in our results this behaviour results from a saturated scattering rate or diffusion constant, the 
temperature dependence coming from the effective number of carriers or compressibility. 
Our calculations establish that this is indeed the mechanism applying at very high temperature $T\gtrsim 0.2 D$ 
but it is far from obvious that it still applies in the lower temperature range $T\lesssim 1000$~K 
relevant to measurements on underdoped cuprates.} 

\begin{figure}
\begin{center}
\includegraphics[width=0.6\columnwidth,keepaspectratio]{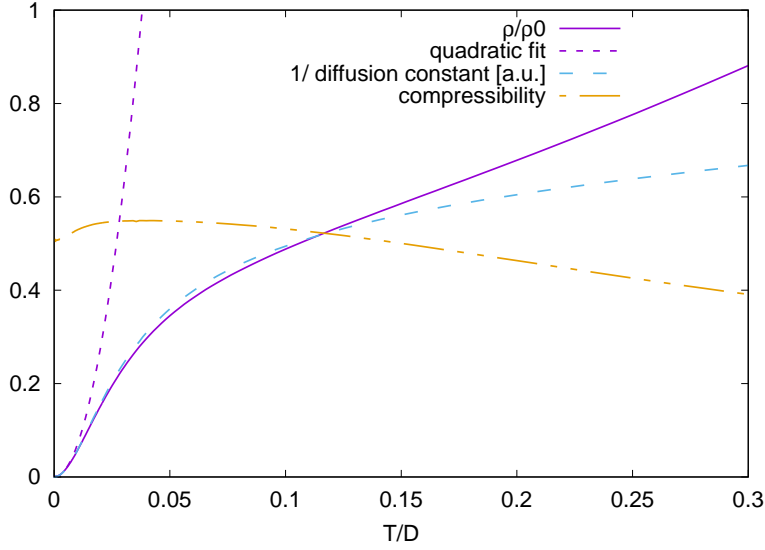}
\caption{{\color{black} Resistivity of the infinite-connectivity Bethe lattice at $U=\infty$ and $n=0.7$ as obtained from a NRG 
solution of the DMFT equations, plotted together with the inverse diffusion constant and the compressibility. The quadratic fit displays the $T^2$ behavior of the inverse diffusion constant and the resistivity in the Fermi liquid regime. In the range $T/D\sim 0.01-0.05$, both the resistivity and the inverse diffusion constant are quasi-linear in the temperature, while the compressibility is essentially constant. }
\label{diffusion_vs_rho_lowT}}
\end{center}
\end{figure}

\ag{In this respect, an interesting observation, with possible qualitative relevance to underdoped cuprates, can be made 
from the DMFT-NRG results, as plotted in Fig.~\ref{diffusion_vs_rho_lowT}. 
In this figure, we display simultaneously the resistivity, the compressibility and the inverse of the diffusion constant as a function 
of temperature. We see that the approximately linear dependence of $\rho(T)$ in the range $T/D\sim 0.01-0.05$ 
(below the `knee') does correspond to a regime in which the diffusion constant varies approximately as $1/T$, while the compressibility 
(effective number of carriers) is essentially constant. This regime results from the crossover between the very low-$T$ Fermi liquid regime 
in which ${\cal D}\propto 1/T^2$ and the high-$T$ asymptotic regime in which ${\cal D}\sim \mathrm{const}$. 
 }
 
\ag{Besides the difficulty of reaching the high-temperature regime in which our calculations apply, an additional
complication in the solid-state context is the presence of additional degrees of freedom, such as phonons and other electron bands, 
which precludes a detailed quantitative comparison.}
For this reason, we expect that an ideal setup in which the predictions and quantitative results of this article 
could be tested is that of fermionic cold atoms in optical lattices 
(see Ref.~\onlinecite{bloch_rmp_2008} for a review and e.g. Ref.~\onlinecite{schneider12} for an 
experimental investigation of transport in the optical lattice). 
Both thermodynamic quantities and transport can now be measured in such systems. 
Furthermore, the high to intermediate temperature range (of order the hopping amplitude $t$) is currently the most easily accessible. 
Finally, the presence of the lattice can be turned on and off,
providing the means for studying the transport in the high-temperature limit either in the continuum or on the lattice.

\acknowledgements
We acknowledge useful discussions with Jure Kokalj and Jak\v{s}a Vu\v{c}i\v{c}evi\'c. 
This work was supported by the European Research Council (ERC-319286 QMAC), and by the Swiss National Science Foundation (NCCR MARVEL).
JM and R\v{Z} acknowledge the support of the Slovenian Research Agency
(ARRS) under P1-0044 and J1-7259. The work at UCSC was supported by the U.S. Department of Energy (DOE), Office of Science, Basic Energy Sciences (BES) under Award DE-FG02-06ER46319.

\appendix
\section{ High-frequency expansions of $\chi(\vk,i\Omega_n)$ \label{chi}}
In this appendix, we show how the RHS of \disp{chicd} can be expanded in a series in $\frac{1}{i\Omega_n}$. The integral appearing in \disp{chicd} is best handled by applying the change of variables $\tau\leftrightarrow (\beta-\tau)$ to the first term in the integrand. 
This yields the expression

\beq
\chi_{O.O}(\vk,i\Omega_n) = -  \sum_{c=0,d=0}^\infty \frac{(-1)^{c+d}}{c!(c+d)!} \langle \hat{H}^{c+d}_1 \hat{O}_{k}  \hat{H}^c_1\hat{O}_{-k}\rangle_{0,c}\frac{2}{1+\delta_{d,0}}\int_0^\beta d\tau \cos(\Omega_n\tau)(\tau-\beta)^c\tau^{c+d}.
\label{chicd2}
\eeq

Expanding the $(\tau-\beta)^c$, plugging in $\hat{H}_1=\tilde{\hat{H}}_1E$, and using the formula

\beq
\int_0^\beta \tau^a \cos(\Omega_n \tau) d\tau = \delta_{n,0}\frac{\beta^{a+1}}{a+1} + (1-\delta_{n,0})\sum_{\substack{j=2\\even}}^a \frac{a!(-1)^{\frac{j}{2}+1}}{(a-j+1)!} \frac{\beta^{a+1-j}}{\Omega_n^j} \;\;\;(a\ge0),
\label{cosint}
\eeq
\disp{chicd2} becomes
\beq
\chi_{O.O}(\vk,i\Omega_n) &=& \sum_{c=0,d=0}^\infty \frac{(-1)^{c+d}}{(c+d)!}\frac{2}{1+\delta_{d,0}} E^{2c+d+\gamma(O)+1} \langle \tilde{\hat{H}}^{c+d}_1 \tilde{\hat{O}}_{k}  \tilde{\hat{H}}^c_1\tilde{\hat{O}}_{-k}\rangle_{0,c}\nn\\
&&\sum_{j=0}^c \frac{(-1)^j}{j!(c-j)!}\sum_{\substack{r=0\\even}}^{2c+d-j} \frac{(2c+d-j)!}{(2c+d-j-r+1)!} \beta^{2c+d+1-r}\left[\frac{(1-\delta_{r,0})(1-\delta_{n,0})}{(i\Omega_n)^r}-\delta_{r,0}\delta_{n,0}\right].
\label{chicd3}
\eeq
Finally, rearranging the sum yields
\beq
\chi_{O.O}(\vk,i\Omega_n) &=& \sum_{\substack{r=0\\even}}^{\infty}\left[\frac{(1-\delta_{r,0})(1-\delta_{n,0})}{(i\Omega_n)^r}-\delta_{r,0}\delta_{n,0}\right]\nn\\
&&\sum_{\substack{c=0,d=0\\2c+d\ge r}}^\infty  \beta^{2c+d+1-r} \frac{(-1)^{c+d}}{(c+d)!}\frac{2}{1+\delta_{d,0}} E^{2c+d+\gamma(O)+1} \langle \tilde{\hat{H}}^{c+d}_1 \tilde{\hat{O}}_{k}  \tilde{\hat{H}}^c_1\tilde{\hat{O}}_{-k}\rangle_{0,c}\nn\\
&&\sum_{j=0}^{min[c,2c+d-r]} \frac{(-1)^j}{j!(c-j)!} \frac{(2c+d-j)!}{(2c+d-j-r+1)!}.
\label{chicd4}
\eeq

\section{High frequency expansion of $G(\vk,i\omega_n)$ \label{Gapp}}
In this appendix, we derive the expansion of $G(\vk,i\omega_n)$ in powers of $\frac{1}{i\omega_n+\mu}$. Expanding the exponentials in \disp{Gvktau}, we obtain
\beq
G(\vk,\tau) = -e^{\mu \tau}\sum_{a=0,b=0}^\infty \frac{(\tau-\beta)^a}{a!}\frac{(-\tau)^b}{b!} \langle \hat{H}_1^a c_{k\si} \hat{H}_1^b c^\dagger_{k\si}\rangle_{0,c}.
s\label{Gktau}
\eeq
Expanding $(\tau-\beta)^a$, plugging in $\hat{H}_1=E \tilde{\hat{H}}_1$, and taking the Fourier transform, we find that
\beq
G(\vk,i\omega_n) = -\sum_{a=0,b=0}^\infty \frac{1}{a!}\frac{(-1)^b}{b!} E^{a+b}\langle \tilde{\hat{H}}_1^a c_{k\si} \tilde{\hat{H}}_1^b c^\dagger_{k\si}\rangle_{0,c}\sum_{j=0}^a {a\choose j} (-\beta)^j \int_0^\beta e^{(i\omega_n+\mu)\tau}\tau^{a+b-j}d\tau.
\label{Gk}
\eeq
Plugging in the integral
\beq
\int_0^\beta e^{i\omega_n\tau} e^{\mu\tau}\tau^a d\tau = \sum_{r=1}^{a+1} \frac{(-1)^r}{(i\omega_n+\mu)^r} \frac{a!}{(a+1-r)!} \beta^{a+1-r}  (e^{\beta \mu}+\delta_{a+1-r,0}),
\label{intG}
\eeq	
we find that
\beq
G(\vk,i\omega_n) = -\sum_{a=0,b=0}^\infty \frac{(-1)^b}{b!} E^{a+b}\langle \tilde{\hat{H}}_1^a c_{k\si} \tilde{\hat{H}}_1^b c^\dagger_{k\si}\rangle_{0,c}\sum_{j=0}^a \frac{(-1)^j}{j!(a-j)!} \sum_{r=1}^{a+b-j+1} &&\frac{(-1)^{r} }{(i\omega_n+\mu)^{r}}\frac{(a+b-j)!}{(a+b-j+1-r)!}\nn\\
&&\beta^{a+b+1-r}(e^{\bar{\mu}}+\delta_{a+b-j+1-r,0}),
\label{Gk2}
\eeq
where we have used the definition $\bar{\mu}=\beta \mu$. Finally, rearranging the series in powers of $\frac{1}{i\omega_n+\mu}$ yields
\barray
G(\vk,i\omega_n) = \sum_{m=1}^\infty \frac{1}{(i\omega_n+\mu)^m}&&\sum_{\substack{a=0,b=0\\a+b\ge m-1}}^\infty \frac{1}{b!} E^{a+b}\langle \tilde{\hat{H}}_1^a c_{k\si} \tilde{\hat{H}}_1^b c^\dagger_{k\si}\rangle_{0,c} \beta^{a+b+1-m} \nn\\
&&\sum_{j=0}^{min(a+b+1-m,a)} \frac{1}{j!(a-j)!}  (-1)^{j+m-b+1}\frac{(a+b-j)!}{(a+b-j-m+1)!}(e^{\bar{\mu}}+\delta_{j,a+b+1-m}).\nn\\
\label{Gk3}
\earray

\section{Maximum Entropy Method \label{MEM}}

Suppose that one has an even distribution $P(x)$, whose first $k +1$ even moments are known. The MEM estimates $P(x)$ by minimizing the functional $S[P(x)]$ w.r.t. $P(x)$. We define this functional as
\beq
S[P(x)] \equiv -\int dx P(x) [\ln P(x) - 1]
\eeq
This functional must be minimized with the constraint that 
\beq
\int dx P(x) x^{2n}=m_{2n} \;\;\;\;\;\;(n=0\ldots k)
\label{MEMmoments}
\eeq
Therefore, we introduce $k+1$ Lagrange multipliers $\lambda_{2n}$ and minimize the overall functional
\beq
I[P(x)] = -\int dx P(x) [\ln P(x) - 1]-\sum_{n=0}^k \lambda_{2n}\left[\int dx\ x^{2n} P(x)-m_{2n}\right].\label{functional}
\eeq
Minimizing w.r.t $P(x)$ yields
\beq
P(x) = \exp[{-\sum_{n=0}^k \lambda_{2n} x^{2n}}]\label{Px}.
\eeq
Plugging \disp{Px} into \disp{functional} yields
\beq
I[P(x)] = \int dx P(x) +\sum_{n=0}^k \lambda_{2n} m_{2n}. \label{functionalsimple}
\eeq
The minimization of \disp{functionalsimple} w.r.t to the Lagrange multipliers is now performed numerically, and ensures that the $\lambda_{2n}$ are chosen such that \disp{MEMmoments} is satisfied. 

\section{Reconstruction of the optical conductivity from its moments using Mori formalism. \label{momentsrec}}

We have developed here a new method (\disp{taurrelaxation}) to reconstruct $f^{(1)}(\frac{\omega}{D})$. In the Mori approach to transport\cite{Mori, Mori2, Lovesey}, the relaxation function $R(\vk,t)$ is given by the Fourier Transform of $\frac{\sigma(\vk,\omega)}{\pi\si_0}$.
\beq
R(\vk,t) = \int d\omega e^{i\omega t} \frac{\sigma(\vk,\omega)}{\pi\si_0}.
\label{relaxation}
\eeq
The Laplace transform of $R(\vk,t)$ is defined as
\beq
R(\vk,s) = \int_0^\infty dt  \ e^{-st}  R(\vk,t).
\label{Laplace}
\eeq
Combining \disp{relaxation} and \disp{Laplace}, $R(\vk,s)$ is related to $\frac{\sigma(\vk,\omega)}{\pi\si_0}$ via the transform
\beq
R(\vk,s) = \int d\omega \frac{\sigma(\vk,\omega)}{\pi\si_0} \frac{1}{s-i\omega}.
\label{Rsig} 
\eeq
Defining $m_n(\vk)=\int d\omega \frac{\sigma(\vk,\omega)}{\pi\si_0} \omega^n$, we normalize $ \frac{\sigma(\vk,\omega)}{\pi\si_0}$ by dividing it by $m_0(\vk) = R(\vk,t=0)$. Then \disp{Rsig} becomes
\beq
\frac{R(\vk,s)}{m_0(\vk)} = \int d\omega \frac{\sigma(\vk,\omega)}{\pi\si_0} \frac{1}{m_0(\vk)} \frac{1}{s-i\omega}.
\label{Rsig2} 
\eeq
$\frac{\sigma(\vk,\omega)}{\pi\si_0}$ can be recovered from $\frac{R(\vk,s)}{m_0(\vk)} $ via the formula
\beq
\frac{\sigma(\vk,\omega)}{\pi\si_0} = \frac{m_0(\vk)}{\pi} \ \Re e \left[\frac{R(\vk,i\omega + \eta)}{m_0(\vk)} \right].
\label{Rsig3}
\eeq
We now write \disp{Rsig2} as a continued fraction
\beq
\frac{R(\vk,s)}{m_0(\vk)} = \frac{1}{s+}  \;\; \frac{\delta_1(\vk)}{s +}\;\; \frac{\delta_2(\vk)}{s+}\;\;\frac{\delta_3(\vk)}{s+\ldots}.
\eeq
The $\delta_n$ can be written in terms of the $m_n$ as
\beq
\delta_1=\tilde{m}_2;\;\;\;\;\delta_2=\frac{\tilde{m}_4}{\tilde{m}_2}-\tilde{m}_2;\;\;\;\;\;\delta_3=\frac{\tilde{m}_6\tilde{m}_2-\tilde{m}_4^2}{\tilde{m}_2(\tilde{m}_4-\tilde{m}_2^2)}\;\ldots,
\label{delta}
\eeq
where $\tilde{m}_n=\frac{m_n}{m_0}$. By going down one step in the continued fraction, it is possible to define a new function $R_1(\vk,s)$,
\beq
R_1(\vk,s) = \frac{1}{s +}\;\; \frac{\delta_2(\vk)}{s+}\;\;\frac{\delta_3(\vk)}{s+\ldots}, 
\label{R1}
\eeq
so that
\beq
\frac{R(\vk,s)}{m_0(\vk)} = \frac{1}{s+\delta_1(\vk)R_1(\vk,s)}.  
\eeq
The residual relaxation function, $R_1(\vk,t)$, is related to $R_1(\vk,s)$ through \disp{Laplace}. One may also define a residual conductivity $\sigma_1(\vk,\omega)$, related to $R_1(\vk,s)$ through \disp{Rsig}. Note that since the leading order term in \disp{R1} is $\frac{1}{s}$, $\frac{\sigma_1(\vk,\omega)}{\pi\si_0}$ is already normalized. It is possible to define an infinite sequence of $R_i(\vk,s)$, 
\beq
R_i(\vk,s) = \frac{1}{s +}\;\; \frac{\delta_{i+1}(\vk)}{s+}\;\;\frac{\delta_{i+2}(\vk)}{s+\ldots}\;\;\;\;\;(i\ge1), 
\label{Ri}
\eeq
so that $\frac{R(\vk,s)}{m_0(\vk)}$ can be written in terms of $R_i(\vk,s)$ as 
\beq
\frac{R(\vk,s)}{m_0(\vk)} = \frac{1}{s+}  \;\; \frac{\delta_1(\vk)}{s +}\ldots\frac{\delta_{i-1}(\vk)}{s+\delta_i(\vk) R_i(\vk,s)}.
\label{RRi}
\eeq
To reconstruct $\frac{\sigma(\vk,\omega)}{\pi\si_0}$ from its first $r+1$ even moments (i.e. all $m_n(\vk)$ are known through $n=2r$), we use the following procedure. Using \disp{delta}, calculate $\delta_1(\vk), \ldots \delta_r(\vk)$. Then, using \disp{RRi}, express $\frac{R(\vk,s)}{m_0(\vk)}$ in terms of $R_{r-1}(\vk,s)$. Estimate $R_{r-1}(\vk,s)$ using $\delta_r(\vk)$. Finally, use \disp{Rsig3} to obtain $\frac{\sigma(\vk,\omega)}{\pi\si_0}$.
Using \disp{Ri}, we express $R_{r-1}(\vk,s)$ as
\beq
R_{r-1}(\vk,s) = \frac{1}{s +}\;\; \frac{\delta_{r}(\vk)}{s+}\;\;\frac{\delta_{r+1}(\vk)}{s+\ldots}.
\label{RRm1}
\eeq
This expansion is valid in the large $s$ limit, and in this limit we keep only the first two leading order terms.
\beq
R_{r-1}(\vk,s) = \frac{1}{s +\frac{\delta_r(\vk)}{s}} \;\;\;(s\gg\delta_r(\vk)).
\label{RRm1larges}
\eeq
The form appropriate for the small $s$ limit is the Drude form
\beq
R_{r-1}(\vk,s) = \frac{1}{s+\tau_r^{-1}(\vk)}\;\;\;(s\ll\tau_r^{-1}(\vk)).
\label{RRm1smalls}
\eeq
Using \disp{Rsig3} and \disp{Laplace}, this corresponds to 
\beq
\frac{\si_{r-1}(\vk,\omega)}{\pi\si_0} = \frac{1}{\pi} \ \frac{\tau_r^{-1}(\vk)}{\omega^2 + [\tau_r^{-1}(\vk)]^2};\;\;\;R_{r-1}(\vk,t) = \exp\left[-\tau_r^{-1}(\vk) \ |t|\right].
\eeq
The two are Fourier Transform paris as required by \disp{relaxation}. The exponential decay is the physically correct long-time dependence of correlation functions, and hence of the relaxation function. The crossover between the short-time and long-time regimes is expected to occur when $s=\tau_r^{-1}(\vk)$, and we therefore estimate $\tau_r^{-1}(\vk)$ by equating the RHS of \disp{RRm1larges} with the RHS of \disp{RRm1smalls} at this value of $s$. Hence, we find that 

\beq
\tau_r^{-1}(\vk)=\sqrt{\delta_r(\vk)}
\label{taurrelaxation}
\eeq 

Therefore, our final expression for $\frac{R(\vk,s)}{m_0(\vk)}$ is
\beq
\frac{R(\vk,s)}{m_0(\vk)} = \frac{1}{s+}  \;\; \frac{\delta_1(\vk)}{s +}\ldots\frac{\delta_{r-2}(\vk)}{s+\frac{\delta_{r-1}(\vk)} {s+\sqrt{\delta_r(\vk)}}}.
\label{finalR}
\eeq 
To obtain $\frac{\sigma(\omega)}{\pi\si_0}$, we plug this form into the RHS of \disp{Rsig3} and take the $k\to0$ limit. Note that $r$ can be varied at will, and that therefore this provides us with a systematic truncation procedure for the reconstruction problem.



\begin{thebibliography}{72}
\expandafter\ifx\csname natexlab\endcsname\relax\def\natexlab#1{#1}\fi
\expandafter\ifx\csname bibnamefont\endcsname\relax
  \def\bibnamefont#1{#1}\fi
\expandafter\ifx\csname bibfnamefont\endcsname\relax
  \def\bibfnamefont#1{#1}\fi
\expandafter\ifx\csname citenamefont\endcsname\relax
  \def\citenamefont#1{#1}\fi
\expandafter\ifx\csname url\endcsname\relax
  \def\url#1{\texttt{#1}}\fi
\expandafter\ifx\csname urlprefix\endcsname\relax\def\urlprefix{URL }\fi
\providecommand{\bibinfo}[2]{#2}
\providecommand{\eprint}[2][]{\url{#2}}

\bibitem[{\citenamefont{Gunnarsson et~al.}(2003)\citenamefont{Gunnarsson,
  Calandra, and Han}}]{gunnarsson_saturation_rmp}
\bibinfo{author}{\bibfnamefont{O.}~\bibnamefont{Gunnarsson}},
  \bibinfo{author}{\bibfnamefont{M.}~\bibnamefont{Calandra}}, \bibnamefont{and}
  \bibinfo{author}{\bibfnamefont{J.~E.} \bibnamefont{Han}},
  \bibinfo{journal}{Rev. Mod. Phys.} \textbf{\bibinfo{volume}{75}},
  \bibinfo{pages}{1085} (\bibinfo{year}{2003}),
  \urlprefix\url{http://link.aps.org/doi/10.1103/RevModPhys.75.1085}.

\bibitem[{\citenamefont{Calandra and Gunnarsson}(2002)}]{CalandraGunnarsson}
\bibinfo{author}{\bibfnamefont{M.}~\bibnamefont{Calandra}} \bibnamefont{and}
  \bibinfo{author}{\bibfnamefont{O.}~\bibnamefont{Gunnarsson}},
  \bibinfo{journal}{Phys. Rev. B} \textbf{\bibinfo{volume}{66}},
  \bibinfo{pages}{205105} (\bibinfo{year}{2002}),
  \urlprefix\url{http://link.aps.org/doi/10.1103/PhysRevB.66.205105}.

\bibitem[{\citenamefont{Calandra and Gunnarsson}(2003)}]{calandra03epl}
\bibinfo{author}{\bibfnamefont{M.}~\bibnamefont{Calandra}} \bibnamefont{and}
  \bibinfo{author}{\bibfnamefont{O.}~\bibnamefont{Gunnarsson}},
  \bibinfo{journal}{Europhys. Lett.,} \textbf{\bibinfo{volume}{61}},
  \bibinfo{pages}{88} (\bibinfo{year}{2003}).

\bibitem[{\citenamefont{Werman and Berg}(2016)}]{berg_badmetal_2016}
\bibinfo{author}{\bibfnamefont{Y.}~\bibnamefont{Werman}} \bibnamefont{and}
  \bibinfo{author}{\bibfnamefont{E.}~\bibnamefont{Berg}},
  \bibinfo{journal}{Phys. Rev. B} \textbf{\bibinfo{volume}{93}},
  \bibinfo{pages}{075109} (\bibinfo{year}{2016}),
  \urlprefix\url{http://link.aps.org/doi/10.1103/PhysRevB.93.075109}.

\bibitem[{\citenamefont{Emery and Kivelson}(1995)}]{emery_kivelson_prl_1995}
\bibinfo{author}{\bibfnamefont{V.~J.} \bibnamefont{Emery}} \bibnamefont{and}
  \bibinfo{author}{\bibfnamefont{S.~A.} \bibnamefont{Kivelson}},
  \bibinfo{journal}{Phys. Rev. Lett.} \textbf{\bibinfo{volume}{74}},
  \bibinfo{pages}{3253} (\bibinfo{year}{1995}).

\bibitem[{\citenamefont{Hussey et~al.}(2004)\citenamefont{Hussey, Takenaka, and
  Takagi}}]{hussey_phil_mag_2004}
\bibinfo{author}{\bibfnamefont{N.}~\bibnamefont{Hussey}},
  \bibinfo{author}{\bibfnamefont{K.}~\bibnamefont{Takenaka}}, \bibnamefont{and}
  \bibinfo{author}{\bibfnamefont{H.}~\bibnamefont{Takagi}},
  \bibinfo{journal}{Phil. Mag.} \textbf{\bibinfo{volume}{84}},
  \bibinfo{pages}{2847} (\bibinfo{year}{2004}).

\bibitem[{\citenamefont{Mirzaei et~al.}(2013)\citenamefont{Mirzaei, Stricker,
  Hancock, Berthod, Georges, van Heumen, Chan, Zhao, {Li}, {Greven}
  et~al.}}]{mirzaei_fermiliquid_cuprates_pnas_2013}
\bibinfo{author}{\bibfnamefont{S.~I.} \bibnamefont{Mirzaei}},
  \bibinfo{author}{\bibfnamefont{D.}~\bibnamefont{Stricker}},
  \bibinfo{author}{\bibfnamefont{J.~N.} \bibnamefont{Hancock}},
  \bibinfo{author}{\bibfnamefont{C.}~\bibnamefont{Berthod}},
  \bibinfo{author}{\bibfnamefont{A.}~\bibnamefont{Georges}},
  \bibinfo{author}{\bibfnamefont{E.}~\bibnamefont{van Heumen}},
  \bibinfo{author}{\bibfnamefont{M.~K.} \bibnamefont{Chan}},
  \bibinfo{author}{\bibfnamefont{X.}~\bibnamefont{Zhao}},
  \bibinfo{author}{\bibfnamefont{Y.}~\bibnamefont{{Li}}},
  \bibinfo{author}{\bibfnamefont{M.}~\bibnamefont{{Greven}}},
  \bibnamefont{et~al.}, \bibinfo{journal}{Proc. Natl. Acad. Sci.}
  \textbf{\bibinfo{volume}{110}}, \bibinfo{pages}{5774} (\bibinfo{year}{2013}).

\bibitem[{\citenamefont{{Bari{\v s}i{\'c}} et~al.}()\citenamefont{{Bari{\v
  s}i{\'c}}, {Chan}, {Veit}, {Dorow}, {Ge}, {Tang}, {Tabis}, {Yu}, {Zhao}, and
  {Greven}}}]{barisic_hidden_2015}
\bibinfo{author}{\bibfnamefont{N.}~\bibnamefont{{Bari{\v s}i{\'c}}}},
  \bibinfo{author}{\bibfnamefont{M.~K.} \bibnamefont{{Chan}}},
  \bibinfo{author}{\bibfnamefont{M.~J.} \bibnamefont{{Veit}}},
  \bibinfo{author}{\bibfnamefont{C.~J.} \bibnamefont{{Dorow}}},
  \bibinfo{author}{\bibfnamefont{Y.}~\bibnamefont{{Ge}}},
  \bibinfo{author}{\bibfnamefont{Y.}~\bibnamefont{{Tang}}},
  \bibinfo{author}{\bibfnamefont{W.}~\bibnamefont{{Tabis}}},
  \bibinfo{author}{\bibfnamefont{G.}~\bibnamefont{{Yu}}},
  \bibinfo{author}{\bibfnamefont{X.}~\bibnamefont{{Zhao}}}, \bibnamefont{and}
  \bibinfo{author}{\bibfnamefont{M.}~\bibnamefont{{Greven}}},
  \bibinfo{note}{arXiv:1507.07885}.

\bibitem[{\citenamefont{Tyler et~al.}(1998)\citenamefont{Tyler, Mackenzie,
  NishiZaki, and Maeno}}]{maeno_badmetal_prb_1998}
\bibinfo{author}{\bibfnamefont{A.~W.} \bibnamefont{Tyler}},
  \bibinfo{author}{\bibfnamefont{A.~P.} \bibnamefont{Mackenzie}},
  \bibinfo{author}{\bibfnamefont{S.}~\bibnamefont{NishiZaki}},
  \bibnamefont{and} \bibinfo{author}{\bibfnamefont{Y.}~\bibnamefont{Maeno}},
  \bibinfo{journal}{Phys. Rev. B} \textbf{\bibinfo{volume}{58}},
  \bibinfo{pages}{R10107} (\bibinfo{year}{1998}),
  \urlprefix\url{http://link.aps.org/doi/10.1103/PhysRevB.58.R10107}.

\bibitem[{\citenamefont{Mackenzie and Maeno}(2003)}]{Mackenzie-2003}
\bibinfo{author}{\bibfnamefont{A.~P.} \bibnamefont{Mackenzie}}
  \bibnamefont{and} \bibinfo{author}{\bibfnamefont{Y.}~\bibnamefont{Maeno}},
  \bibinfo{journal}{Rev. Mod. Phys.} \textbf{\bibinfo{volume}{75}},
  \bibinfo{pages}{657} (\bibinfo{year}{2003}).

\bibitem[{\citenamefont{Deng et~al.}(2013)\citenamefont{Deng, Mravlje,
  \ifmmode~\check{Z}\else \v{Z}\fi{}itko, Ferrero, Kotliar, and
  Georges}}]{Badmetal}
\bibinfo{author}{\bibfnamefont{X.}~\bibnamefont{Deng}},
  \bibinfo{author}{\bibfnamefont{J.}~\bibnamefont{Mravlje}},
  \bibinfo{author}{\bibfnamefont{R.}~\bibnamefont{\ifmmode~\check{Z}\else
  \v{Z}\fi{}itko}}, \bibinfo{author}{\bibfnamefont{M.}~\bibnamefont{Ferrero}},
  \bibinfo{author}{\bibfnamefont{G.}~\bibnamefont{Kotliar}}, \bibnamefont{and}
  \bibinfo{author}{\bibfnamefont{A.}~\bibnamefont{Georges}},
  \bibinfo{journal}{Phys. Rev. Lett.} \textbf{\bibinfo{volume}{110}},
  \bibinfo{pages}{086401} (\bibinfo{year}{2013}),
  \urlprefix\url{http://link.aps.org/doi/10.1103/PhysRevLett.110.086401}.

\bibitem[{\citenamefont{Georges et~al.}(1996)\citenamefont{Georges, Kotliar,
  Krauth, and Rozenberg}}]{georges_rmp_1996}
\bibinfo{author}{\bibfnamefont{A.}~\bibnamefont{Georges}},
  \bibinfo{author}{\bibfnamefont{G.}~\bibnamefont{Kotliar}},
  \bibinfo{author}{\bibfnamefont{W.}~\bibnamefont{Krauth}}, \bibnamefont{and}
  \bibinfo{author}{\bibfnamefont{M.~J.} \bibnamefont{Rozenberg}},
  \bibinfo{journal}{Rev. Mod. Phys.} \textbf{\bibinfo{volume}{68}},
  \bibinfo{pages}{13} (\bibinfo{year}{1996}).

\bibitem[{\citenamefont{Merino and McKenzie}(2000)}]{MerinoMcKenzie}
\bibinfo{author}{\bibfnamefont{J.}~\bibnamefont{Merino}} \bibnamefont{and}
  \bibinfo{author}{\bibfnamefont{R.~H.} \bibnamefont{McKenzie}},
  \bibinfo{journal}{Phys. Rev. B} \textbf{\bibinfo{volume}{61}},
  \bibinfo{pages}{7996} (\bibinfo{year}{2000}),
  \urlprefix\url{http://link.aps.org/doi/10.1103/PhysRevB.61.7996}.

\bibitem[{\citenamefont{Xu et~al.}(2013)\citenamefont{Xu, Haule, and
  Kotliar}}]{xu_hidden_prl_2013}
\bibinfo{author}{\bibfnamefont{W.}~\bibnamefont{Xu}},
  \bibinfo{author}{\bibfnamefont{K.}~\bibnamefont{Haule}}, \bibnamefont{and}
  \bibinfo{author}{\bibfnamefont{G.}~\bibnamefont{Kotliar}},
  \bibinfo{journal}{Phys. Rev. Lett.} \textbf{\bibinfo{volume}{111}},
  \bibinfo{pages}{036401} (\bibinfo{year}{2013}),
  \urlprefix\url{http://link.aps.org/doi/10.1103/PhysRevLett.111.036401}.

\bibitem[{\citenamefont{{P{\' a}lsson} and
  {Kotliar}}(1998)}]{palsson_thermo_1998_prl}
\bibinfo{author}{\bibfnamefont{G.}~\bibnamefont{{P{\' a}lsson}}}
  \bibnamefont{and}
  \bibinfo{author}{\bibfnamefont{G.}~\bibnamefont{{Kotliar}}},
  \bibinfo{journal}{Phys. Rev. Lett.} \textbf{\bibinfo{volume}{80}},
  \bibinfo{pages}{4775} (\bibinfo{year}{1998}).

\bibitem[{\citenamefont{Palsson}(2001)}]{palsson_phd}
\bibinfo{author}{\bibfnamefont{G.}~\bibnamefont{Palsson}}, Ph.D. thesis,
  \bibinfo{school}{Rutgers University} (\bibinfo{year}{2001}).

\bibitem[{\citenamefont{Vu\ifmmode \check{c}\else \v{c}\fi{}i\ifmmode
  \check{c}\else \v{c}\fi{}evi\ifmmode~\acute{c}\else \'{c}\fi{}
  et~al.}(2015)\citenamefont{Vu\ifmmode \check{c}\else \v{c}\fi{}i\ifmmode
  \check{c}\else \v{c}\fi{}evi\ifmmode~\acute{c}\else \'{c}\fi{},
  Tanaskovi\ifmmode~\acute{c}\else \'{c}\fi{}, Rozenberg, and
  Dobrosavljevi\ifmmode~\acute{c}\else \'{c}\fi{}}}]{vucicevic_badmetal_prl}
\bibinfo{author}{\bibfnamefont{J.}~\bibnamefont{Vu\ifmmode \check{c}\else
  \v{c}\fi{}i\ifmmode \check{c}\else \v{c}\fi{}evi\ifmmode~\acute{c}\else
  \'{c}\fi{}}},
  \bibinfo{author}{\bibfnamefont{D.}~\bibnamefont{Tanaskovi\ifmmode~\acute{c}\else
  \'{c}\fi{}}}, \bibinfo{author}{\bibfnamefont{M.~J.} \bibnamefont{Rozenberg}},
  \bibnamefont{and}
  \bibinfo{author}{\bibfnamefont{V.}~\bibnamefont{Dobrosavljevi\ifmmode~\acute{c}\else
  \'{c}\fi{}}}, \bibinfo{journal}{Phys. Rev. Lett.}
  \textbf{\bibinfo{volume}{114}}, \bibinfo{pages}{246402}
  (\bibinfo{year}{2015}),
  \urlprefix\url{http://link.aps.org/doi/10.1103/PhysRevLett.114.246402}.

\bibitem[{\citenamefont{{Hartnoll}}(2009)}]{hartnoll_lectures}
\bibinfo{author}{\bibfnamefont{S.~A.} \bibnamefont{{Hartnoll}}},
  \bibinfo{journal}{Classical and Quantum Gravity}
  \textbf{\bibinfo{volume}{26}}, \bibinfo{eid}{224002} (\bibinfo{year}{2009}),
  \eprint{arXiv:0903.3246}.

\bibitem[{\citenamefont{Hartnoll}(2015)}]{Hartnoll}
\bibinfo{author}{\bibfnamefont{S.}~\bibnamefont{Hartnoll}},
  \bibinfo{journal}{Nature Phys.} \textbf{\bibinfo{volume}{11}},
  \bibinfo{pages}{54} (\bibinfo{year}{2015}).

\bibitem[{\citenamefont{Pakhira and McKenzie}(2015)}]{pakhira15}
\bibinfo{author}{\bibfnamefont{N.}~\bibnamefont{Pakhira}} \bibnamefont{and}
  \bibinfo{author}{\bibfnamefont{R.~H.} \bibnamefont{McKenzie}},
  \bibinfo{journal}{Phys. Rev. B} \textbf{\bibinfo{volume}{91}},
  \bibinfo{pages}{075124} (\bibinfo{year}{2015}),
  \urlprefix\url{http://link.aps.org/doi/10.1103/PhysRevB.91.075124}.

\bibitem[{\citenamefont{Plischke}(1974)}]{Plischke}
\bibinfo{author}{\bibfnamefont{M.}~\bibnamefont{Plischke}},
  \bibinfo{journal}{J. Stat. Phys.} \textbf{\bibinfo{volume}{11}},
  \bibinfo{pages}{159} (\bibinfo{year}{1974}).

\bibitem[{\citenamefont{Kubo and Tada}(1983)}]{KuboTada}
\bibinfo{author}{\bibfnamefont{K.}~\bibnamefont{Kubo}} \bibnamefont{and}
  \bibinfo{author}{\bibfnamefont{M.}~\bibnamefont{Tada}},
  \bibinfo{journal}{Progress of Theoretical Physics}
  \textbf{\bibinfo{volume}{69}}, \bibinfo{pages}{1345} (\bibinfo{year}{1983}),
  \eprint{http://ptp.oxfordjournals.org/content/69/5/1345.full.pdf+html},
  \urlprefix\url{http://ptp.oxfordjournals.org/content/69/5/1345.abstract}.

\bibitem[{\citenamefont{Kubo and Tada}(1984)}]{KuboTada2}
\bibinfo{author}{\bibfnamefont{K.}~\bibnamefont{Kubo}} \bibnamefont{and}
  \bibinfo{author}{\bibfnamefont{M.}~\bibnamefont{Tada}},
  \bibinfo{journal}{Progress of Theoretical Physics}
  \textbf{\bibinfo{volume}{71}}, \bibinfo{pages}{479} (\bibinfo{year}{1984}),
  \eprint{http://ptp.oxfordjournals.org/content/71/3/479.full.pdf+html},
  \urlprefix\url{http://ptp.oxfordjournals.org/content/71/3/479.abstract}.

\bibitem[{\citenamefont{Pan and Wang}(1991)}]{PanWang}
\bibinfo{author}{\bibfnamefont{K.-K.} \bibnamefont{Pan}} \bibnamefont{and}
  \bibinfo{author}{\bibfnamefont{Y.-L.} \bibnamefont{Wang}},
  \bibinfo{journal}{Phys. Rev. B} \textbf{\bibinfo{volume}{43}},
  \bibinfo{pages}{3706} (\bibinfo{year}{1991}),
  \urlprefix\url{http://link.aps.org/doi/10.1103/PhysRevB.43.3706}.

\bibitem[{\citenamefont{Bartkowiak and Chao}(1992)}]{Bartkowiak}
\bibinfo{author}{\bibfnamefont{M.}~\bibnamefont{Bartkowiak}} \bibnamefont{and}
  \bibinfo{author}{\bibfnamefont{K.~A.} \bibnamefont{Chao}},
  \bibinfo{journal}{Phys. Rev. B} \textbf{\bibinfo{volume}{46}},
  \bibinfo{pages}{9228} (\bibinfo{year}{1992}),
  \urlprefix\url{http://link.aps.org/doi/10.1103/PhysRevB.46.9228}.

\bibitem[{\citenamefont{Putikka et~al.}(1998)\citenamefont{Putikka, Luchini,
  and Singh}}]{PLS}
\bibinfo{author}{\bibfnamefont{W.~O.} \bibnamefont{Putikka}},
  \bibinfo{author}{\bibfnamefont{M.~U.} \bibnamefont{Luchini}},
  \bibnamefont{and} \bibinfo{author}{\bibfnamefont{R.~R.~P.}
  \bibnamefont{Singh}}, \bibinfo{journal}{Phys. Rev. Lett.}
  \textbf{\bibinfo{volume}{81}}, \bibinfo{pages}{2966} (\bibinfo{year}{1998}),
  \urlprefix\url{http://link.aps.org/doi/10.1103/PhysRevLett.81.2966}.

\bibitem[{\citenamefont{Scarola et~al.}(2009)\citenamefont{Scarola, Pollet,
  Oitmaa, and Troyer}}]{Scarolaetal}
\bibinfo{author}{\bibfnamefont{V.~W.} \bibnamefont{Scarola}},
  \bibinfo{author}{\bibfnamefont{L.}~\bibnamefont{Pollet}},
  \bibinfo{author}{\bibfnamefont{J.}~\bibnamefont{Oitmaa}}, \bibnamefont{and}
  \bibinfo{author}{\bibfnamefont{M.}~\bibnamefont{Troyer}},
  \bibinfo{journal}{Phys. Rev. Lett.} \textbf{\bibinfo{volume}{102}},
  \bibinfo{pages}{135302} (\bibinfo{year}{2009}),
  \urlprefix\url{http://link.aps.org/doi/10.1103/PhysRevLett.102.135302}.

\bibitem[{\citenamefont{De~Leo et~al.}(2011)\citenamefont{De~Leo, Bernier,
  Kollath, Georges, and Scarola}}]{hiTDMFT}
\bibinfo{author}{\bibfnamefont{L.}~\bibnamefont{De~Leo}},
  \bibinfo{author}{\bibfnamefont{J.-S.} \bibnamefont{Bernier}},
  \bibinfo{author}{\bibfnamefont{C.}~\bibnamefont{Kollath}},
  \bibinfo{author}{\bibfnamefont{A.}~\bibnamefont{Georges}}, \bibnamefont{and}
  \bibinfo{author}{\bibfnamefont{V.~W.} \bibnamefont{Scarola}},
  \bibinfo{journal}{Phys. Rev. A} \textbf{\bibinfo{volume}{83}},
  \bibinfo{pages}{023606} (\bibinfo{year}{2011}),
  \urlprefix\url{http://link.aps.org/doi/10.1103/PhysRevA.83.023606}.

\bibitem[{\citenamefont{Pairault et~al.}(2000)\citenamefont{Pairault,
  S\'en\'echal, and Tremblay}}]{Pairault}
\bibinfo{author}{\bibfnamefont{S.}~\bibnamefont{Pairault}},
  \bibinfo{author}{\bibfnamefont{D.}~\bibnamefont{S\'en\'echal}},
  \bibnamefont{and} \bibinfo{author}{\bibfnamefont{A.-M.}
  \bibnamefont{Tremblay}}, \bibinfo{journal}{Eur. Phys. J. B}
  \textbf{\bibinfo{volume}{16}}, \bibinfo{pages}{85} (\bibinfo{year}{2000}).

\bibitem[{\citenamefont{Khatami et~al.}(2014)\citenamefont{Khatami,
  Perepelitsky, Rigol, and Shastry}}]{Ehsan}
\bibinfo{author}{\bibfnamefont{E.}~\bibnamefont{Khatami}},
  \bibinfo{author}{\bibfnamefont{E.}~\bibnamefont{Perepelitsky}},
  \bibinfo{author}{\bibfnamefont{M.}~\bibnamefont{Rigol}}, \bibnamefont{and}
  \bibinfo{author}{\bibfnamefont{B.~S.} \bibnamefont{Shastry}},
  \bibinfo{journal}{Phys. Rev. E} \textbf{\bibinfo{volume}{89}},
  \bibinfo{pages}{063301} (\bibinfo{year}{2014}),
  \urlprefix\url{http://link.aps.org/doi/10.1103/PhysRevE.89.063301}.

\bibitem[{\citenamefont{Khatami et~al.}(2013)\citenamefont{Khatami, Hansen,
  Perepelitsky, Rigol, and Shastry}}]{hiTECFL}
\bibinfo{author}{\bibfnamefont{E.}~\bibnamefont{Khatami}},
  \bibinfo{author}{\bibfnamefont{D.}~\bibnamefont{Hansen}},
  \bibinfo{author}{\bibfnamefont{E.}~\bibnamefont{Perepelitsky}},
  \bibinfo{author}{\bibfnamefont{M.}~\bibnamefont{Rigol}}, \bibnamefont{and}
  \bibinfo{author}{\bibfnamefont{B.~S.} \bibnamefont{Shastry}},
  \bibinfo{journal}{Phys. Rev. B} \textbf{\bibinfo{volume}{87}},
  \bibinfo{pages}{161120} (\bibinfo{year}{2013}),
  \urlprefix\url{http://link.aps.org/doi/10.1103/PhysRevB.87.161120}.

\bibitem[{\citenamefont{Shastry}(2011)}]{ECFL}
\bibinfo{author}{\bibfnamefont{B.~S.} \bibnamefont{Shastry}},
  \bibinfo{journal}{Phys. Rev. Lett.} \textbf{\bibinfo{volume}{107}},
  \bibinfo{pages}{056403} (\bibinfo{year}{2011}),
  \urlprefix\url{http://link.aps.org/doi/10.1103/PhysRevLett.107.056403}.

\bibitem[{\citenamefont{Hansen and Shastry}(2013)}]{ECFL2ndorder}
\bibinfo{author}{\bibfnamefont{D.}~\bibnamefont{Hansen}} \bibnamefont{and}
  \bibinfo{author}{\bibfnamefont{B.~S.} \bibnamefont{Shastry}},
  \bibinfo{journal}{Phys. Rev. B} \textbf{\bibinfo{volume}{87}},
  \bibinfo{pages}{245101} (\bibinfo{year}{2013}),
  \urlprefix\url{http://link.aps.org/doi/10.1103/PhysRevB.87.245101}.

\bibitem[{\citenamefont{Shastry and Perepelitsky}(2016)}]{ECFLcutoff}
\bibinfo{author}{\bibfnamefont{B.~S.} \bibnamefont{Shastry}} \bibnamefont{and}
  \bibinfo{author}{\bibfnamefont{E.}~\bibnamefont{Perepelitsky}},
  \bibinfo{journal}{Phys. Rev. B} \textbf{\bibinfo{volume}{94}},
  \bibinfo{pages}{045138} (\bibinfo{year}{2016}),
  \urlprefix\url{http://link.aps.org/doi/10.1103/PhysRevB.94.045138}.

\bibitem[{\citenamefont{Basko et~al.}(2006)\citenamefont{Basko, Aleiner, and
  Altshuler}}]{MBLT1}
\bibinfo{author}{\bibfnamefont{D.}~\bibnamefont{Basko}},
  \bibinfo{author}{\bibfnamefont{I.}~\bibnamefont{Aleiner}}, \bibnamefont{and}
  \bibinfo{author}{\bibfnamefont{B.}~\bibnamefont{Altshuler}},
  \bibinfo{journal}{Annals of Physics} \textbf{\bibinfo{volume}{321}},
  \bibinfo{pages}{1126 } (\bibinfo{year}{2006}), ISSN
  \bibinfo{issn}{0003-4916},
  \urlprefix\url{http://www.sciencedirect.com/science/article/pii/S0003491605002630}.

\bibitem[{\citenamefont{Pal and Huse}(2010)}]{MBLT2}
\bibinfo{author}{\bibfnamefont{A.}~\bibnamefont{Pal}} \bibnamefont{and}
  \bibinfo{author}{\bibfnamefont{D.~A.} \bibnamefont{Huse}},
  \bibinfo{journal}{Phys. Rev. B} \textbf{\bibinfo{volume}{82}},
  \bibinfo{pages}{174411} (\bibinfo{year}{2010}),
  \urlprefix\url{http://link.aps.org/doi/10.1103/PhysRevB.82.174411}.

\bibitem[{\citenamefont{Maldague}(1977)}]{maldague_fsumrule}
\bibinfo{author}{\bibfnamefont{P.~F.} \bibnamefont{Maldague}},
  \bibinfo{journal}{Phys. Rev. B} \textbf{\bibinfo{volume}{16}},
  \bibinfo{pages}{2437} (\bibinfo{year}{1977}),
  \urlprefix\url{http://link.aps.org/doi/10.1103/PhysRevB.16.2437}.

\bibitem[{\citenamefont{Bari et~al.}(1970)\citenamefont{Bari, Adler, and
  Lange}}]{Bari_fsumrule}
\bibinfo{author}{\bibfnamefont{R.~A.} \bibnamefont{Bari}},
  \bibinfo{author}{\bibfnamefont{D.}~\bibnamefont{Adler}}, \bibnamefont{and}
  \bibinfo{author}{\bibfnamefont{R.~V.} \bibnamefont{Lange}},
  \bibinfo{journal}{Phys. Rev. B} \textbf{\bibinfo{volume}{2}},
  \bibinfo{pages}{2898} (\bibinfo{year}{1970}),
  \urlprefix\url{http://link.aps.org/doi/10.1103/PhysRevB.2.2898}.

\bibitem[{\citenamefont{Sadakata and Hanamura}(1973)}]{Sadakata_fsumrule}
\bibinfo{author}{\bibfnamefont{I.}~\bibnamefont{Sadakata}} \bibnamefont{and}
  \bibinfo{author}{\bibfnamefont{E.}~\bibnamefont{Hanamura}},
  \bibinfo{journal}{Journal of the Physical Society of Japan}
  \textbf{\bibinfo{volume}{34}}, \bibinfo{pages}{882} (\bibinfo{year}{1973}),
  \eprint{http://dx.doi.org/10.1143/JPSJ.34.882},
  \urlprefix\url{http://dx.doi.org/10.1143/JPSJ.34.882}.

\bibitem[{\citenamefont{Jakli\v{c} and
  Prelov\v{s}ek}(2000)}]{prelovsek_fsumrule}
\bibinfo{author}{\bibfnamefont{J.}~\bibnamefont{Jakli\v{c}}} \bibnamefont{and}
  \bibinfo{author}{\bibfnamefont{P.}~\bibnamefont{Prelov\v{s}ek}},
  \bibinfo{journal}{Advances in Physics} \textbf{\bibinfo{volume}{49}},
  \bibinfo{pages}{1} (\bibinfo{year}{2000}),
  \urlprefix\url{http://dx.doi.org/10.1080/000187300243381}.

\bibitem[{\citenamefont{Shastry}(2006)}]{shastry_fsumrule}
\bibinfo{author}{\bibfnamefont{B.~S.} \bibnamefont{Shastry}},
  \bibinfo{journal}{Phys. Rev. B} \textbf{\bibinfo{volume}{73}},
  \bibinfo{pages}{085117} (\bibinfo{year}{2006}),
  \urlprefix\url{http://link.aps.org/doi/10.1103/PhysRevB.73.085117}.

\bibitem[{\citenamefont{Kokalj}(2015)}]{Kokalj}
\bibinfo{author}{\bibfnamefont{J.}~\bibnamefont{Kokalj}}
  (\bibinfo{year}{2015}), \bibinfo{note}{arXiv: 1512.00875}.

\bibitem[{\citenamefont{Kohn}(1964)}]{kohn_twist}
\bibinfo{author}{\bibfnamefont{W.}~\bibnamefont{Kohn}}, \bibinfo{journal}{Phys.
  Rev.} \textbf{\bibinfo{volume}{133}}, \bibinfo{pages}{A171}
  (\bibinfo{year}{1964}),
  \urlprefix\url{http://link.aps.org/doi/10.1103/PhysRev.133.A171}.

\bibitem[{\citenamefont{Shastry and Sutherland}(1990)}]{shastry_sutherland}
\bibinfo{author}{\bibfnamefont{B.~S.} \bibnamefont{Shastry}} \bibnamefont{and}
  \bibinfo{author}{\bibfnamefont{B.}~\bibnamefont{Sutherland}},
  \bibinfo{journal}{Phys. Rev. Lett.} \textbf{\bibinfo{volume}{65}},
  \bibinfo{pages}{243} (\bibinfo{year}{1990}),
  \urlprefix\url{http://link.aps.org/doi/10.1103/PhysRevLett.65.243}.

\bibitem[{\citenamefont{{Khurana}}(1990)}]{khurana_vertex}
\bibinfo{author}{\bibfnamefont{A.}~\bibnamefont{{Khurana}}},
  \bibinfo{journal}{Phys. Rev. Lett.} \textbf{\bibinfo{volume}{64}},
  \bibinfo{pages}{1990} (\bibinfo{year}{1990}).

\bibitem[{\citenamefont{Jakli\ifmmode~\check{c}\else \v{c}\fi{} and
  Prelov\ifmmode~\check{s}\else
  \v{s}\fi{}ek}(1994)}]{jaklic_prelovsek_prb_1994}
\bibinfo{author}{\bibfnamefont{J.}~\bibnamefont{Jakli\ifmmode~\check{c}\else
  \v{c}\fi{}}} \bibnamefont{and}
  \bibinfo{author}{\bibfnamefont{P.}~\bibnamefont{Prelov\ifmmode~\check{s}\else
  \v{s}\fi{}ek}}, \bibinfo{journal}{Phys. Rev. B}
  \textbf{\bibinfo{volume}{50}}, \bibinfo{pages}{7129} (\bibinfo{year}{1994}),
  \urlprefix\url{http://link.aps.org/doi/10.1103/PhysRevB.50.7129}.

\bibitem[{\citenamefont{Jakli\ifmmode~\check{c}\else \v{c}\fi{} and
  Prelov\ifmmode~\check{s}\else
  \v{s}\fi{}ek}(1995)}]{jaklic_prelovsek_prb_1995}
\bibinfo{author}{\bibfnamefont{J.}~\bibnamefont{Jakli\ifmmode~\check{c}\else
  \v{c}\fi{}}} \bibnamefont{and}
  \bibinfo{author}{\bibfnamefont{P.}~\bibnamefont{Prelov\ifmmode~\check{s}\else
  \v{s}\fi{}ek}}, \bibinfo{journal}{Phys. Rev. B}
  \textbf{\bibinfo{volume}{52}}, \bibinfo{pages}{6903} (\bibinfo{year}{1995}),
  \urlprefix\url{http://link.aps.org/doi/10.1103/PhysRevB.52.6903}.

\bibitem[{\citenamefont{Gull et~al.}(2011)\citenamefont{Gull, Millis,
  Lichtenstein, Rubtsov, Troyer, and Werner}}]{gull_rmp_2011}
\bibinfo{author}{\bibfnamefont{E.}~\bibnamefont{Gull}},
  \bibinfo{author}{\bibfnamefont{A.~J.} \bibnamefont{Millis}},
  \bibinfo{author}{\bibfnamefont{A.~I.} \bibnamefont{Lichtenstein}},
  \bibinfo{author}{\bibfnamefont{A.~N.} \bibnamefont{Rubtsov}},
  \bibinfo{author}{\bibfnamefont{M.}~\bibnamefont{Troyer}}, \bibnamefont{and}
  \bibinfo{author}{\bibfnamefont{P.}~\bibnamefont{Werner}},
  \bibinfo{journal}{Rev. Mod. Phys.} \textbf{\bibinfo{volume}{83}},
  \bibinfo{pages}{349} (\bibinfo{year}{2011}).

\bibitem[{\citenamefont{Kadanoff and Martin}(1963)}]{KadanoffMartin}
\bibinfo{author}{\bibfnamefont{L.~P.} \bibnamefont{Kadanoff}} \bibnamefont{and}
  \bibinfo{author}{\bibfnamefont{P.~C.} \bibnamefont{Martin}},
  \bibinfo{journal}{Annals of Physics} \textbf{\bibinfo{volume}{24}},
  \bibinfo{pages}{419} (\bibinfo{year}{1963}).

\bibitem[{\citenamefont{Fetter and Walecka}(2003)}]{FW}
\bibinfo{author}{\bibfnamefont{A.~L.} \bibnamefont{Fetter}} \bibnamefont{and}
  \bibinfo{author}{\bibfnamefont{J.~D.} \bibnamefont{Walecka}},
  \emph{\bibinfo{title}{Quantum Theory of Many-Particle Systems}}
  (\bibinfo{publisher}{Dover Publications, INC. Mineola, NY},
  \bibinfo{year}{2003}).

\bibitem[{\citenamefont{Perepelitsky}(2013)}]{Edward}
\bibinfo{author}{\bibfnamefont{E.}~\bibnamefont{Perepelitsky}}
  (\bibinfo{year}{2013}), \bibinfo{note}{arXiv: 1310.3797}.

\bibitem[{\citenamefont{Metzner}(1991)}]{Metzner}
\bibinfo{author}{\bibfnamefont{W.}~\bibnamefont{Metzner}},
  \bibinfo{journal}{Phys. Rev. B} \textbf{\bibinfo{volume}{43}},
  \bibinfo{pages}{8549} (\bibinfo{year}{1991}),
  \urlprefix\url{http://link.aps.org/doi/10.1103/PhysRevB.43.8549}.

\bibitem[{\citenamefont{Limelette et~al.}(2003)\citenamefont{Limelette,
  Wzietek, Florens, Georges, Costi, Pasquier, J\'erome, M\'ezi\`ere, and
  Batail}}]{limelette_prl_2003}
\bibinfo{author}{\bibfnamefont{P.}~\bibnamefont{Limelette}},
  \bibinfo{author}{\bibfnamefont{P.}~\bibnamefont{Wzietek}},
  \bibinfo{author}{\bibfnamefont{S.}~\bibnamefont{Florens}},
  \bibinfo{author}{\bibfnamefont{A.}~\bibnamefont{Georges}},
  \bibinfo{author}{\bibfnamefont{T.~A.} \bibnamefont{Costi}},
  \bibinfo{author}{\bibfnamefont{C.}~\bibnamefont{Pasquier}},
  \bibinfo{author}{\bibfnamefont{D.}~\bibnamefont{J\'erome}},
  \bibinfo{author}{\bibfnamefont{C.}~\bibnamefont{M\'ezi\`ere}},
  \bibnamefont{and} \bibinfo{author}{\bibfnamefont{P.}~\bibnamefont{Batail}},
  \bibinfo{journal}{Phys. Rev. Lett.} \textbf{\bibinfo{volume}{91}},
  \bibinfo{pages}{016401} (\bibinfo{year}{2003}),
  \urlprefix\url{http://link.aps.org/doi/10.1103/PhysRevLett.91.016401}.

\bibitem[{\citenamefont{Kurosaki et~al.}(2005)\citenamefont{Kurosaki, Shimizu,
  Miyagawa, Kanoda, and Saito}}]{Kurosakietal}
\bibinfo{author}{\bibfnamefont{Y.}~\bibnamefont{Kurosaki}},
  \bibinfo{author}{\bibfnamefont{Y.}~\bibnamefont{Shimizu}},
  \bibinfo{author}{\bibfnamefont{K.}~\bibnamefont{Miyagawa}},
  \bibinfo{author}{\bibfnamefont{K.}~\bibnamefont{Kanoda}}, \bibnamefont{and}
  \bibinfo{author}{\bibfnamefont{G.}~\bibnamefont{Saito}},
  \bibinfo{journal}{Phys. Rev. Lett.} \textbf{\bibinfo{volume}{95}},
  \bibinfo{pages}{177001} (\bibinfo{year}{2005}),
  \urlprefix\url{http://link.aps.org/doi/10.1103/PhysRevLett.95.177001}.

\bibitem[{\citenamefont{Pruschke et~al.}(1993)\citenamefont{Pruschke, Cox, and
  Jarrell}}]{PCJ}
\bibinfo{author}{\bibfnamefont{T.}~\bibnamefont{Pruschke}},
  \bibinfo{author}{\bibfnamefont{D.~L.} \bibnamefont{Cox}}, \bibnamefont{and}
  \bibinfo{author}{\bibfnamefont{M.}~\bibnamefont{Jarrell}},
  \bibinfo{journal}{EPL (Europhysics Letters)} \textbf{\bibinfo{volume}{21}},
  \bibinfo{pages}{593} (\bibinfo{year}{1993}),
  \urlprefix\url{http://stacks.iop.org/0295-5075/21/i=5/a=015}.

\bibitem[{\citenamefont{Jarrell et~al.}(1995)\citenamefont{Jarrell, Freericks,
  and Pruschke}}]{JFP}
\bibinfo{author}{\bibfnamefont{M.}~\bibnamefont{Jarrell}},
  \bibinfo{author}{\bibfnamefont{J.~K.} \bibnamefont{Freericks}},
  \bibnamefont{and} \bibinfo{author}{\bibfnamefont{T.}~\bibnamefont{Pruschke}},
  \bibinfo{journal}{Phys. Rev. B} \textbf{\bibinfo{volume}{51}},
  \bibinfo{pages}{11704} (\bibinfo{year}{1995}),
  \urlprefix\url{http://link.aps.org/doi/10.1103/PhysRevB.51.11704}.

\bibitem[{\citenamefont{Merino et~al.}(2008)\citenamefont{Merino, Dumm,
  Drichko, Dressel, and McKenzie}}]{Merino_2008}
\bibinfo{author}{\bibfnamefont{J.}~\bibnamefont{Merino}},
  \bibinfo{author}{\bibfnamefont{M.}~\bibnamefont{Dumm}},
  \bibinfo{author}{\bibfnamefont{N.}~\bibnamefont{Drichko}},
  \bibinfo{author}{\bibfnamefont{M.}~\bibnamefont{Dressel}}, \bibnamefont{and}
  \bibinfo{author}{\bibfnamefont{R.~H.} \bibnamefont{McKenzie}},
  \bibinfo{journal}{Phys. Rev. Lett.} \textbf{\bibinfo{volume}{100}},
  \bibinfo{pages}{086404} (\bibinfo{year}{2008}),
  \urlprefix\url{http://link.aps.org/doi/10.1103/PhysRevLett.100.086404}.

\bibitem[{\citenamefont{Kuramoto and Watanabe}(1987)}]{Kuramoto}
\bibinfo{author}{\bibfnamefont{Y.}~\bibnamefont{Kuramoto}} \bibnamefont{and}
  \bibinfo{author}{\bibfnamefont{T.}~\bibnamefont{Watanabe}},
  \bibinfo{journal}{Physica B} \textbf{\bibinfo{volume}{148}},
  \bibinfo{pages}{80} (\bibinfo{year}{1987}).

\bibitem[{\citenamefont{M\"uller-Hartmann}(1989)}]{Muller-Hartmann}
\bibinfo{author}{\bibfnamefont{E.}~\bibnamefont{M\"uller-Hartmann}},
  \bibinfo{journal}{Z. Phys. B} \textbf{\bibinfo{volume}{74}},
  \bibinfo{pages}{507} (\bibinfo{year}{1989}).

\bibitem[{mom()}]{momentsweb}
\bibinfo{note}{See
  \url{https://www.cpht.polytechnique.fr/cpht/correl/documents/repository_hiT.htm}
  for the higher order moments.}

\bibitem[{\citenamefont{Mori}(1965{\natexlab{a}})}]{Mori}
\bibinfo{author}{\bibfnamefont{H.}~\bibnamefont{Mori}},
  \bibinfo{journal}{Progr. Theor. Phys.} \textbf{\bibinfo{volume}{33}},
  \bibinfo{pages}{423} (\bibinfo{year}{1965}{\natexlab{a}}).

\bibitem[{\citenamefont{Mori}(1965{\natexlab{b}})}]{Mori2}
\bibinfo{author}{\bibfnamefont{H.}~\bibnamefont{Mori}},
  \bibinfo{journal}{Progr. Theor. Phys.} \textbf{\bibinfo{volume}{34}},
  \bibinfo{pages}{339} (\bibinfo{year}{1965}{\natexlab{b}}).

\bibitem[{\citenamefont{Lovesey and Meserve}(1972)}]{Lovesey}
\bibinfo{author}{\bibfnamefont{S.~W.} \bibnamefont{Lovesey}} \bibnamefont{and}
  \bibinfo{author}{\bibfnamefont{R.~A.} \bibnamefont{Meserve}},
  \bibinfo{journal}{Phys. Rev. Lett.} \textbf{\bibinfo{volume}{28}},
  \bibinfo{pages}{614} (\bibinfo{year}{1972}),
  \urlprefix\url{http://link.aps.org/doi/10.1103/PhysRevLett.28.614}.

\bibitem[{\citenamefont{Bulla et~al.}(2008)\citenamefont{Bulla, Costi, and
  Pruschke}}]{bulla2008}
\bibinfo{author}{\bibfnamefont{R.}~\bibnamefont{Bulla}},
  \bibinfo{author}{\bibfnamefont{T.}~\bibnamefont{Costi}}, \bibnamefont{and}
  \bibinfo{author}{\bibfnamefont{T.}~\bibnamefont{Pruschke}},
  \bibinfo{journal}{Rev. Mod. Phys.} \textbf{\bibinfo{volume}{80}},
  \bibinfo{pages}{395} (\bibinfo{year}{2008}).

\bibitem[{\citenamefont{\v{Z}itko and Pruschke}(2009)}]{zitko2009}
\bibinfo{author}{\bibfnamefont{R.}~\bibnamefont{\v{Z}itko}} \bibnamefont{and}
  \bibinfo{author}{\bibfnamefont{T.}~\bibnamefont{Pruschke}},
  \bibinfo{journal}{Phys. Rev. B} \textbf{\bibinfo{volume}{79}},
  \bibinfo{pages}{085106} (\bibinfo{year}{2009}).

\bibitem[{\citenamefont{\ifmmode~\check{Z}\else \v{Z}\fi{}itko
  et~al.}(2013)\citenamefont{\ifmmode~\check{Z}\else \v{Z}\fi{}itko, Hansen,
  Perepelitsky, Mravlje, Georges, and Shastry}}]{DMFT-ECFL}
\bibinfo{author}{\bibfnamefont{R.}~\bibnamefont{\ifmmode~\check{Z}\else
  \v{Z}\fi{}itko}}, \bibinfo{author}{\bibfnamefont{D.}~\bibnamefont{Hansen}},
  \bibinfo{author}{\bibfnamefont{E.}~\bibnamefont{Perepelitsky}},
  \bibinfo{author}{\bibfnamefont{J.}~\bibnamefont{Mravlje}},
  \bibinfo{author}{\bibfnamefont{A.}~\bibnamefont{Georges}}, \bibnamefont{and}
  \bibinfo{author}{\bibfnamefont{B.~S.} \bibnamefont{Shastry}},
  \bibinfo{journal}{Phys. Rev. B} \textbf{\bibinfo{volume}{88}},
  \bibinfo{pages}{235132} (\bibinfo{year}{2013}),
  \urlprefix\url{http://link.aps.org/doi/10.1103/PhysRevB.88.235132}.

\bibitem[{\citenamefont{Shastry}(2009)}]{shastry_electrothermal}
\bibinfo{author}{\bibfnamefont{B.~S.} \bibnamefont{Shastry}},
  \bibinfo{journal}{Reports on Progress in Physics}
  \textbf{\bibinfo{volume}{72}}, \bibinfo{pages}{016501}
  (\bibinfo{year}{2009}),
  \urlprefix\url{http://stacks.iop.org/0034-4885/72/i=1/a=016501}.

\bibitem[{\citenamefont{Chaikin and Beni}(1976)}]{chaikin_beni_1976}
\bibinfo{author}{\bibfnamefont{P.~M.} \bibnamefont{Chaikin}} \bibnamefont{and}
  \bibinfo{author}{\bibfnamefont{G.}~\bibnamefont{Beni}},
  \bibinfo{journal}{Phys. Rev. B} \textbf{\bibinfo{volume}{13}},
  \bibinfo{pages}{647} (\bibinfo{year}{1976}).

\bibitem[{\citenamefont{Bari\v{s}i\'c et~al.}(2013)\citenamefont{Bari\v{s}i\'c,
  Chan, Li, Yu, Zhao, Dressel, Smontara, and Greven}}]{barisic13}
\bibinfo{author}{\bibfnamefont{N.}~\bibnamefont{Bari\v{s}i\'c}},
  \bibinfo{author}{\bibfnamefont{M.~K.} \bibnamefont{Chan}},
  \bibinfo{author}{\bibfnamefont{Y.}~\bibnamefont{Li}},
  \bibinfo{author}{\bibfnamefont{G.}~\bibnamefont{Yu}},
  \bibinfo{author}{\bibfnamefont{X.}~\bibnamefont{Zhao}},
  \bibinfo{author}{\bibfnamefont{M.}~\bibnamefont{Dressel}},
  \bibinfo{author}{\bibfnamefont{A.}~\bibnamefont{Smontara}}, \bibnamefont{and}
  \bibinfo{author}{\bibfnamefont{M.}~\bibnamefont{Greven}},
  \bibinfo{journal}{PNAS} \textbf{\bibinfo{volume}{110}},
  \bibinfo{pages}{12235} (\bibinfo{year}{2013}).

\bibitem[{\citenamefont{Takagi et~al.}(1992)\citenamefont{Takagi, Batlogg, Kao,
  Kwo, Cava, Krajewski, and Peck}}]{takagi_1992}
\bibinfo{author}{\bibfnamefont{H.}~\bibnamefont{Takagi}},
  \bibinfo{author}{\bibfnamefont{B.}~\bibnamefont{Batlogg}},
  \bibinfo{author}{\bibfnamefont{H.~L.} \bibnamefont{Kao}},
  \bibinfo{author}{\bibfnamefont{J.}~\bibnamefont{Kwo}},
  \bibinfo{author}{\bibfnamefont{R.~J.} \bibnamefont{Cava}},
  \bibinfo{author}{\bibfnamefont{J.~J.} \bibnamefont{Krajewski}},
  \bibnamefont{and} \bibinfo{author}{\bibfnamefont{W.~F.} \bibnamefont{Peck}},
  \bibinfo{journal}{Phys. Rev. Lett.} \textbf{\bibinfo{volume}{69}},
  \bibinfo{pages}{2975} (\bibinfo{year}{1992}),
  \urlprefix\url{http://link.aps.org/doi/10.1103/PhysRevLett.69.2975}.

\bibitem[{\citenamefont{Bloch et~al.}(2008)\citenamefont{Bloch, Dalibard, and
  Zwerger}}]{bloch_rmp_2008}
\bibinfo{author}{\bibfnamefont{I.}~\bibnamefont{Bloch}},
  \bibinfo{author}{\bibfnamefont{J.}~\bibnamefont{Dalibard}}, \bibnamefont{and}
  \bibinfo{author}{\bibfnamefont{W.}~\bibnamefont{Zwerger}},
  \bibinfo{journal}{Rev. Mod. Phys.} \textbf{\bibinfo{volume}{80}},
  \bibinfo{pages}{885} (\bibinfo{year}{2008}),
  \urlprefix\url{http://link.aps.org/doi/10.1103/RevModPhys.80.885}.

\bibitem[{\citenamefont{Schneider et~al.}(2012)\citenamefont{Schneider,
  Hackermüller, Ronzheimer, Will, Braun, Best, Bloch, Demler, Mandt, Rasch
  et~al.}}]{schneider12}
\bibinfo{author}{\bibfnamefont{U.}~\bibnamefont{Schneider}},
  \bibinfo{author}{\bibfnamefont{L.}~\bibnamefont{Hackermüller}},
  \bibinfo{author}{\bibfnamefont{J.~P.} \bibnamefont{Ronzheimer}},
  \bibinfo{author}{\bibfnamefont{S.}~\bibnamefont{Will}},
  \bibinfo{author}{\bibfnamefont{S.}~\bibnamefont{Braun}},
  \bibinfo{author}{\bibfnamefont{T.}~\bibnamefont{Best}},
  \bibinfo{author}{\bibfnamefont{I.}~\bibnamefont{Bloch}},
  \bibinfo{author}{\bibfnamefont{E.}~\bibnamefont{Demler}},
  \bibinfo{author}{\bibfnamefont{S.}~\bibnamefont{Mandt}},
  \bibinfo{author}{\bibfnamefont{D.}~\bibnamefont{Rasch}},
  \bibnamefont{et~al.}, \bibinfo{journal}{Nat. Phys.}
  \textbf{\bibinfo{volume}{8}}, \bibinfo{pages}{213} (\bibinfo{year}{2012}).

\end{thebibliography}
\end{document}